\documentclass[english,15pt]{article}
\usepackage[T1]{fontenc}
\usepackage[latin1]{inputenc}
\usepackage{geometry}
\usepackage{float}
\usepackage{mathrsfs}
\usepackage{amsmath}
\usepackage{amssymb}
\usepackage{graphicx}
\usepackage{hyperref}
\usepackage[color=yellow]{todonotes}
\hypersetup{
    colorlinks=true,
    linkcolor=blue,
    filecolor=magenta,      
    urlcolor=cyan,
    citecolor = red,
}

\usepackage{yhmath}
\makeatletter

\usepackage{bm}
\numberwithin{equation}{section}

\floatstyle{ruled}
\usepackage{algorithm}
\usepackage{algpseudocode}
\usepackage{caption}
\usepackage{needspace}
\DeclareCaptionFormat{algor}{%
  \needspace{6\baselineskip}\hrulefill\par\offinterlineskip\vskip1pt%
    \textbf{#1#2:} #3\offinterlineskip\hrulefill}
\DeclareCaptionStyle{algori}{singlelinecheck=off,format=algor,labelsep=space}

\usepackage{algorithm}
 \usepackage{algorithmicx}

\usepackage{amsthm}

\usepackage{mathrsfs}
\usepackage{amssymb}

\usepackage{amsfonts}

\usepackage{epsfig}

\usepackage{bm}

\usepackage{mathrsfs}

\usepackage{enumerate}

\@ifundefined{definecolor}{\@ifundefined{definecolor}
 {\@ifundefined{definecolor}
 {\usepackage{color}}{}
}{}
}{}

\usepackage{subfig}\usepackage[all]{xy}
\usepackage{booktabs}
\usepackage{multirow}

\newtheorem{rem}{Remark}[section]
\newtheorem{exam}{Example}[section]

\newcounter{hypA}

\newcounter{hypB}

\newcounter{hypD}

\usepackage[nameinlink, capitalise]{cleveref}

\providecommand{\algorithmname}{Algorithm}
\floatname{algorithm}{\protect\algorithmname}

\usepackage{babel}\date{}

\makeatother

\usepackage{diagbox}

\begin{document}

\begin{center}

{\Large \textbf{Two Localization Strategies for Sequential MCMC Data Assimilation with Applications to Nonlinear Non-Gaussian Geophysical Models}}

\vspace{0.5cm}

Hamza Ruzayqat${}^1$, Hristo G.~Chipilski${}^2$, and Omar Knio${}^1$\\
\smallskip
{\footnotesize ${}^1$Applied Mathematics and Computational Science Program, \\ Computer, Electrical and Mathematical Sciences and Engineering Division, \\ King Abdullah University of Science and Technology, Thuwal, 23955-6900, KSA.} \\
{\footnotesize ${}^2$Department of Scientific Computing, Florida State University, \\ 400 Dirac Science Library, Tallahassee, FL 32306-4120, USA.} \\
{\footnotesize E-Mail:\,}  \texttt{\emph{\footnotesize hamza.ruzayqat@kaust.edu.sa}},  \texttt{\emph{\footnotesize hchipilski@fsu.edu}},  \texttt{\emph{\footnotesize omar.knio@kaust.edu.sa}} 

\begin{abstract}
We present a localized data assimilation (DA) scheme based on sequential Markov Chain Monte Carlo (SMCMC) \cite{smcmc}, a provably convergent method for filtering high-dimensional, nonlinear, and potentially non-Gaussian state-space models. Unlike particle filters, which are exact for such models, SMCMC does not assign weights to samples, avoiding weight degeneracy in small-ensemble regimes. We design two localization approaches within the SMCMC framework that exploit observation sparsity to reduce effective degrees of freedom. The first variant collects observed blocks into a single reduced domain and runs parallel MCMC chains over this combined region. The second further reduces per-chain state dimension by decomposing the observed region into independent blocks with compact halos, applying Gaspari--Cohn tapering to down-weight distant observations. When the observation model is linear Gaussian, the approximate filtering density reduces to a Gaussian mixture from which independent samples can be drawn exactly; otherwise, we employ an MCMC kernel. We test on high-dimensional ($d \sim 10^4 - 10^5$) state-space models, including a linear Gaussian model and a nonlinear multilayer shallow water equation with linear and nonlinear observation operators. We consider Gaussian and Student-$t$ observation noise, showing that LSMCMC handles heavy-tailed errors that cause ensemble Kalman methods to diverge. Observations include synthetic and real SWOT (NASA) and ocean drifter (NOAA) data. We compare the two variants against each other and the local ensemble transform Kalman filter (LETKF).

\bigskip

\\
\noindent \textbf{MSC classes}:	62M20, 60G35, 60J20, 94A12, 93E11, 65C40
\\
\noindent \textbf{Corresponding author}: Hamza Ruzayqat. E-mail:
\href{mailto:hamza.ruzayqat@kaust.edu.sa}{hamza.ruzayqat@kaust.edu.sa}  \\
\noindent \textbf{Code:} \url{https://github.com/ruzayqat/LSMCMC}
\end{abstract}
\end{center}

\section{Introduction}
Data assimilation (DA) plays a crucial role in understanding and predicting complex systems states as accurately as possible. By combining partial, noisy observational data with numerical models, DA aims to predict the conditional probability distribution, or the filter distribution, of underlying hidden state variables that cannot be directly observed. In practice, data assimilation is widely utilized across various fields. In finance, it helps in predicting market trends and managing risks by improving the accuracy of financial models. In engineering, it is used to optimize systems and processes by refining predictions based on real-time data. In geophysical sciences, data assimilation is crucial for applications such as numerical weather prediction, where it provides vital information for forecasting weather patterns, understanding atmospheric conditions, and modeling oceanographic phenomena. These applications are particularly useful for paleoclimate reconstructions, ensuring maritime safety, and predicting natural hazards like tsunamis \cite{tsunami}. For a comprehensive introduction on DA and stochastic filtering see \cite{bain,cappe,jazwinski_1970,cohn_1997}. For a deeper exploration of specific applications and methodologies see \cite{carrassi, ghil, kalnay_mote_da_2024}.

Ensemble methods, such as the ensemble Kalman filter (EnKF) \cite{evensen1, houtekamer, evensen2} and sequential Monte Carlo methods such as particle filters (PFs) \cite{PF1, delm13, leeuwen}, are powerful DA tools that have gained significant attention in recent work; see \cite{vetra} for a review of DA methods. These probabilistic methods, which can be formalized within a sequential Bayesian inference framework, use multiple ensemble members or particles to represent and propagate the system's state dynamics, in conjunction with the available observations. Despite their inherent computational efficiency and their wide usage, ensemble Kalman methods can be inaccurate when dealing with strongly nonlinear and non-Gaussian models. Additionally, they have a tendency to significantly underestimate the overall uncertainty in the system in small ensemble regimes \cite{pires, storto, vetra}. PFs, on the other hand, are exact methods (see e.g. \cite{delmoral2004}) that assign weights to the ensemble members (also called particles) that reflect the likelihood of the observations. As the number of particles approaches infinity, the PF estimates the conditional distribution of the hidden state exactly, regardless of the type of noise present in the system or whether the underlying dynamical model is linear or non-linear. The computational complexity of PFs is fixed per-time-step update. However, in order to prevent the so-called weight degeneracy (see e.g. \cite{bengtsson}), the number of particles must grow exponentially with the dimensionality of the state space. This is the primary limitation of standard PF approaches, as they are typically only practical for low-dimensional problems. Several works in the literature have explored different techniques such as tempering, lagging and nudging to tackle this issue; see e.g. \cite{ades, laggedPF, cotter, kantas}. 

Another exact method for filtering is Markov chain Monte Carlo (MCMC) \cite{cappe}. MCMC can be used to sample from the filter distribution, however, its computational complexity grows linearly with time, which makes it impractical for real-time applications. A possible solution to the linear growth of complexity is to use sequential MCMC. At each assimilation time step, one builds an MCMC chain initialized from one random sample at the previous time step. Sequential MCMC has been used in dynamic contexts in several works. For instance, in the work of \cite{septier}, a sequential MCMC based on Hamiltonian Monte Carlo (HMC) \cite{HMC} was presented. The authors use HMC to build a chain of $N$ samples from the posterior initialized from a uniformly picked sample from the previous time step. Another example is the work of \cite{carmi}, where the authors use Sequential MCMC for dynamical cluster tracking.
 
In \cite{smcmc}, a more practical filtering approach is presented where sequential MCMC chains are used to target the filter distribution at each time-step update. The method was mainly developed for state-space models (SSMs) with unknown random observational locations, but it can be used for general SSMs. The initial idea behind their scheme first appeared in \cite{berzuini} and was later used for point process filtering in \cite{martin}. The authors in \cite{smcmc} demonstrate that when high accuracy is needed and the state dimension is very large, the sequential MCMC filtering scheme (which we refer to as SMCMC from now on) is much more accurate and efficient when compared to ensemble methods. When the forward model noise covariance matrix $Q_k$ (and the observation noise covariance $R_k$) are dense, evaluating the transition and likelihood densities requires $\mathcal{O}(d^2)$ and $\mathcal{O}((d_y^k)^2)$ operations per sample, respectively, due to the matrix--vector products in the quadratic forms (here $d$ is the dimension of the discretized state vector $\mathbf{Z}_{t_k}\in\mathbb{R}^d$ and $d_y^k$ the number of scalar observations at assimilation time $t_k$. In many practical applications, $Q_k$ and $R_k$ are diagonal, so the quadratic forms reduce to element-wise operations and the overall cost per update becomes $\mathcal{O}(N_a d)$, where $N_a$ denotes the number of MCMC (analysis) samples and $d$ the state dimension. More generally, if $Q_k$ and $R_k$ possess exploitable structure (e.g. Toeplitz/circulant or block-circulant) for stationary kernels on regular grids, separable Kronecker structure on tensor-product grids, or hierarchical off-diagonal low-rank structure, then matrix--vector products can be carried out in $\mathcal{O}(d\log d)$ (or near-linear) time using FFT-based or hierarchical algorithms, leading to an overall complexity of $\mathcal{O}(N_a d\log d)$ per update.

In many real-world applications, observations on the system are either very sparse in the domain of interest or localized in particular regions. Many localization techniques have been developed in the literature to leverage this, thereby improving the overall performance. In \cite{gaspari}, the authors introduced a distance-based tapering function which is compactly supported that can be multiplied by the covariance matrix or its estimation to reduce the influence of observations that are spatially far from the state variables being updated. Even when most of the state is densely observed, there is still a need to suppress the spurious long-range correlations introduced by finite-sample covariance estimates, which is another motivation to use localization. This type of techniques have been largely used in the literature for improving ensemble methods performance, such as EnKF or ensemble transform Kalman filter; see e.g. \cite{houtekamer, houtekamer01, hamill, whitaker02, anderson03, ott, miyoshi, greybush}. Whereas fixed localization functions have been widely used, it was shown that adaptive localization, which dynamically adjusts the localization radius length scale based on the spatial correlation structure, is very effective in practice, e.g. \cite{anderson, anderson12}. Furthermore, localization and adaptive localization have been used in PFs as well. An approach is to apply localization directly to the weights and resampling process aiming to reduce the path degeneracy and improving the effective sample size. The localization is applied to the particles update ensuring that only observations within a distance influence each particle. However, this is not straightforward because of the so-called ``gluing'' problem: after local resampling, the locally selected particles must be recombined into globally consistent states, which can introduce discontinuities at block boundaries and remains largely unsolved; see \cite{poterjoy, farchi}.

In this work, we build upon the SMCMC approach in \cite{smcmc} and propose two localized variants. Both exploit the spatial sparsity of observations by partitioning the domain into subdomains and restricting the MCMC updates to regions where observations exist, thereby reducing the effective state dimension from $d$ to $d' < d$. Since the main cost of the SMCMC filter comes from evaluating pointwise the transition and likelihood densities, which involves vector-matrix multiplications, both variants dramatically reduce this cost while maintaining high accuracy. The first variant collects all observed subdomains into a combined reduced domain and runs parallel MCMC chains over it. The second variant further reduces the per-chain dimension by decomposing the observed region into independent blocks, each augmented with a compact halo, and applying Gaspari--Cohn observation-noise tapering to smoothly down-weight distant observations. The cost order of the new algorithms is $O(N_ad')$ with $d' < d$. 

The contributions of this article are as follows:
\begin{itemize}
\item We introduce two localization techniques for the SMCMC filter based on the sparsity of the observations: (a) a joint localization scheme (Algorithm~\ref{alg:loc_smcmc_v1}) that collects all observed subdomains into a single reduced domain and runs parallel MCMC chains over this combined region, and (b) a halo-based per-block localization (Algorithm~\ref{alg:loc_smcmc_v2}) that decomposes the problem into independent, fully parallel MCMC chains over each observed block with Gaspari--Cohn observation-noise tapering to smoothly down-weight distant observations; see Figures~\ref{fig:v1_illustration} and~\ref{fig:v2_illustration} for an illustration.

\item We carefully distinguish between the number of forecast samples $N_f$ and the number of analysis (MCMC) samples $N_a$, and show that when the observation model is linear and Gaussian, the filtering density is a Gaussian mixture from which $N_a$ independent samples can be drawn exactly without MCMC (Example~\ref{ex:linear_gaussian}).

\item For nonlinear observation models, we employ an MCMC kernel (see Example~\ref{ex:nonlinear_mcmc}) and MCMC chains are run in parallel.

\item We apply the method on several challenging SSMs with both linear and nonlinear observations models, including the nonlinear multilayer shallow water equations. We further demonstrate robustness to non-Gaussian observation noise by employing Cauchy-distributed (Student-$t$, $\nu{=}1$) errors, motivated by the heavy-tailed error distributions observed in real ocean drifter data \cite{elipot2016}.

\item We demonstrate the methodology on filtering problems with two types of observations: (a) synthetic and real data from the SWOT mission led by NASA \cite{nasa, swot}, and (b) real ocean drifter observations from NOAA \cite{noaa}.
\end{itemize}

This article is structured as follows.  Section~\ref{sec:model_prel} describes the class of SSMs considered in this paper. In Section~\ref{sec:smcmc} we provide a brief review of the SMCMC filter and discuss the direct Gaussian mixture sampling for linear-Gaussian observation models and the choice of MCMC kernel for nonlinear or non-Gaussian observation models. Section~\ref{sec:lsmcmc} proposes the localization SMCMC algorithm in two variants: a joint scheme and a halo-based per-block scheme with Gaspari--Cohn tapering. In Section~\ref{sec:numerics} we demonstrate the methodology on several SSMs. Finally, in Section~\ref{sec:conclusion} we conclude with a summary of key results and a discussion of possible future directions. 

\section{Preliminaries}\label{sec:model}
\label{sec:model_prel}
Consider a two-dimensional grid $\mathsf{G}$ of size $N_g\in \mathbb{N}$. $\mathsf{G}$ is typically a connected region such as a square or a rectangle with $N_g = N_g^x \times N_g^y$, where $N_g^x, N_g^y \in \mathbb{N}$. Assume there is an unknown continuous time stochastic process $(\mathbf{Z}_t)_{t\geq 0}$, with $\mathbf{Z}_t\in\mathbb{R}^d$, and $d\in \mathbb{N}$ is a multiple of $N_g^x N_g^y$ (depending on the number of prognostic variables defined on each grid point). We will consider noisy observations that are arriving at times, $\mathsf{T}:=\{t_1,t_2,\dots\}\subset \mathbb{N}$, $0<t_1<t_2<\cdots$, at a collection of known locations. Let $m_k\in \mathbb{N}$ be the number of observed locations at time $t_k$. Then, at the observed location $j\in \{1, \cdots, m_k\}$, the observation is denoted by $Y_{t_k}^{(j)} \in \mathbb{R}^{s}$, $s\in \mathbb{N}$ the number of observed variables per observational location. We denote by $\mathbf{Y}_{t_k}\in \mathbb{R}^{d_y^k}$, $d_y^k = sm_k$, the collection of observations at time $t_k$. We have a superscript $k$ attached to $d_y$ to indicate that the number of available observations per assimilation cycle may change from time to time. We will assume the time-lag between two consecutive observations to be $L\in \mathbb{N}$, that is $t_{k+1} - t_k = L\,\Delta t$, where $\Delta t$ is the fixed model time step.

The modeling approach for the process $\mathbf{Z}_t$ is based on a continuous-time framework, motivated by applications in atmospheric and ocean sciences. In these domains, physical quantities such as height, wind, or water velocity are described using continuous-time, spatially-varying physical models represented by systems of partial differential equations (PDEs). The continuous PDE state is formally infinite-dimensional; throughout this article $d$ denotes the dimension of the \emph{spatially discretized} state on which the algorithm operates. To account for model uncertainty, the dynamics of $\mathbf{Z}_t$ are modeled using a stochastic process, where the noise can be incorporated either continuously (as in stochastic PDEs) or discretely in time.

The framework requires that at observation times $t_k \in \mathsf{T}$, the process $(\mathbf{Z}_{t_k})_{k\geq 1}$ is Markov with a known, positive transition density $f_k(\mathbf{z}_{t_{k-1}},\mathbf{z}_{t_k})$. The initial state $\mathbf{Z}_0$ is assumed known. Each observation $\mathbf{Y}_{t_k}$ depends only on $\mathbf{Z}_{t_k}$ through a positive likelihood density $g_k(\mathbf{z}_{t_k},\mathbf{y}_{t_k})$. Here, the comma-separated arguments denote conditional densities, i.e., $g_k(\mathbf{z}_{t_k},\mathbf{y}_{t_k})\equiv g_k(\mathbf{y}_{t_k}\mid\mathbf{z}_{t_k})$ and $f_k(\mathbf{z}_{t_{k-1}},\mathbf{z}_{t_k})\equiv f_k(\mathbf{z}_{t_k}\mid\mathbf{z}_{t_{k-1}})$; this is equivalent to the explicit conditional notation used elsewhere. The subscript $k$ in $f_k$ and $g_k$ allows for time-inhomogeneous dynamics or observation processes.

We assume that $f_k$ (or a suitable approximation) can be evaluated pointwise. This includes stochastic differential equations and their time-discretization approximations, stochastic PDEs, or PDEs with discrete-time additive noise. A state-space model that fits this setup and applies to all examples in this article is presented below.

\textbf{Hidden signal:}
We will consider $\mathbf{Z}_t$ to be a vector containing hidden state variables at positions defined on $\mathsf{G}$. Let $\Phi:\mathbb{R}^d \times \mathbb{R}_+^2 \to \mathbb{R}^d$ be some linear or nonlinear function defined on its domain, then for $k\in \mathbb{N}$:
\begin{equation}
\label{eq:signal_unobserved}
\mathbf{Z}_t  = \Phi(\mathbf{Z}_{t_{k-1}},t_{k-1}; t), \qquad t \in (t_{k-1},t_k),
\end{equation}
and at time of observation $t_k$:
\begin{equation}
\label{eq:signal_observed}
\mathbf{Z}_{t_k} = \Phi(\mathbf{Z}_{t_{k-1}}, t_{k-1}; t_k) + \mathbf{W}_{t_k},
\end{equation}
where $\mathbf{W}_{t_k}\stackrel{\textrm{\emph{i.i.d.}}}{\sim}\mathcal{N}_d(0,Q_k)$ is an i.i.d.~sequence of $d-$dimensional Gaussian random variables of zero mean and covariance matrix $Q_k$. This does not imply that the physical system is noise-free between observations. $\mathbf{W}_{t_k}$ represents model error accumulated over the assimilation window $(t_{k-1},t_k]$  as is standard in discrete-time state-space formulations. The same total process noise can equivalently be distributed across the $L$ forward sub-steps without changing the filtering recursion.

\textbf{Observations:} In addition, the following observational model is considered
\begin{equation}
\label{eq:obs_model}
\mathbf{Y}_{t_k} = \mathscr{O}_{t_k}(\mathbf{Z}_{t_k}) + \mathbf{V}_{t_k},
\end{equation}
where $\mathscr{O}_{t_k}:\mathbb{R}^d\to \mathbb{R}^{d_y^k}$ is an $\mathbb{R}^{d_y^k}$-vector valued function and $\{\mathbf{V}_{t_k}\}$ is an i.i.d.~noise sequence with known density. In most experiments, $\mathbf{V}_{t_k} \sim \mathcal{N}_{d_y^k}(0,R_k)$; however, the LSMCMC framework accommodates arbitrary observation noise distributions, and in Section~\ref{subsec:nlgamma_experiment} we consider heavy-tailed Cauchy ($t_1$) errors to demonstrate this generality.

Throughout this work, we restrict ourselves to the case where both $Q_k$ and $R_k$ are \emph{diagonal} matrices. This is a natural assumption in many applications where the noise components are independent across spatial locations and observed variables. Under this assumption, evaluating the transition density $f_k$ and the likelihood $g_k$ involves only element-wise operations, reducing the per-sample cost of the SMCMC algorithm to $\mathcal{O}(d + d_y^k)$. This diagonal assumption is a practical modeling choice rather than a general property. Genuine model error (e.g., from unresolved parameterizations) can be spatially and cross-variable correlated, so $Q_k$ need not be diagonal. In many geophysical applications, however, these correlations are local, resulting in a covariance matrix that is banded or approximately banded, with only a limited number of significant off-diagonal entries near the main diagonal. Hence, the localized blocks introduced in Section~\ref{sec:lsmcmc} can be enlarged with additional halo rings to accommodate a banded $Q_k$ while remaining computationally efficient. Furthermore, when $Q_k$ admits a spectral factorization (e.g., circulant or Toeplitz structure from stationary spatial correlations), matrix-vector products involving $Q_k^{-1}$ in the transition density $f_k$ can be evaluated with FFTs at $\mathcal{O}(d\log d)$ rather than $\mathcal{O}(d^2)$ cost, preserving near-linear per-sample complexity even for correlated model error.

\section{SMCMC Filter}
\label{sec:smcmc}
Inference of the hidden state is performed using conditional distributions given the available observations. One can either consider the whole path trajectory $\mathbb{P}(\mathbf{Z}_{t_1},\ldots,\mathbf{Z}_{t_k}|(\mathbf{Y}_{t_p})_{p\leq {k}})$ (smoothing) or just the marginal $\mathbb{P}(\mathbf{Z}_{t_k}|(\mathbf{Y}_{t_p})_{p\leq {k}})$ (filtering).
For $k\in\mathbb{N}$ we define the smoothing density (the dependence on observations is dropped from the notation)
\begin{equation}
\label{eq:smoothing_density}
\begin{aligned}
\Pi_k(\mathbf{z}_{t_{1}},\ldots, \mathbf{z}_{t_k}) &\propto \prod_{p=1}^k f_p(\mathbf{z}_{t_{p-1}},\mathbf{z}_{t_p}) g_p(\mathbf{z}_{t_p},\mathbf{y}_{t_p}) \\
&\propto f_k(\mathbf{z}_{t_{k-1}},\mathbf{z}_{t_k}) g_k(\mathbf{z}_{t_k},\mathbf{y}_{t_k}) \Pi_{k-1}(\mathbf{z}_{t_{1:k-1}}). 
\end{aligned}
\end{equation}
Let $\pi_k(\mathbf{z}_{t_k})$ be the marginal in the $\mathbf{z}_{t_k}$ coordinate of the smoothing distribution $\Pi_k$. It easily follows that the filter density can be obtained recursively
\begin{equation}
\label{eq:filter_recursion}
\pi_k(\mathbf{z}_{t_k}) \propto g_k(\mathbf{z}_{t_k},\mathbf{y}_{t_k})\int_{\mathbb{R}^d}f_k(\mathbf{z}_{t_{k-1}},\mathbf{z}_{t_k})\pi_{k-1}(\mathbf{z}_{t_{k-1}})d\mathbf{z}_{t_{k-1}}.    
\end{equation}
The filtering problem we are considering is inherently discrete-time in nature, as the observations are obtained at discrete time instances. However, we can still obtain the conditional probability distribution $\mathbb{P}(\mathbf{Z}_t | (\mathbf{Y}_{t_p})_{p \leq k})$ for any time $t\in (t_k, t_{k+1})$. This can be achieved by integrating the previous filter distribution $\pi_{k-1}$ with the corresponding transition density, and then applying the Chapman-Kolmogorov equations. This allows us to propagate the filter distribution continuously in time, even though the observations are discrete.

The method of \cite{smcmc} efficiently approximates \eqref{eq:filter_recursion} directly via a sequence of MCMC chains each initialized from a previously obtained approximation of $\pi_{k-1}$. 
At time $t_1$ their algorithm targets $\pi_1(\mathbf{z}_{t_1})$ exactly by running an MCMC kernel with invariant distribution $\pi_1(\mathbf{z}_{t_1})$ for $N_a$ steps, that is to draw $N_a$ samples from
\begin{equation}
\label{eq:pi_1}
\pi_1(\mathbf{z}_{t_1}) \propto g_1(\mathbf{z}_{t_1},\mathbf{y}_{t_1})f_1(\mathbf{Z}_0,\mathbf{z}_{t_1}).
\end{equation}
At later times $t_k$, $k\geq 2$, the filter distribution $\pi_{k-1}$ is approximated by an empirical measure constructed from $N_f$ forecast samples carried over from the previous step. In the original SMCMC algorithm of \cite{smcmc}, all $N_a$ samples from the previous MCMC chain serve as forecast samples, i.e., $N_f = N_a$. We introduce a simple but effective modification: we retain only a subset of $N_f \leq N_a$ samples for the forecast step. This decoupling is motivated by the observation that, during the MCMC chain at each cycle, the auxiliary ancestor index $j$ (see \eqref{eq:filter_approx_sum} and \eqref{eq:filter_approx} below) selects among the $N_f$ previous samples, and many MCMC iterates may re-use the same index, so only a fraction of the forecast samples actually influence the posterior. Keeping $N_f$ small therefore economizes on the forecast step, which requires running the forward model $N_f$ times, while still allowing a long MCMC chain ($N_a \gg N_f$) to thoroughly explore the posterior. This is particularly beneficial when the forward model is computationally expensive. A conceptual distinction from ensemble-Kalman methods is that SMCMC analysis members are fresh posterior draws at each cycle and do not carry continuous identities or trajectories through time. The filter tracks the evolving filtering \emph{distribution} (its moments and samples) rather than individual member histories, so reducing $N_a$ to $N_f$ does not sever any otherwise-continuous member trajectory. This reduction from $N_a$ to $N_f$ members can be done by random grouping and averaging. This averages independent posterior \emph{samples} at the same grid points (a Monte-Carlo variance-reduction step), not spatially distinct fields; nonetheless, averaging draws from a non-Gaussian or multimodal posterior can blur sharp features and degrade balance. Our implementation therefore also supports a \emph{subsampling} reduction that retains $N_f$ random draws without averaging, which preserves the full posterior and dynamical balance at the cost of higher sampling variance; the two reductions give statistically indistinguishable RMSE in our experiments, and subsampling is recommended for balance-critical (e.g.\ high-resolution atmospheric) applications.

Concretely, the empirical measure of the $N_f$ forecast samples is 
\begin{equation*}
S_{k-1}^{N_f}(\mathbf{z}_{t_{k-1}}):=\frac{1}{N_f} \sum_{i=1}^{N_f} \delta_{\{\mathbf{Z}_{t_{k-1}}^{(i)}\}}(\mathbf{z}_{t_{k-1}}),
\end{equation*}
where $\delta_{\{\mathbf{Z}_{t_k}\}}(\mathbf{z})$ denotes the Dirac delta measure centered at $\mathbf{Z}_{t_k}$. Substituting $S_{k-1}^{N_f}$ into \eqref{eq:filter_recursion} yields the approximate filtering density
\begin{equation}
\label{eq:filter_approx_sum}
\pi_k^{N_f} (\mathbf{z}_{t_k})\propto g_k(\mathbf{z}_{t_k},\mathbf{y}_{t_k}) \frac{1}{N_f}\sum_{j=1}^{N_f} f_k(\mathbf{Z}_{t_{k-1}}^{(j)}, \mathbf{z}_{t_k} ).
\end{equation}
This density is the marginal of the joint distribution
\begin{equation}
\label{eq:filter_approx}
\pi_k^{N_f}(\mathbf{z}_{t_k},j) \propto g_k(\mathbf{z}_{t_k},\mathbf{y}_{t_k}) f_k(\mathbf{Z}_{t_{k-1}}^{(j)},\mathbf{z}_{t_k})p(j),
\end{equation}
where $j\in\{1,\ldots,N_f\}$ is an auxiliary ancestor index with uniform prior $p(j)=1/N_f$. The SMCMC algorithm then draws $N_a$ samples from $\pi_k^{N_f}(\mathbf{z}_{t_k},j)$ using an MCMC kernel, with $N_a \geq N_f$. Here, ``sequential'' refers to cycling the filter over assimilation times $t_k$; the sampler used \emph{within} each cycle is itself an intrinsically sequential Markov-chain procedure.

For a given function $\varphi:\mathbb{R}^d\rightarrow\mathbb{R}$ that is $\pi_k$-integrable, one is interested in estimating expectations w.r.t.\ the filtering distribution at observation times $(t_k)_{k\geq 1}$. These expectations are given by $\pi_k(\varphi):= \int_{\mathbb{R}^d}\varphi(\mathbf{z}_{t_k}) \pi_k(\mathbf{z}_{t_k})\, d\mathbf{z}_{t_k}$. After obtaining $N_a$ samples from $\pi_k^{N_f}$ in \eqref{eq:filter_approx}, the expectation $\pi_k(\varphi)$ can be estimated through 
\begin{equation*}
\widehat{\pi}_k^{N_a}(\varphi) := \frac{1}{N_a} \sum_{i=1}^{N_a} \varphi(\mathbf{Z}_{t_k}^{(i)}),
\end{equation*}
which as $N_a\to \infty$ converges to the true expectation $\pi_k(\varphi)$ almost surely \cite{smcmc}. In this paper, we consider $\varphi(z)=z$.

\subsection*{Choice of MCMC Kernel}
\label{subsec:mcmc_kernels}

The choice of the MCMC kernel used to sample from $\pi_k^{N_f}$ in \eqref{eq:filter_approx} (or its localized version \eqref{eq:filter_local}) depends critically on the structure of the observation model $\mathscr{O}_{t_k}$. We describe two important cases below.

We emphasize that when the observation model is linear and Gaussian, the filtering density takes the form of a Gaussian mixture from which $N_a$ independent samples can be drawn directly without any MCMC iterations, as shown in the following example. This observation, which was not exploited in the original SMCMC framework of \cite{smcmc}, eliminates burn-in and inter-sample correlation in this important special case.

\begin{exam}[Linear-Gaussian observation model -- direct sampling]
\label{ex:linear_gaussian}
Consider the SSM \eqref{eq:signal_observed}--\eqref{eq:obs_model} with a linear observation operator $\mathscr{O}_{t_k}(\mathbf{z})=C\mathbf{z}$ and Gaussian noises in both the state and observation equations. That is,
\begin{equation}
\label{eq:linear_ssm_ex1}
\begin{aligned}
\mathbf{Z}_{t_k} &= \Phi(\mathbf{Z}_{t_{k-1}}, t_{k-1}; t_k) + \mathbf{W}_{t_k}, \qquad \mathbf{W}_{t_k} \sim \mathcal{N}_d(0, Q_k), \\
\mathbf{Y}_{t_k} &= C\mathbf{Z}_{t_k} + \mathbf{V}_{t_k}, \qquad\qquad\quad\;\;\; \mathbf{V}_{t_k} \sim \mathcal{N}_{d_y^k}(0, R_k),
\end{aligned}
\end{equation}
where $C\in \mathbb{R}^{d_y^k\times d}$. In this case, the transition density for a given ancestor $j \in \{1,\cdots,N_f\}$ is $f_k(\mathbf{Z}_{t_{k-1}}^{(j)}, \mathbf{z}_{t_k}) = \mathcal{N}_d(\mathbf{z}_{t_k};\boldsymbol{\mu}_k^{(j)}, Q_k)$, where $\boldsymbol{\mu}_k^{(j)} := \Phi(\mathbf{Z}_{t_{k-1}}^{(j)}, t_{k-1}; t_k)$, and the likelihood is $g_k(\mathbf{z}_{t_k}, \mathbf{y}_{t_k}) = \mathcal{N}_{d_y^k}(\mathbf{y}_{t_k}; C\mathbf{z}_{t_k}, R_k)$. Since both densities are Gaussian, the product $f_k \cdot g_k$ is proportional to a Gaussian in $\mathbf{z}_{t_k}$ for each fixed $j$. Completing the square yields the conditional $\pi_k^{N_f}(\mathbf{z}_{t_k} \mid j) = \mathcal{N}_d(\mathbf{z}_{t_k}; \boldsymbol{m}_k^{(j)}, \Sigma_k)$, where
\begin{equation}
\label{eq:gm_posterior}
\begin{aligned}
\Sigma_k &= \left(Q_k^{-1} + C^\top R_k^{-1} C\right)^{-1}, \\
\boldsymbol{m}_k^{(j)} &= \Sigma_k\!\left(Q_k^{-1}\boldsymbol{\mu}_k^{(j)} + C^\top R_k^{-1}\mathbf{y}_{t_k}\right).
\end{aligned}
\end{equation}
Marginalizing over $j$ with the uniform prior $p(j) = 1/N_f$, the approximate filter distribution is a Gaussian mixture
\begin{equation}
\label{eq:gaussian_mixture}
\pi_k^{N_f}(\mathbf{z}_{t_k}) = \frac{1}{N_f}\sum_{j=1}^{N_f} w_k^{(j)}\,\mathcal{N}_d\!\left(\mathbf{z}_{t_k}\,;\, \boldsymbol{m}_k^{(j)},\, \Sigma_k\right),
\end{equation}
where the mixture weights $w_k^{(j)} \propto \mathcal{N}_{d_y^k}(\mathbf{y}_{t_k}\,;\, C\boldsymbol{\mu}_k^{(j)},\, C Q_k C^\top + R_k)$ are the marginal likelihoods of each component. In high dimension and \emph{without} localization, these weights generically degenerate (almost all mass concentrates on a single component), so one effectively samples from a single Gaussian. Localization alleviates this by reducing the effective dimension entering each weight and restoring weight diversity across the mixture. Sampling $N_a$ samples from this distribution is straightforward: first draw $j$ from the categorical distribution with probabilities proportional to $\{w_k^{(j)}\}_{j=1}^{N_f}$, then draw $\mathbf{z}_{t_k} \sim \mathcal{N}_d(\boldsymbol{m}_k^{(j)}, \Sigma_k)$. In other words, we are sampling from the filter approximation in \eqref{eq:filter_approx_sum} instead of \eqref{eq:filter_approx}.

Consequently, when the observation model is linear-Gaussian, no MCMC iterations are needed; one can draw independent and exact samples from $\pi_k^{N_f}$ at each assimilation step, which eliminates burn-in and correlation between successive samples. The covariance $\Sigma_k$ and its Cholesky factor can be precomputed once since they do not depend on $j$.
\end{exam}

\begin{exam}[Non-Gaussian likelihood -- MCMC sampling]
\label{ex:nonlinear_mcmc}
When the observation operator $\mathscr{O}_{t_k}$ is nonlinear-Gaussian, or when it is linear but the observation noise is non-Gaussian, the product $f_k \cdot g_k$ is no longer Gaussian in $\mathbf{z}_{t_k}$. Consequently, the closed-form mixture representation in Example~\ref{ex:linear_gaussian} is no longer available (unless one adopts specific parameterizations based on transport maps; see \cite{chipilski_2025}). In particular, even when $\mathscr{O}_{t_k}$ is linear, a non-Gaussian error distribution prevents the likelihood from combining with the Gaussian transition densities into a single tractable distribution, thereby precluding direct sampling. In either setting, one must instead resort to MCMC sampling from the joint distribution $\pi_k^{N_f}(\mathbf{z}_{t_k}, j)$ in \eqref{eq:filter_approx}.

Our implementation supports four MCMC kernels: Random Walk Metropolis with Gibbs (RWM-Gibbs), Preconditioned Crank--Nicolson (pCN) \cite{pCN}, Metropolis-Adjusted Langevin Algorithm (MALA) \cite{MALA}, and Hamiltonian Monte Carlo (HMC) \cite{HMC}.  In all four kernels, the auxiliary variable $j$ can be updated jointly or via a separate Gibbs step, and step-size parameters are adaptively tuned during a burn-in phase to achieve a target acceptance rate; see \cite{septier} for a related discussion.
\end{exam}

\subsection*{Cost of SMCMC Algorithm}
The SMCMC method proved particularly effective and efficient in high-dimensional problems. Since $Q_k$ is assumed diagonal (see Section~\ref{sec:model_prel}), evaluating $f_k$ requires only element-wise operations over $d$ components, respectively. Therefore, the cost of each MCMC step is $\mathcal{O}(d)$ (assuming $d_y^k \ll d)$, and the total cost of the SMCMC algorithm per assimilation cycle is $\mathcal{O}(d\,N_a)$. On a multi-core machine, the element-wise operations are naturally parallelized across cores.


In light of \eqref{eq:signal_unobserved}--\eqref{eq:obs_model}, the SMCMC filtering method of \cite{smcmc} is summarized in Algorithm~\ref{alg:smcmc} in the Appendix.

\section{Localized Sequential MCMC Filter}
\label{sec:lsmcmc}
In many practical applications, the observational locations tend to be very sparse or highly localized in space. As an example of such applications is the set of data obtained from the SWOT mission led by NASA. It aims to provide high-resolution, detailed measurements across large swaths of oceans, lakes, and rivers with unprecedented accuracy collected via innovative interferometric radars. This set of data is expected to enhance predictive DA models by providing accurate initial and boundary conditions as well as accurate observations, thus improving forecasts for climate, weather, and hydrological systems. Another example is the set of ocean data collected via the so-called ocean drifters, which are floating devices equipped with sensors that collect vital data on ocean currents, water velocities, and temperatures as they drift with the flow of the water. These drifters are deployed across the globe, but their distribution is often sparse, especially in remote regions like the polar seas or vast stretches of the open ocean. 

In these situations, filtering the entire signal $\mathbf{Z}_{t_k}$ all at once would be computationally impractical. Instead, a more attractive approach would be to focus the filtering on the specific regions and time points where observations are available, rather than attempting to filter the entire signal as a whole. 
The basic idea behind our localization scheme is to perform a domain localization where the domain $\mathsf{G}$ is partitioned into $\Gamma$ subdomains $\mathsf{G}=\bigcup_{i=1}^{\Gamma} G_i$, and then, to find all subdomains $G_i$'s that contain observational locations and update the state variables in these subdomains. Note that this partition can be time-dependent; in other words, one can increase or decrease the number of subdomains $\Gamma$ at time $t$ to adopt the distribution of observations coming at this particular time. Let $\overline{\mathbf{x}}_{t_k} \subset \mathsf{G}$ denote the collection of grid points inside the subdomains that contain observational locations at time $t_k$. Let $\mathbf{z}_{t_k}(\overline{\mathbf{x}}_{t_k})\in \mathbb{R}^{d_k}$, $d_k < d$, be a sub-vector of the whole hidden signal $\mathbf{z}_{t_k}$ with variables at locations $\overline{\mathbf{x}}_{t_k}$.

\begin{figure}[h!]
\centering
\includegraphics[width=0.65\textwidth]{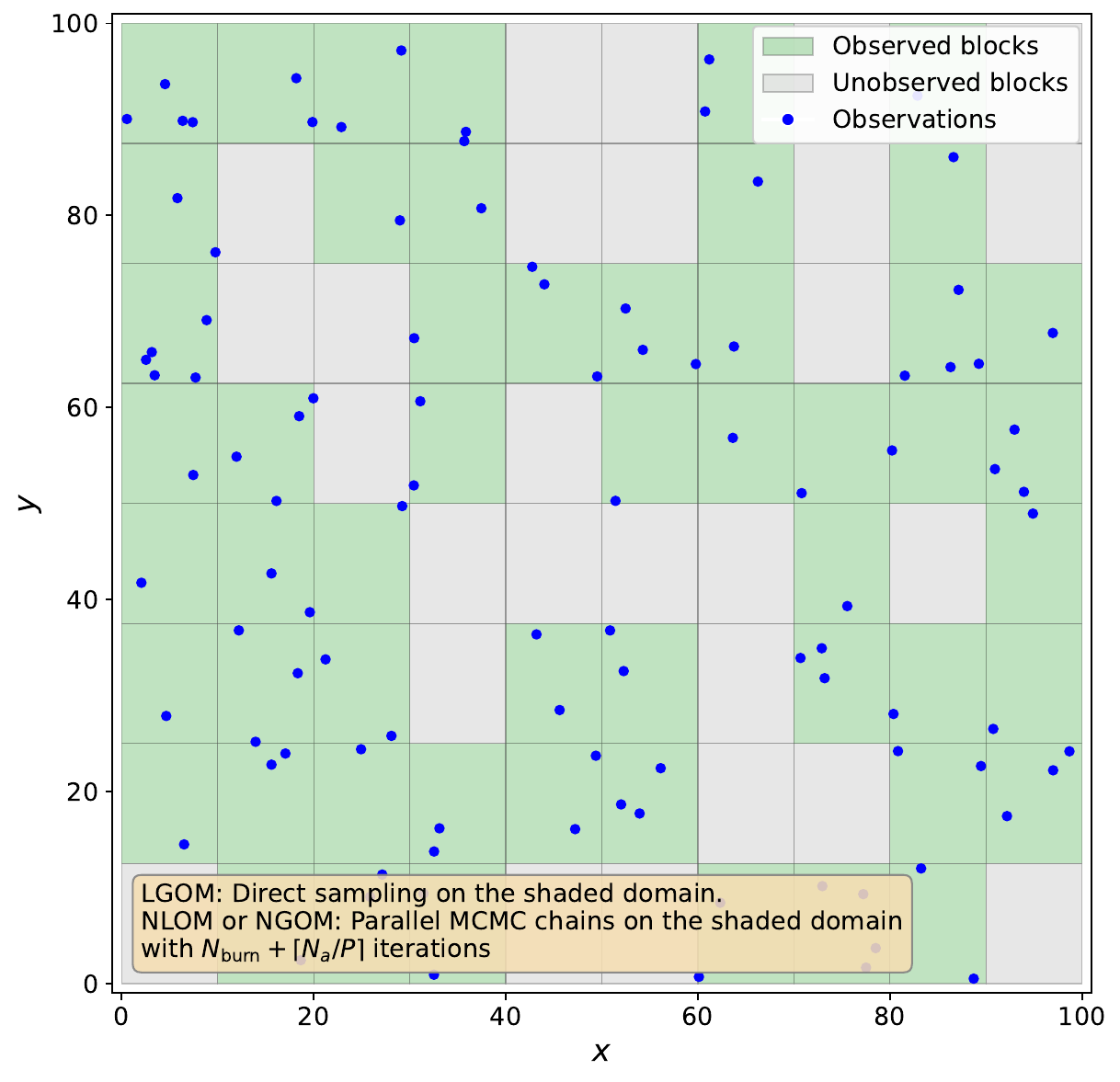}
\caption{Variant~1 (joint observed-block localization) illustrated on a $100\times 100$ grid with 80 blocks and randomly placed observations. Here LGOM = Linear Gaussian Observation Model; NLGOM = Non-Linear Gaussian Observation Model; NGOM = Non-Gaussian Observation Model. Observed blocks (green) are updated jointly; unobserved blocks (gray) retain their forecast values. For an LGOM the posterior is sampled directly; for an NLGOM or NGOM $P$ parallel MCMC chains of length $N_{\text{burn}}+\lceil N_a/P \rceil$ are run on the combined shaded domain.}
\label{fig:v1_illustration}
\end{figure}

\begin{figure}[h!]
\centering
\includegraphics[width=0.65\textwidth]{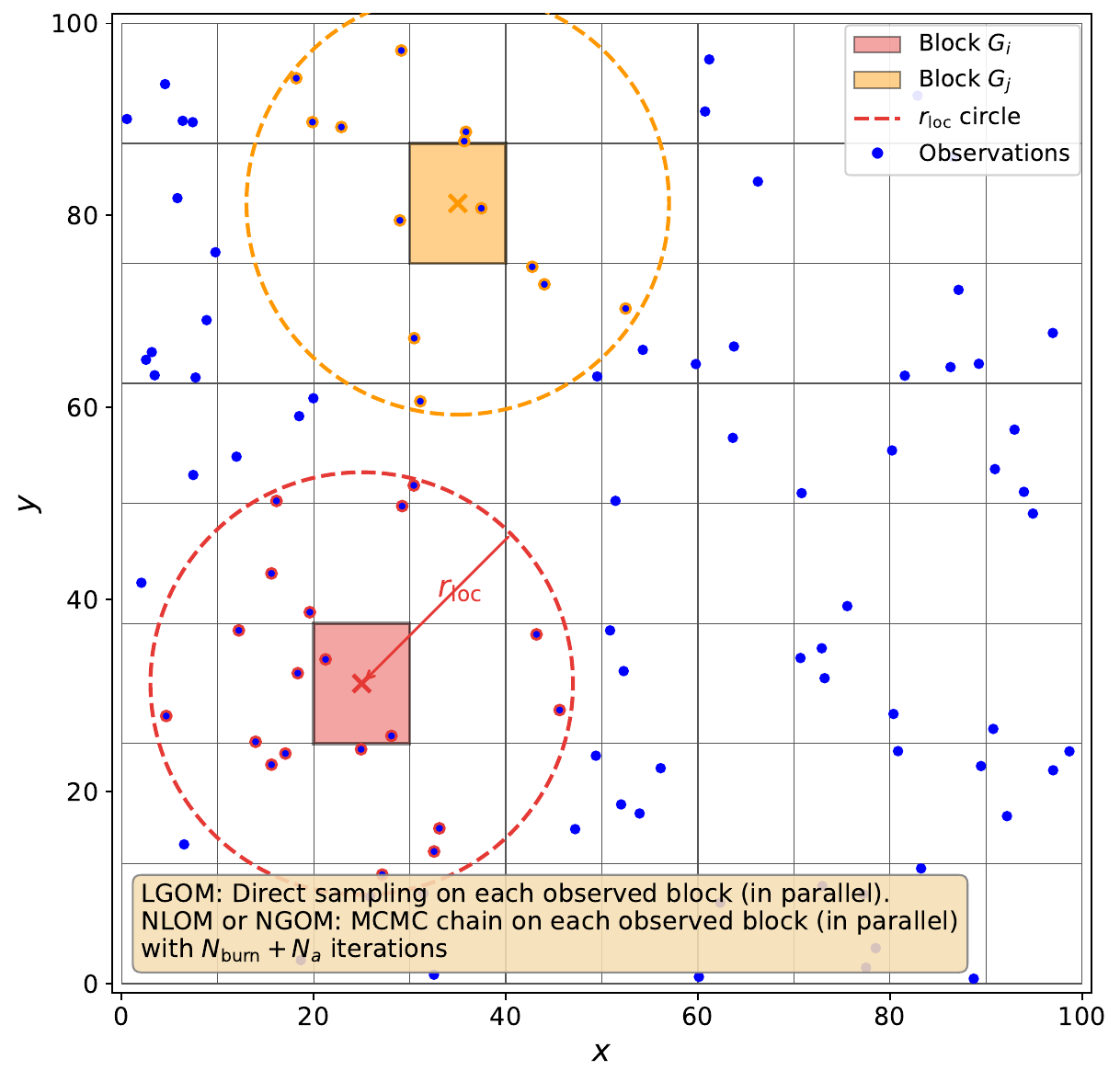}
\caption{Variant~2 (per-block halo localization) illustrated on the same $100\times 100$ grid and observation layout as Figure~\ref{fig:v1_illustration} (LGOM, NLGOM, and NGOM defined therein). Each observed block is updated independently using observations that fall within a halo of radius $r_{\mathrm{loc}}$. For an LGOM the blocks are sampled directly in parallel; for an NLGOM or NGOM each block runs an MCMC chain of length $N_{\text{burn}}+N_a$ in parallel.}
\label{fig:v2_illustration}
\end{figure}

We present two variants of the localized SMCMC filter, which we illustrate in Figures~\ref{fig:v1_illustration} and~\ref{fig:v2_illustration}. Both variants share the common idea of restricting the MCMC updates to the observed region, but they differ in how they decompose the problem and blend the resulting updates.

\subsection{Variant 1: Joint Observed-Block Localization}
\label{subsec:v1_localization}

The first variant collects all subdomains that contain observations into a single combined observed region and generates posterior samples over this reduced domain. Throughout, a tilde ($\,\widetilde{\cdot}$\,) denotes a density restricted to the observed reduced domain $\overline{\mathbf{x}}_{t_k}$, i.e.\ evaluated only on the sub-vector $\mathbf{z}_{t_k}(\overline{\mathbf{x}}_{t_k})$ of state components inside the observed subdomains. Specifically, at the first observation time $t_1$, one draws $N_a$ samples $\{\mathbf{Z}_{t_1}^{(i)}(\overline{\mathbf{x}}_{t_1})\}_{1\leq i\leq N_a}$ from the distribution
\begin{equation}
\label{eq:loc_pi_1}
\widetilde{\pi}_1(\mathbf{z}_{t_1}(\overline{\mathbf{x}}_{t_1})) \propto \widetilde{g}_1(\mathbf{z}_{t_1}(\overline{\mathbf{x}}_{t_1}),\mathbf{y}_{t_1})~\widetilde{f}_1(\mathbf{z}_0(\overline{\mathbf{x}}_{t_1}),\mathbf{z}_{t_1}(\overline{\mathbf{x}}_{t_1})).
\end{equation}
At subsequent times $t_k$, $k>1$, the localized filtering distribution is
\begin{equation}
\label{eq:filter_sum_local}
\widetilde{\pi}_k^{N_f}(\mathbf{z}_{t_k}(\overline{\mathbf{x}}_{t_k})) \propto \widetilde{g}_k(\mathbf{z}_{t_k}(\overline{\mathbf{x}}_{t_k}),\mathbf{y}_{t_k})\;\frac{1}{N_f}\sum_{j=1}^{N_f} \widetilde{f}_k\!\left(\mathbf{Z}_{t_{k-1}}^{(j)}(\overline{\mathbf{x}}_{t_k}),\,\mathbf{z}_{t_k}(\overline{\mathbf{x}}_{t_k})\right).
\end{equation}
When the observation model is linear-Gaussian, \eqref{eq:filter_sum_local} is a Gaussian mixture from which $N_a$ independent samples can be drawn directly (Example~\ref{ex:linear_gaussian}). Otherwise, one introduces the ancestor index~$j$ and draws $N_a$ samples $\{\mathbf{Z}_{t_k}^{(i)}(\overline{\mathbf{x}}_{t_k}),\, j_i\}_{1\leq i \leq N_a}$ from the joint distribution
\begin{equation}
\label{eq:filter_local}
\widetilde{\pi}_k^{N_f}(\mathbf{z}_{t_k}(\overline{\mathbf{x}}_{t_k}),j) \propto \widetilde{g}_k(\mathbf{z}_{t_k}(\overline{\mathbf{x}}_{t_k}),\mathbf{y}_{t_k}) ~\widetilde{f}_k(\mathbf{z}_{t_{k-1}}^{(j)}(\overline{\mathbf{x}}_{t_k}),\mathbf{z}_{t_k}(\overline{\mathbf{x}}_{t_k}))~p(j),
\end{equation}
using an appropriate MCMC kernel (Example~\ref{ex:nonlinear_mcmc}). The density $\widetilde{f}_k$ can be thought of as the restriction of the transition density $f_k$ on subdomains $G_i$'s with locations $\overline{\mathbf{x}}_{t_k}$, and $\widetilde{g}_k$ is the likelihood density resulting from a modified observational model $\widetilde{\mathscr{O}}_{t_k}:\mathbb{R}^{d_k}\to \mathbb{R}^{d_y^k}$. For instance, if $\mathscr{O}_{t_k}=C\mathbf{z}_{t_k}$ for some matrix $C\in \mathbb{R}^{d_y^k\times d}$, then one would modify $C$ by keeping only the columns that correspond to $\mathbf{z}_{t_k}(\overline{\mathbf{x}}_{t_k})$ and have $\widetilde{\mathscr{O}}_{t_k}=\widetilde{C}\mathbf{z}_{t_k}(\overline{\mathbf{x}}_{t_k})$, where $\widetilde{C}\in \mathbb{R}^{d_y^k\times d_k}$ is the new modified matrix. After obtaining the samples $\mathbf{z}_{t_k}(\overline{\mathbf{x}}_{t_k})$, one updates the rest of the variables (that is, update $\mathbf{z}_{t_k}(\overline{\mathbf{x}}_{t_k}^c)$, where $\overline{\mathbf{x}}_{t_k}^c$ refers to the complement $\mathsf{G}\setminus \overline{\mathbf{x}}_{t_k}$) according to the dynamics in the forward model. In light of \eqref{eq:signal_unobserved}--\eqref{eq:obs_model}, for a given $j\in \{1,\cdots,N_f\}$, one would have 
%
\begin{equation}
\widetilde{f}_k\left(\mathbf{Z}_{t_{k-1}}^{(j)}(\overline{\mathbf{x}}_{t_k}),\mathbf{z}_{t_k}(\overline{\mathbf{x}}_{t_k})\right) \propto \exp\left\{-\frac{1}{2}\bigl[\mathbf{z}_{t_k}(\overline{\mathbf{x}}_{t_k})-\overline{\boldsymbol{\mu}}_k^{(j)}\bigr]^\top\widetilde{Q_k^{-1}}\bigl[\mathbf{z}_{t_k}(\overline{\mathbf{x}}_{t_k})-\overline{\boldsymbol{\mu}}_k^{(j)}\bigr]\right\},
\end{equation}
where $\overline{\boldsymbol{\mu}}_k^{(j)}:=\Phi\bigl(\mathbf{Z}_{t_{k-1}}^{(j)},t_{k-1};t_k\bigr)(\overline{\mathbf{x}}_{t_k})$ and $\widetilde{Q_k^{-1}}$ is the matrix with rows and columns of $Q_k^{-1}$ that correspond to the grid subdomains that contain the locations $\overline{\mathbf{x}}_{t_k}$. The steps of this variant are summarized in Algorithm~\ref{alg:loc_smcmc_v1}. We note that the localization described here is not limited to SSMs with Gaussian noise, but it can be applied to any SSM as long as the noise covariance matrix $Q_k$ and its inverse can be computed or estimated. 

%
%
\begin{flushleft}
\needspace{4\baselineskip}
\captionsetup[algorithm]{style=algori}
\captionof{algorithm}{LSMCMC -- Variant 1 (Joint Observed-Block Localization)}
\label{alg:loc_smcmc_v1}
\textbf{Input:} Matrices $Q_k$, $Q_k^{-1}$; partition $\mathsf{G}=\bigcup_{i=1}^{\Gamma} G_i$; initial state $\breve{\mathbf{Z}}_0=\mathbf{z}_0$; observations $\{\mathbf{y}_{t_k}\}_{k\geq 1}$ and their locations; number of forecast samples $N_f$; number of analysis samples $N_a \geq N_f$; time-lag $L$. Set $\tau_k:=(t_k-t_{k-1})/L$.

\begin{enumerate}
\item \textbf{Initialize} ($k=1$):
\begin{enumerate}
\item Forecast: for $l=0,\cdots,L-1$, compute $\breve{\mathbf{Z}}_{(l+1)\tau_1} = \Phi(\breve{\mathbf{Z}}_{l\tau_1},l\tau_1;(l+1)\tau_1)$.
\item Identify the subdomains containing observations at $t_1$; collect $\overline{\mathbf{x}}_{t_1}$ and extract sub-matrix $\widetilde{Q_1^{-1}}$.
\item Add noise: $\tilde{\mathbf{Z}}_{t_1}:=\breve{\mathbf{Z}}_{t_1}+ \mathbf{W}_{t_1}$, $\mathbf{W}_{t_1}\sim \mathcal{N}_d(0,Q_1)$.
\item Generate $N_a$ samples $\{\mathbf{Z}_{t_1}^{(i)}(\overline{\mathbf{x}}_{t_1})\}_{i=1}^{N_a}$ from \eqref{eq:loc_pi_1}---either by direct sampling when the SSM is as in Example~\ref{ex:linear_gaussian}, or by running an MCMC kernel initialized at $\tilde{\mathbf{Z}}_{t_1}(\overline{\mathbf{x}}_{t_1})$ otherwise (Example~\ref{ex:nonlinear_mcmc}). The unobserved components $\overline{\mathbf{x}}_{t_1}^c$ retain the forecast value $\tilde{\mathbf{Z}}_{t_1}(\overline{\mathbf{x}}_{t_1}^c)$.
\item Estimate the filter mean $\widehat{\mathbf{m}}_{t_1}$:
 $\widehat{\mathbf{m}}_{t_1}(\overline{\mathbf{x}}_{t_1}) \leftarrow \frac{1}{N_a}\sum_{i=1}^{N_a} \mathbf{Z}_{t_1}^{(i)}(\overline{\mathbf{x}}_{t_1})$ and
$\widehat{\mathbf{m}}_{t_1}(\overline{\mathbf{x}}_{t_1}^c) \leftarrow \tilde{\mathbf{Z}}_{t_1}(\overline{\mathbf{x}}_{t_1}^c)$.
\item Reduce the $N_a$ samples to $N_f$ analysis members (e.g.\ by random grouping and averaging) and update only the observed cells $\overline{\mathbf{x}}_{t_1}$ for the next forecast step.
\end{enumerate}

\item \textbf{For} $k=2,\ldots,T$:
\begin{enumerate}
\item Forecast: for each $i=1,\cdots, N_f$, propagate 
$\breve{\mathbf{Z}}_{(l+1)\tau_k+t_{k-1}}^{(i)} = \Phi(\breve{\mathbf{Z}}_{l\tau_k+t_{k-1}}^{(i)},l\tau_k+t_{k-1};(l+1)\tau_k+t_{k-1})$
for $l=0,\ldots,L-1$, with $\breve{\mathbf{Z}}^{(i)}_{t_{k-1}}= \mathbf{Z}^{(i)}_{t_{k-1}}$.
\item Identify subdomains with observations at $t_k$; collect $\overline{\mathbf{x}}_{t_k}$ and extract sub-matrix $\widetilde{Q_k^{-1}}$.
\item Add noise: $\tilde{\mathbf{Z}}_{t_k}^{(i)}:=\breve{\mathbf{Z}}_{t_k}^{(i)}+ \mathbf{W}_{t_k}^{(i)}$, $\mathbf{W}_{t_k}^{(i)}\sim \mathcal{N}_d(0,Q_k)$, for $i=1,\ldots,N_f$.
\item Generate $N_a$ samples $\{\mathbf{Z}_{t_k}^{(i)}(\overline{\mathbf{x}}_{t_k})\}_{i=1}^{N_a}$ from $\widetilde{\pi}_k^{N_f}$ in \eqref{eq:filter_sum_local}---either by direct sampling when the SSM is as in Example~\ref{ex:linear_gaussian} or by running an MCMC kernel otherwise (Example~\ref{ex:nonlinear_mcmc}).  The unobserved components $\overline{\mathbf{x}}_{t_k}^c$ are not sampled; each forecast member retains its forecast state at those locations.
\item Estimate the filter mean $\widehat{\mathbf{m}}_{t_k}$: $\widehat{\mathbf{m}}_{t_k}(\overline{\mathbf{x}}_{t_k}) \leftarrow \frac{1}{N_a}\sum_{i=1}^{N_a} \mathbf{Z}_{t_k}^{(i)}(\overline{\mathbf{x}}_{t_k})$ and
$\widehat{\mathbf{m}}_{t_k}(\overline{\mathbf{x}}_{t_k}^c) \leftarrow \frac{1}{N_f}\sum_{i=1}^{N_f} \tilde{\mathbf{Z}}_{t_k}^{(i)}(\overline{\mathbf{x}}_{t_k}^c).
$
\item Reduce the $N_a$ samples to $N_f$ analysis members (e.g.\ by random grouping and averaging) and update only the observed cells $\overline{\mathbf{x}}_{t_k}$ of the ensemble for the next forecast step.
\end{enumerate}
\end{enumerate}

\textbf{Output:} Return $\{\widehat{\mathbf{m}}_{t_k}\}_{k\in\{1,\cdots,T\}}$.

\vspace{-0.1cm}
\hrulefill
\vspace{0.2cm}
\end{flushleft}

\subsection{Variant 2: Halo-Based Per-Block Localization}
\label{subsec:halo_localization}

While the joint approach of Variant~1 reduces the state dimension from $d$ to $d_k$, the effective dimension can still be large when many subdomains contain observations. Variant~2, which we call \emph{halo-based per-block localization}, processes each observed subdomain independently, thereby decomposing the problem into many smaller, fully parallel MCMC chains.

For each observed subdomain $G_i$, we define an extended neighborhood (or \emph{halo}) $\widehat{G}_i \supseteq G_i$ that includes all grid points within a prescribed radius $r_h$ (measured in grid points) of the centroid $\mathbf{c}_i$ of $G_i$. The halo provides spatial context: the local likelihood is evaluated using all observations whose locations fall within $\widehat{G}_i$, with observation noise inflated via Gaspari--Cohn tapering so that distant observations have progressively less influence. Let $\widehat{\mathbf{x}}_{t_k}^{(i)}$ and $\mathbf{y}_{t_k}^{(i)}$ denote the grid points and local observations in $\widehat{G}_i$, respectively. The local target distribution for block $i$ is defined over the halo state:
\begin{equation}
\label{eq:halo_target}
\widehat{\pi}_k^{(i)}(\mathbf{z}_{t_k}(\widehat{\mathbf{x}}_{t_k}^{(i)}), j) \propto \widehat{g}_k^{(i)}(\mathbf{z}_{t_k}(\widehat{\mathbf{x}}_{t_k}^{(i)}), \mathbf{y}_{t_k}^{(i)}) ~\widehat{f}_k^{(i)}(\mathbf{Z}_{t_{k-1}}^{(j)}(\widehat{\mathbf{x}}_{t_k}^{(i)}), \mathbf{z}_{t_k}(\widehat{\mathbf{x}}_{t_k}^{(i)}))~p(j),
\end{equation}
where $\widehat{f}_k^{(i)}$ is the restriction of the transition density to the halo and $\widehat{g}_k^{(i)}$ is the corresponding local likelihood evaluated with GC-tapered observation noise \eqref{eq:gc_taper}.

To smoothly downweight the influence of distant observations without introducing discontinuities, we taper the observation-error variance using the Gaspari--Cohn function $S(\cdot)$ \cite{gaspari}. For each local observation $j$ at location $\mathbf{x}_j \in \widehat{G}_i$, the effective observation noise standard deviation is inflated as
\begin{equation}
\label{eq:gc_taper}
\tilde{\sigma}_{y,j}^{(i)} = \frac{\sigma_{y,j}}{\sqrt{S\!\left(\dfrac{\|\mathbf{x}_j - \mathbf{c}_i\|}{r_h/2}\right)}},
\end{equation}
where $\mathbf{c}_i$ is the centroid of $G_i$ and the Gaspari--Cohn length scale is set to half the halo radius, $r_h/2$, so that the taper reaches zero exactly at the halo boundary $\|\mathbf{x}_j-\mathbf{c}_i\|=r_h$. The Gaspari--Cohn function $S$ is compactly supported with $S(0)=1$ and $S(r)=0$ for $r\geq 2$, and is defined by
\begin{equation}
\label{eq:gc_function}
S(r) = \begin{cases}
1 - \dfrac{5}{3}r^2 + \dfrac{5}{8}r^3 + \dfrac{1}{2}r^4 - \dfrac{1}{4}r^5, & 0 \leq r \leq 1, \\[6pt]
4 - 5r + \dfrac{5}{3}r^2 + \dfrac{5}{8}r^3 - \dfrac{1}{2}r^4 + \dfrac{1}{12}r^5 - \dfrac{2}{3r}, & 1 < r \leq 2, \\[6pt]
0, & r > 2.
\end{cases}
\end{equation}
Observations at the block center retain their nominal precision, while observations near the halo boundary have greatly inflated noise and thus negligible influence. This tapering avoids sharp boundaries in the analysis without requiring any post-hoc blending of overlapping updates. With this choice the argument $r=\|\mathbf{x}_j-\mathbf{c}_i\|/(r_h/2)$ ranges over $[0,2]$ as an observation moves from the centroid to the halo boundary, so the weight $S(r)$ decays smoothly from $S(0)=1$ to $S(2)=0$ and each observation's influence vanishes continuously at the boundary rather than dropping abruptly. For the boundary-most observations, where $S(r)\to 0$, we floor $S$ at a small $\epsilon>0$ to keep the division in \eqref{eq:gc_taper} well defined, which caps the inflated noise at $\sigma_{y,j}/\sqrt{\epsilon}$.

After sampling, only the state variables at the block cells $G_i$ are retained. For an LGOM, the diagonal prior allows the non-block halo variables to be integrated out analytically, so that $N_a$ samples are drawn directly in dimension $|G_i|$. For an NLGOM or NGOM, the MCMC chain operates on the full halo state in dimension $|\widehat{G}_i|$, and only the block-cell components are kept from the resulting samples. Because the partition $\{G_i\}$ is non-overlapping, each grid point is updated by at most one block, and grid points that do not belong to any observed block retain their forecast values. While the cores $G_i$ do not overlap, each is surrounded by a halo $\widehat{G}_i$, so an observation near a block boundary is still assimilated through the neighboring block's halo and is not discarded; the Gaspari--Cohn taper only down-weights it smoothly with distance, as intended for distance-based localization. The halo radius $r_h$ controls this down-weighting trade-off. 

The key advantage of the per-block halo scheme is that the problems for distinct blocks $i$ are independent and can be solved in parallel. Each block's effective sampling dimension is $|G_i|$ for an LGOM or $|\widehat{G}_i|$ for an NLGOM or NGOM, both much smaller than $d$, and the halo provides spatial context through its local observations at essentially no extra cost because the prior covariance is diagonal. In our implementation, we use a thread pool to parallelize over all observed blocks simultaneously. The total cost per assimilation step is then $\mathcal{O}([\kappa(|\widehat{G}_i|) + \kappa_y(\widehat{d}_{y,\max})]N_a / P )$, where $\widehat{d}_{y,\max} = \max_i |\mathbf{y}_{t_k}^{(i)}|$ and $P$ is the number of parallel threads. The halo radius $r_h$ controls the trade-off between accuracy (larger halos capture more distant observations) and efficiency (smaller halos yield fewer local observations and a smaller MCMC state for an NLGOM or NGOM). The steps of this variant are summarized in Algorithm~\ref{alg:loc_smcmc_v2}.

%
%
\begin{flushleft}
\needspace{4\baselineskip}
\captionsetup[algorithm]{style=algori}
\captionof{algorithm}{LSMCMC -- Variant 2 (Halo-Based Per-Block Localization)}
\label{alg:loc_smcmc_v2}
\textbf{Input:} Matrices $Q_k$, $Q_k^{-1}$; partition $\mathsf{G}=\bigcup_{i=1}^{\Gamma} G_i$; halo radius $r_h$; initial state $\breve{\mathbf{Z}}_0=\mathbf{z}_0$; observations $\{\mathbf{y}_{t_k}\}_{k\geq 1}$ and their locations; number of forecast samples $N_f$; number of analysis samples $N_a \geq N_f$; time-lag $L$. Set $\tau_k:=(t_k-t_{k-1})/L$.

\begin{enumerate}
\item \textbf{Initialize} ($k=1$):
\begin{enumerate}
\item Forecast: for $l=0,\cdots,L-1$, compute $\breve{\mathbf{Z}}_{(l+1)\tau_1} = \Phi(\breve{\mathbf{Z}}_{l\tau_1},l\tau_1;(l+1)\tau_1)$.
\item Add noise: $\tilde{\mathbf{Z}}_{t_1}:=\breve{\mathbf{Z}}_{t_1}+ \mathbf{W}_{t_1}$, $\mathbf{W}_{t_1}\sim \mathcal{N}_d(0,Q_1)$.
\item Identify all observed subdomains $\{G_i : \widehat{G}_i \cap \text{obs.\ locations} \neq \emptyset\}$. For each such $G_i$, construct halo $\widehat{G}_i$, extract local observations $\mathbf{y}_{t_1}^{(i)}$, and compute GC-tapered observation noise via \eqref{eq:gc_taper}.
\item \textbf{In parallel} over observed blocks $i$: draw $N_a$ samples from $\widehat{\pi}_1^{(i)}$ in \eqref{eq:halo_target} (direct sampling in dimension $|G_i|$ for LGOM; MCMC in dimension $|\widehat{G}_i|$ for NLGOM or NGOM), using halo observations with GC-tapered noise; retain only block cells $G_i$. Compute the local posterior mean $\overline{\mathbf{z}}_{t_1}^{(i)}(G_i)$.
\item For each observed block $i$, reduce the $N_a$ local samples to $N_f$ analysis members (e.g.\ by random grouping and averaging) and overwrite only the block cells $G_i$ of the $N_f$ forecast members. Grid points outside all observed blocks retain their forecast values.
\end{enumerate}

\item \textbf{For} $k=2,\ldots,T$:
\begin{enumerate}
\item Forecast: for each $i=1,\cdots, N_f$, propagate 
$\breve{\mathbf{Z}}_{(l+1)\tau_k+t_{k-1}}^{(i)} = \Phi(\breve{\mathbf{Z}}_{l\tau_k+t_{k-1}}^{(i)},l\tau_k+t_{k-1};(l+1)\tau_k+t_{k-1})$
for $l=0,\ldots,L-1$, with $\breve{\mathbf{Z}}^{(i)}_{t_{k-1}}= \mathbf{Z}^{(i)}_{t_{k-1}}$.
\item Add noise: $\tilde{\mathbf{Z}}_{t_k}^{(i)}:=\breve{\mathbf{Z}}_{t_k}^{(i)}+ \mathbf{W}_{t_k}^{(i)}$, $\mathbf{W}_{t_k}^{(i)}\sim \mathcal{N}_d(0,Q_k)$, for $i=1,\ldots,N_f$.
\item Identify observed subdomains at $t_k$. For each, construct halo $\widehat{G}_i$, extract local observations $\mathbf{y}_{t_k}^{(i)}$, and compute GC-tapered observation noise via \eqref{eq:gc_taper}.
\item \textbf{In parallel} over observed blocks $i$: draw $N_a$ samples from $\widehat{\pi}_k^{(i)}$ in \eqref{eq:halo_target} (direct for LGOM in $|G_i|$; MCMC for NLGOM or NGOM in $|\widehat{G}_i|$), using halo observations with GC-tapered noise; retain only block cells $G_i$. Compute the local posterior mean $\overline{\mathbf{z}}_{t_k}^{(i)}(G_i)$.
\item For each observed block $i$, reduce the $N_a$ local samples to $N_f$ analysis members (e.g.\ by random grouping and averaging) and overwrite only the block cells $G_i$ of the $N_f$ forecast members. Grid points outside all observed blocks retain their forecast values.
\end{enumerate}
\end{enumerate}

\textbf{Output:} Return $\{\widehat{\pi}_k^{N_a}(\varphi)\}_{k\in\{1,\cdots,T\}}$.

\vspace{-0.1cm}
\hrulefill
\vspace{0.2cm}
\end{flushleft}

\begin{rem}
\label{rem:M>1}
When additional computing resources are available, one may run $M>1$ independent LSMCMC (V1 or V2) experiments in parallel and average their posterior estimates to reduce Monte Carlo variance: $\widehat{\pi}_{k}^{N_a}(\varphi) \leftarrow \frac{1}{M}\sum_{m=1}^{M}\widehat{\pi}_{k}^{(m)}(\varphi)$, where $\widehat{\pi}_{k}^{(m)}$ denotes the estimate from run $m$. 
\end{rem}

\begin{rem}
In many applications, the covariance matrix is often time-independent, and therefore, its inverse or the inverse of its Cholesky factor is computed only once at the beginning. In this case, we drop the subscript $k$ in $Q_k$ and $Q_k^{-1}$ in the above algorithms.
\end{rem}

\section{Numerical Simulations}
\label{sec:numerics}

We demonstrate the proposed LSMCMC schemes on four challenging state-space models: (i)~a linear Gaussian model on a $120 \times 120$ grid ($d=14{,}400$) with SWOT-lile observations, (ii)~the nonlinear multilayer shallow-water equations (MLSWE) on a $70 \times 80$ grid ($d=67{,}200$) with a linear-Gaussian observation operator, (iii)~the MLSWE with a nonlinear (arctan) observation operator and Gaussian observation noise, and (iv)~the MLSWE with the same nonlinear observation operator but non-Gaussian (Cauchy / Student-$t$, $\nu{=}1$) observation noise. In each case we compare LSMCMC Variants~1 and~2 against the LETKF \cite{hunt_2007} with ensemble size $K$, which is used operationally at several numerical weather prediction centers \cite{schraff_2016,frolov_2024}. All experiments were run on a Linux workstation with 52 CPU cores. LETKF is run in parallel for fair comparison.

\subsection{Linear Gaussian State and Observation Model}\label{subsec:lg_experiment}

\subsubsection{Problem setup}
Consider the state-space model $\mathbf{Z}_{t_k} = a\, \mathbf{Z}_{t_{k-1}} + \sigma_z\, \mathbf{W}_{t_k}$, $\mathbf{Y}_{t_k} = C_k \mathbf{Z}_{t_k} + \sigma_y\, \mathbf{V}_{t_k}$, where $\mathbf{Z}_{t_k}\in\mathbb{R}^d$ with $d = 120 \times 120 = 14{,}400$, $a = 0.25$, $\sigma_z = 0.05$, $\sigma_y = 0.05$, and $Q = \sigma_z^2 I_d$, $R_k = \sigma_y^2 I_{d_y^k}$. The observation matrix $C_k\in\mathbb{R}^{d_y^k \times d}$ selects grid points along diagonal SWOT-like swaths that sweep periodically across the domain, simulating the observation pattern of the SWOT satellite altimeter (see Figure~\ref{fig:lg_obs}). At each cycle, approximately $10{-}15\%$ of grid points are observed. Because both the state and observation models are linear-Gaussian, the exact Kalman filter (KF) posterior is available analytically, providing a ground-truth benchmark. We run $T=100$ assimilation cycles.

\subsubsection{Parameter choices}
We partition the grid into $\Gamma = 900$ blocks for V1 and $\Gamma = 2{,}400$ blocks for V2. These choices are tuned to get the best RMSE and CPU times. In the linear-Gaussian case, the filtering density is a Gaussian mixture (Example~\ref{ex:linear_gaussian}), so LSMCMC draws $N_a$ independent samples from the exact posterior without MCMC iterations. Table~\ref{tab:lg_params} summarizes the key parameters.

\begin{table}[ht]
\centering
\caption{Parameters for the linear Gaussian experiment.}
\label{tab:lg_params}
\small
\begin{tabular}{lccc}
\toprule
\multicolumn{4}{c}{State dimension $d = 14{,}400$ \quad | \quad Assimilation cycles $T = 100$} \\
\midrule
Parameter & LSMCMC V1 & LSMCMC V2 & LETKF \\
\midrule
Blocks $\Gamma$ & 900 & 2{,}400 & --- \\
Halo / loc.\ radius & --- & $r_h = 1.0$ & $h_{\text{loc}} = 1.0$ \\
$N_f/N_a$ & 50/500 & 50/500 & $K=50$ \\
RTPP / RTPS & --- / --- & --- / 0.02 & 1.02 / --- \\
\bottomrule
\end{tabular}
\end{table}

\subsubsection{Results}
\begin{figure}[h!]
\centering
\includegraphics[width=0.65\textwidth]{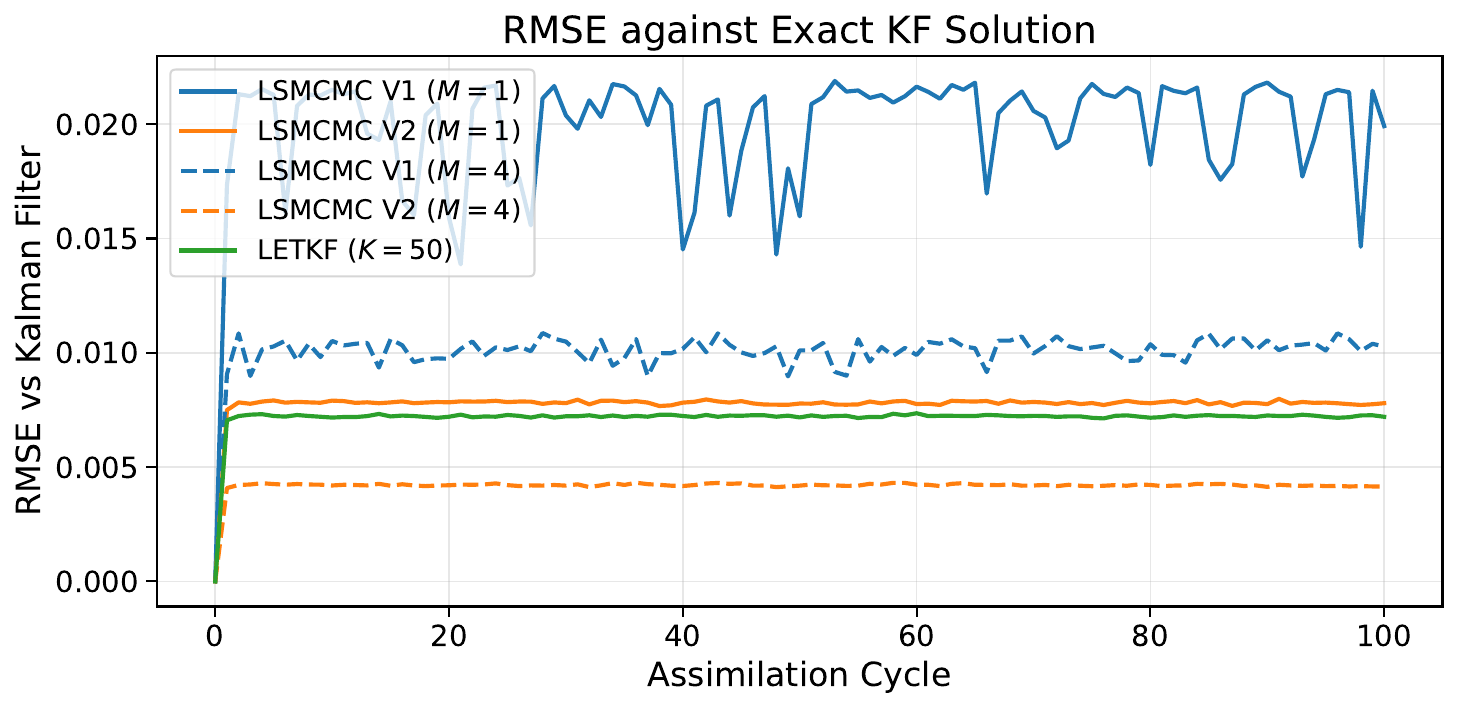}
\caption{Linear Gaussian experiment ($d{=}14{,}400$): RMSE against the exact KF solution over 100 assimilation cycles for LSMCMC V1 and V2 with $M{=}1$ (solid) and $M{=}4$ (dashed), and LETKF with $K{=}50$.}
\label{fig:lg_rmse}
\end{figure}

Figure~\ref{fig:lg_rmse} shows the RMSE of each method against the exact KF posterior mean. In the linear Gaussian setting, the KF posterior mean is the minimum-variance estimator, so this RMSE directly measures how close each approximate filter is to the optimal Bayesian solution (it is the most meaningful accuracy metric in this paper). With a single run ($M{=}1$; see Remark~\ref{rem:M>1}), V2 already achieves an RMSE of $0.0071$, outperforming V1 ($0.0203$) and very close to LETKF ($0.0072$), thanks to finer partitioning ($\Gamma{=}2{,}400$) and halo localization. Averaging $M{=}4$ independent runs further reduces the error: V2 reaches $\mathbf{0.0042}$ at 15 seconds and V1 improves to $0.0101$ at 4.5 seconds. Table~\ref{tab:lg_results} reports the RMSE and total wall-clock time.

\begin{table}[ht]
\centering
\caption{Linear Gaussian: RMSE vs KF and wall-clock time. For $M{=}4$, the reported time is the total time for all four runs which are independent and embarrassingly parallel.}
\label{tab:lg_results}
\small
\begin{tabular}{lcc}
\toprule
Method & RMSE vs KF & Total Time (s) \\
\midrule
LETKF ($K{=}50$)       & 0.0072 & 13 \\
LSMCMC V1 ($M{=}1$)   & 0.0203 & \textbf{4} \\
LSMCMC V2 ($M{=}1$)   & 0.0071 & 13 \\
LSMCMC V1 ($M{=}4$)   & 0.0101 & $4.5$ \\
LSMCMC V2 ($M{=}4$)   & \textbf{0.0042} & $15$ \\
\bottomrule
\end{tabular}
\end{table}

\begin{figure}[h!]
\centering
\includegraphics[width=0.98\textwidth]{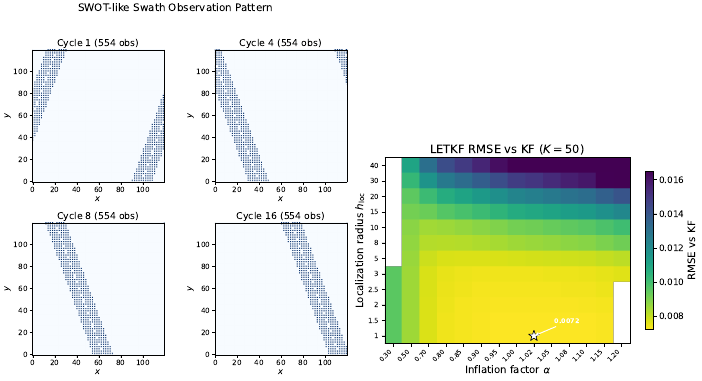}
\caption{(Left) SWOT-like swath observation pattern at selected cycles showing the diagonal sweep across the $120\times120$ grid. (Right) LETKF sensitivity to localization radius $h_{\text{loc}}$ and RTPP inflation factor $\alpha$. The star marks the best configuration ($h_{\text{loc}}{=}1.0$, $\alpha{=}1.02$, RMSE${=}0.0072$).}
\label{fig:lg_obs}
\label{fig:lg_letkf_sensitivity}
\end{figure}

\begin{figure}[h!]
\centering
\includegraphics[width=0.65\textwidth]{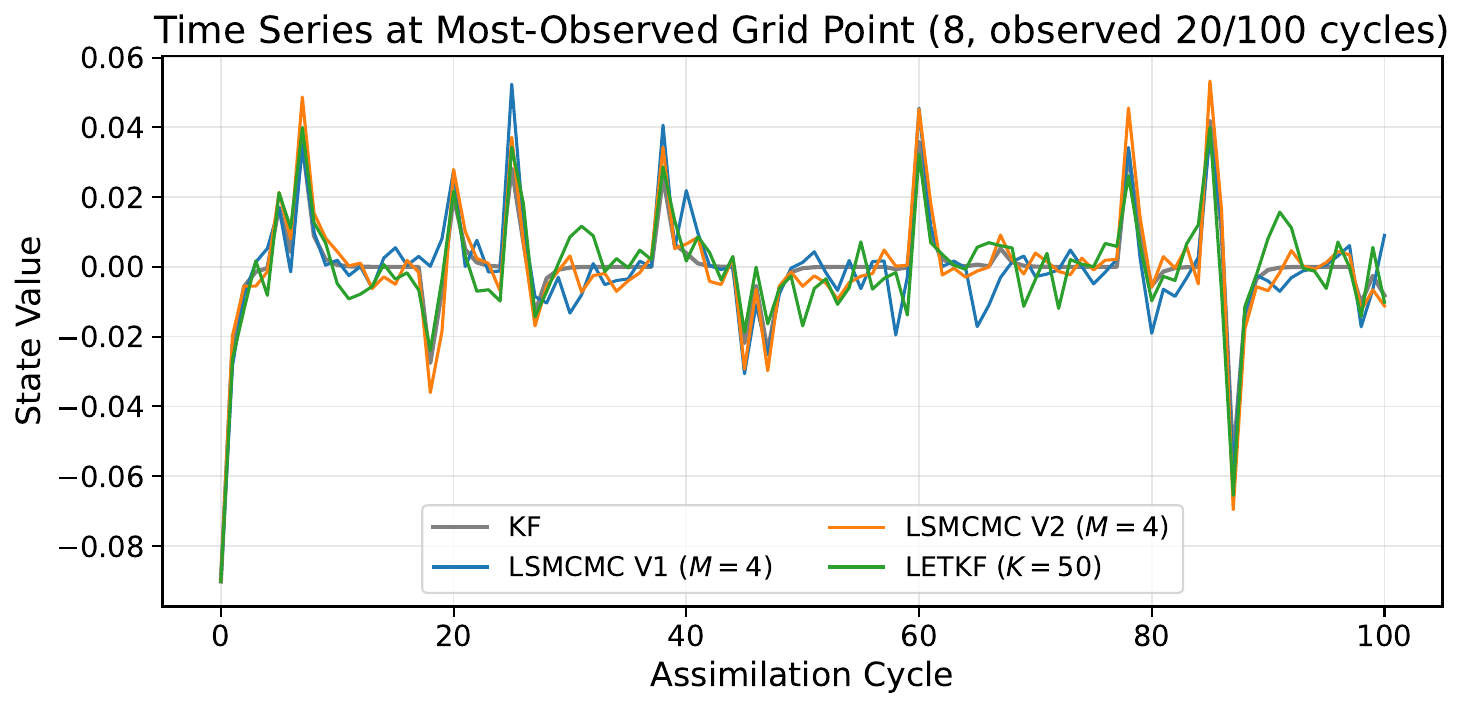}
\caption{Linear Gaussian: time series at the most-observed grid point comparing LSMCMC V1 and V2 ($M{=}4$) and LETKF ($K{=}50$) against the KF.}
\label{fig:lg_coord50}
\end{figure}

\begin{figure}[h!]
\centering
\includegraphics[width=0.99\textwidth]{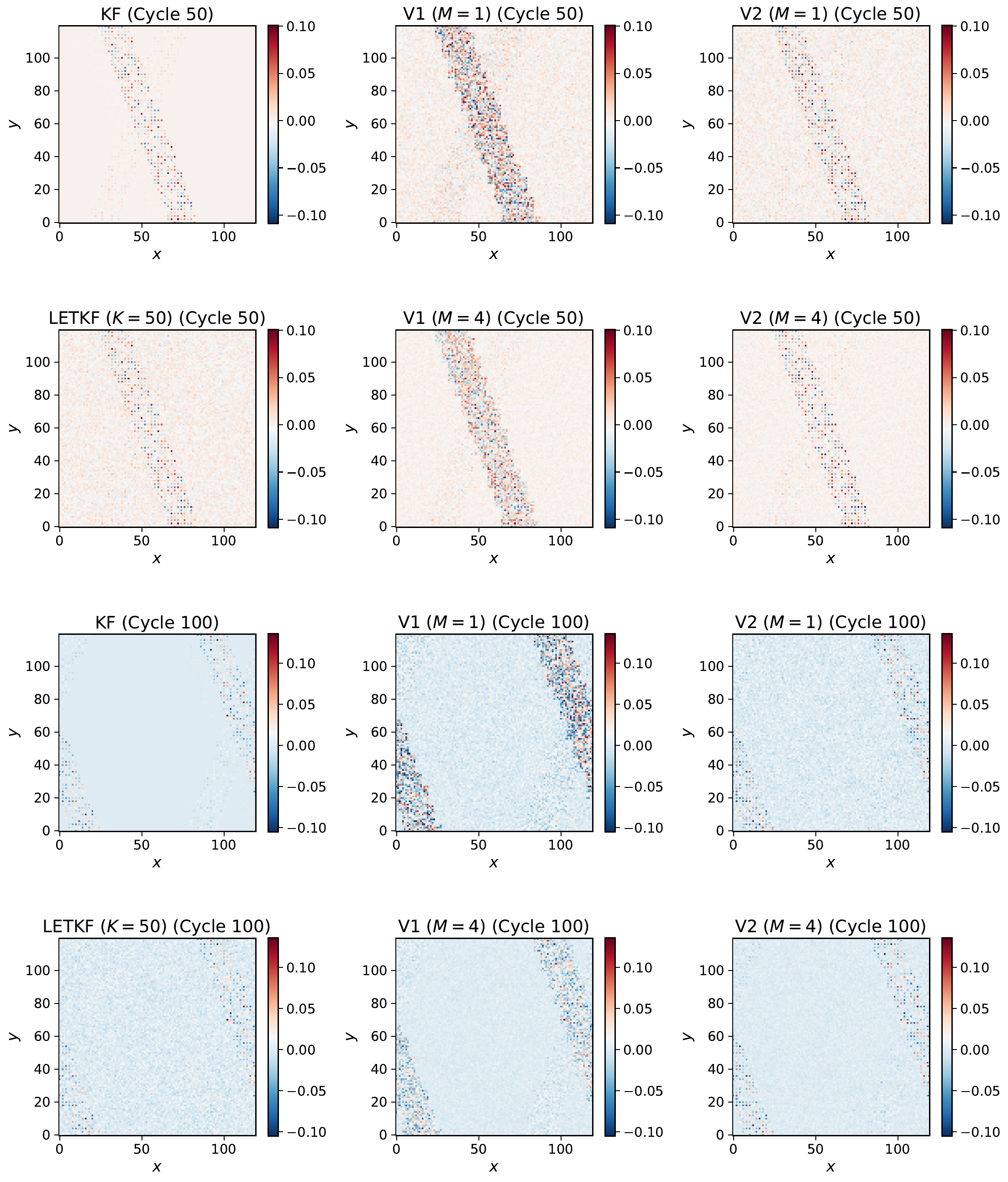}
\caption{Linear Gaussian: analysis fields at cycle~50 (rows~1--2) and cycle~100 (rows~3--4). Each block shows the KF reference, V1 ($M{=}1$), and V2 ($M{=}1$) in the first row, and LETKF ($K{=}50$), V1 ($M{=}4$), and V2 ($M{=}4$) in the second row.}
\label{fig:lg_snapshot}
\end{figure}

The LETKF result is obtained after tuning the localization radius and inflation factor over a grid of $12 \times 14 = 168$ configurations (Figure~\ref{fig:lg_letkf_sensitivity}); small localization radii ($h_{\text{loc}} \leq 3$) and mild inflation ($\alpha \approx 1.0$) perform best, with RMSE degrading rapidly as the localization radius grows.  Figure~\ref{fig:lg_coord50} compares the filter estimates at the most-observed grid point, confirming that both V1 and V2 track the exact KF solution closely throughout the experiment.  Figure~\ref{fig:lg_snapshot} displays the spatial analysis fields at cycles~50 and~100; V2 ($M{=}4$) produces the closest visual match to the KF reference across the entire domain.

\subsection{MLSWE with Linear Observation Model}\label{subsec:ldata_experiment}

\subsubsection{Problem setup}
We consider the 3-layer isopycnal MLSWE model over the North Atlantic ($60^\circ\text{W}{-}20^\circ\text{W}$, $10^\circ\text{N}{-}45^\circ\text{N}$) on a $70 \times 80$ Cartesian grid. The state vector includes 12 fields per grid cell (the layer thickness $h_k$, $u_k$, $v_k$, $T_k$ for $k=0,1,2$), giving a state dimension $d = 12 \times 5{,}600 = 67{,}200$. We emphasise that the first field of each layer is the \emph{layer thickness} $h_k$, not the sea-surface height (SSH). The assimilated surface field is the surface-layer thickness $h_0 = H_{\mathrm{rest},0} + \alpha_0\,\eta$, whereas the physical SSH anomaly $\eta = \big(\sum_k h_k\big) - H_b$ is recovered only in post-processing (and shown in the field figures below). The model is integrated with an RK4 time stepper at $\Delta t = 75$\,s. Boundary conditions are prescribed from HYCOM reanalysis, and the bathymetry is derived from ETOPO1 \cite{etopo1}. Because the isopycnal shallow-water equations do not resolve atmospheric heat exchange, the state transition model includes a Newtonian SST nudging term $-\lambda(T - T_{\mathrm{ref}})$ in the surface-layer temperature tendency, where $T_{\mathrm{ref}}$ is interpolated from HYCOM reanalysis and the relaxation rate $\lambda \approx 2.8\times10^{-4}$\,s$^{-1}$ (time scale $\tau\approx 1$\,h). This prevents the model SST from drifting unrealistically in the absence of full thermodynamic forcing.

\begin{figure}[h!]
\centering
\includegraphics[width=0.75\textwidth]{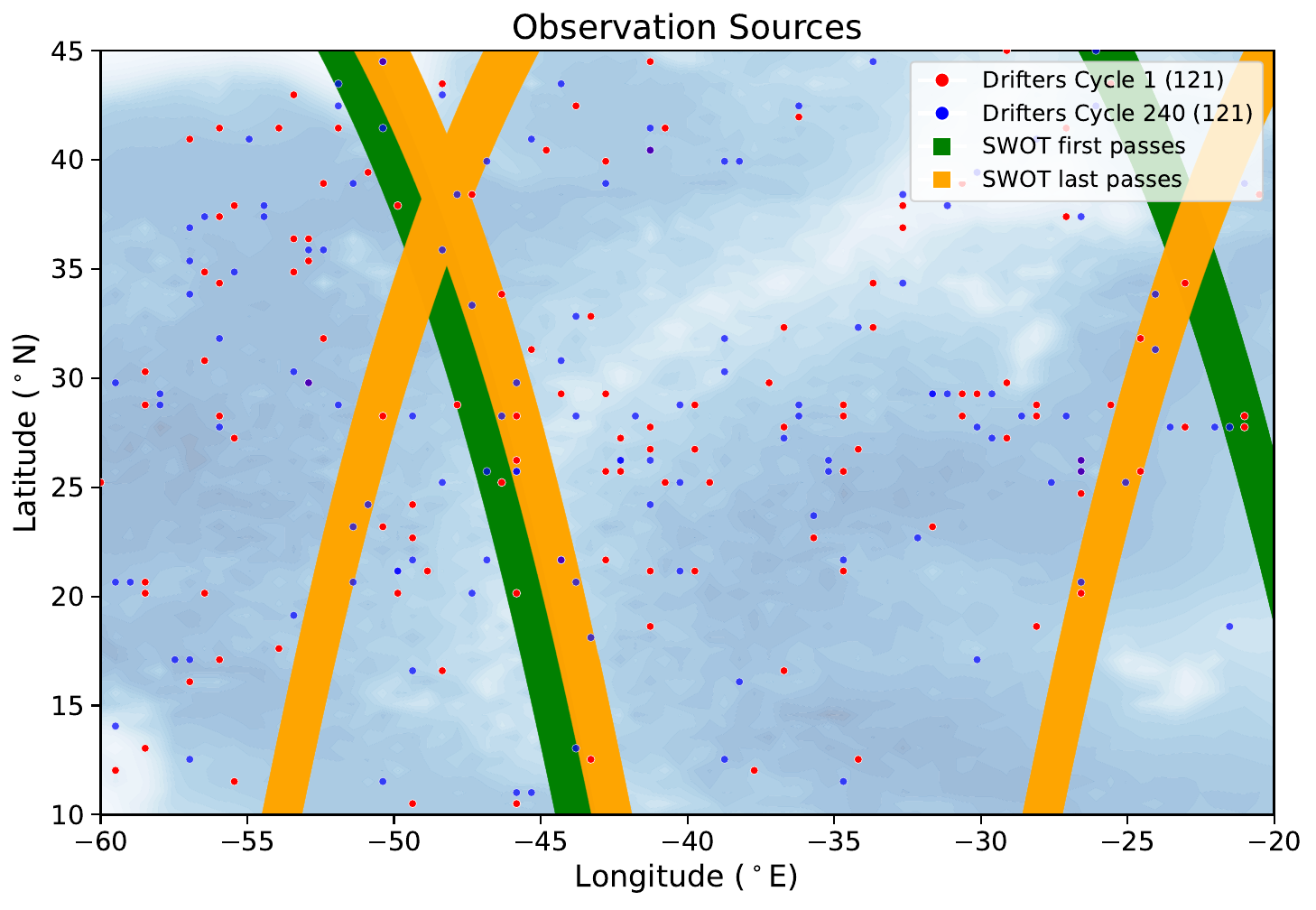}
\caption{Observation sources: surface drifter positions at cycles~1 and~240 (red and blue dots), and SWOT satellite swaths from first and last passes (green and orange). Bathymetry contours are shown in the background.}
\label{fig:obs_coverage}
\end{figure}

The observation model is linear:
$
\mathbf{Y}_{t_k} = C_k \mathbf{Z}_{t_k} + \mathbf{V}_{t_k}$, $\mathbf{V}_{t_k} \sim \mathcal{N}(0, R_k),
$
where $C_k \in \mathbb{R}^{d_{y_k}\times d}$ extracts velocity, SSH, and SST at observed locations. Observations come from two real-world sources:
\begin{enumerate}
\item \textbf{Surface drifters} from the NOAA Global Drifter Program (GDP) \cite{noaa}. Because the GDP hourly product \cite{noaa_hourly} does not yet cover the SWOT science-orbit period (August 2024), we use the quality-controlled 6-hourly interpolated dataset \cite{noaa} and further interpolate each drifter track to hourly resolution via cubic splines, yielding velocity measurements at $\sim$350--620 locations per cycle. The SST observations assimilated in this experiment are the in-situ sea-surface-temperature measurements carried by the same GDP drifters, co-located with the drifter velocity fixes; thus velocity and SST share the same drifter locations.
\item \textbf{SWOT SSH} from the Surface Water and Ocean Topography mission \cite{swot, nasa}, binned onto the model grid with 15\% spatial coverage per cycle.
\end{enumerate}
Figure~\ref{fig:obs_coverage} illustrates the spatial coverage of these two observation sources, including the surface drifter positions and SWOT satellite swaths over the model domain.

Observation noise standard deviations are $\sigma_{\text{vel}} = 0.10$\,m/s, $\sigma_{\text{SSH}} = 0.25$\,m, and $\sigma_{\text{SST}} = 0.40$\,K. These values reflect the dominant error sources in each data stream: the GDP velocity error budget is dominated by the cubic-spline interpolation from 6-hourly fixes to hourly resolution ($\sim 0.05$--$0.10$\,m/s; \cite{elipot2016}); the SWOT SSH noise after spatial averaging onto the $0.5^\circ$ grid is approximately $0.15$--$0.25$\,m, accounting for residual mesoscale representation error; and the SST noise of $0.40$\,K encompasses both instrument error and the sub-grid variability of the SST field.

The forward model noise (process noise) covariance $Q$ is diagonal with standard deviations $\sigma_x^{\text{vel}} = 0.15$\,m/s, $\sigma_x^{\text{SSH}} = 0.20$\,m, and $\sigma_x^{\text{SST}} = 1.0$\,K. Some of these values are chosen to be moderately larger than the observation noise so that the filter assigns sufficient weight to the observations while preventing filter divergence ($\sigma_x^{\text{vel}} / \sigma_{\text{vel}} = 1.5$, and the large $\sigma_x^{\text{SST}} = 1.0$\,K compensates for the absence of atmospheric heat fluxes in the isopycnal model, which cannot internally maintain realistic SST without the nudging term). Process noise is applied only to the surface layer; deep layers receive zero process noise because no observations are available at depth and the model dynamics propagate surface corrections downward. We run 240 assimilation cycles (1-hour intervals, 10-day window starting 2024-08-01). With the model time step $\Delta t = 75$\,s and a $1$-hour assimilation window, there are $L = 3600/75 = 48$ forward RK4 sub-steps between consecutive observation times. To quantify the per-window signal-to-noise ratio (SNR), we measure the RMS increment of the nature-run state over one assimilation window and divide it by the corresponding observation-noise standard deviation: the increments are $0.031$\,m/s (velocity), $0.18$\,m (SSH), and $0.054$\,K (SST), giving SNR values of $\text{SNR}_{\text{vel}} \approx 0.31$, $\text{SNR}_{\text{SSH}} \approx 0.73$, and $\text{SNR}_{\text{SST}} \approx 0.14$. These moderate-to-small ratios indicate that a single window's dynamical signal is comparable to or smaller than the observation noise, so accumulating information across many cycles is essential.

Since the true ocean state is unknown, RMSE must be computed against observations rather than a known truth. The RMSE reported below for the real-data experiment is therefore computed against the assimilated observations, in observation space, by comparing the analysis mean to the observed values at the observation locations.

\subsubsection{Parameter choices}
Table~\ref{tab:ldata_params} summarizes the configuration for all three methods. V1 uses $\Gamma = 50$ blocks and solves a single combined posterior, while V2 uses $\Gamma = 1{,}400$ fine blocks with halo radius $r_h = 2.5$ grid cells. Both LSMCMC variants use $N_f = 25$ forecast samples and draw $N_a = 500$ analysis samples from the exact Gaussian mixture (Example~\ref{ex:linear_gaussian}). LETKF uses $K = 25$ ensemble members (matching ~$N_f$) with localization $h_{\text{loc}} = 60$\,km and RTPS relaxation $\alpha = 0.90$.

\begin{table}[ht]
\centering
\caption{Parameters for the MLSWE linear data model experiment.}
\label{tab:ldata_params}
\small
\begin{tabular}{lccc}
\toprule
\multicolumn{4}{c}{State dimension $d = 67{,}200$ \quad | \quad Assimilation cycles $T = 240$} \\
\midrule
Parameter & LSMCMC V1 & LSMCMC V2 & LETKF \\
\midrule
Blocks $\Gamma$ & 50 & 1{,}400 & --- \\
Halo radius $r_h$ & --- & 2.5 & --- \\
$N_f$ / $N_a$ / $K$ & 25 / 500 & 25 / 500 & --- / --- / 25 \\
Loc.\ scale (km) & --- & --- & 60 \\
RTPP / RTPS $\alpha$ & --- / --- & --- / 0.5 & 0.90 / --- \\
$\sigma_{\text{vel}},\sigma_{\text{SSH}},\sigma_{\text{SST}}$ & \multicolumn{3}{c}{0.10, 0.25, 0.40} \\
$\sigma_x^{\text{vel}},\sigma_x^{\text{SSH}},\sigma_x^{\text{SST}}$ & \multicolumn{2}{c}{0.15, 0.20, 1.0} & 0.03, 0.20, 0.27 \\
\bottomrule
\end{tabular}
\end{table}

\subsubsection{LETKF sensitivity analysis}

We perform a sensitivity analysis for LETKF with $K=25$ and $K=50$ ensemble members. For each $K$, we sweep over the localization half-width $h_{\text{loc}} \in \{20, 40, 60,$ $80, 100, 200, 300\}$\,km and RTPP inflation $\alpha \in \{0.5, 0.7, 0.9, 1.0, 1.1\}$. The best configuration for $K=25$ is $h_{\text{loc}}=20$\,km, $\alpha = 0.9$ (vel RMSE $= 0.0078$\,m/s), and for $K = 50$ it is $h_{\text{loc}}=60$\,km, $\alpha = 0.9$ (vel RMSE $= 0.0093$\,m/s). Notably, small localization radii ($h_{\text{loc}} = 20$\,km) perform best for small ensembles, while larger ensembles tolerate wider localization radii. At $h_{\text{loc}} \geq 200$\,km, LETKF starts to diverge due to spurious long-range correlations.

\subsubsection{Results}
\begin{figure}[h!]
\centering
\includegraphics[width=0.99\textwidth]{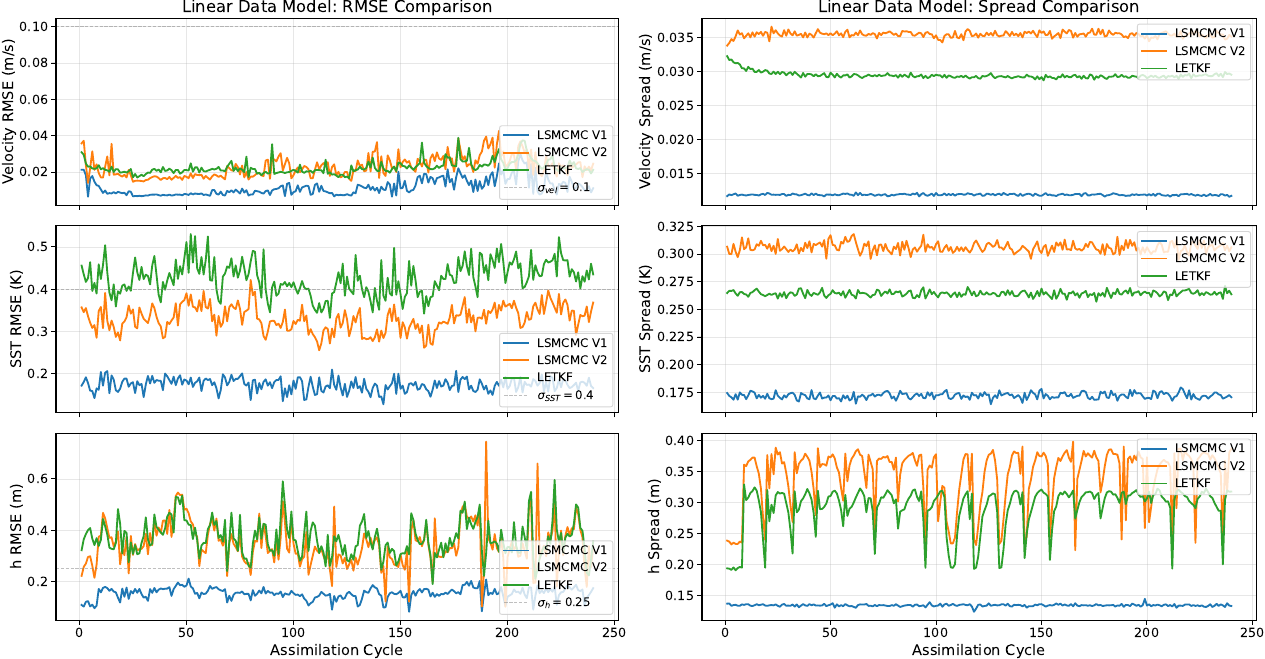}
\caption{MLSWE with linear observations: velocity, SST, and $h$ RMSE (left) and spread (right) comparison of LSMCMC V1, V2, and LETKF over 240 cycles. Dashed lines indicate observation noise standard deviations.}
\label{fig:ldata_rmse}
\end{figure}

Table~\ref{tab:ldata_results} and Figure~\ref{fig:ldata_rmse} report the mean RMSE over 240 cycles together with the cycle-mean forecast ensemble spread. LSMCMC V1 ($\Gamma{=}50$) attains the best RMSE on all three fields (velocity $0.0117$\,m/s, SST $0.174$\,K, $h$ $0.155$\,m). The principal result is calibration: after recalibrating the process-noise inflation, the forecast spread is well matched to the RMSE for velocity and SST, with V1 reaching spread/RMSE ratios near unity ($1.02$ for velocity, $0.99$ for SST, $0.86$ for $h$). LETKF, by contrast, remains under-dispersive for SST (spread/RMSE $\approx 0.62$): because its spread and analysis accuracy are both governed by the same process-noise inflation, shrinking the spread to match the RMSE simultaneously degrades the RMSE, so no single setting yields a calibrated, low-error SST estimate. Throughout this section, the assimilated scalar is the layer-thickness variable $h$; SSH anomaly is obtained diagnostically as $\eta=(\sum_k h_k)-H_b$.

\begin{table}[ht]
\centering
\caption{MLSWE linear data model: mean RMSE over 240 cycles together with the cycle-mean forecast ensemble spread, reported as RMSE\,/\,Spread in each cell. Spread is the cycle-mean forecast ensemble standard deviation. Best (lowest) RMSE per field in bold.}
\label{tab:ldata_results}
\footnotesize
\setlength{\tabcolsep}{3pt}
\begin{tabular}{lcccc}
\toprule
Method & Vel RMSE/Spread (m/s) & SST RMSE/Spread (K) & h RMSE/Spread (m) & Time/cycle (s) \\
\midrule
LETKF & 0.0227 / 0.0294 & 0.426 / 0.264 & 0.369 / 0.292 & 1.2 \\
LSMCMC V1  & \textbf{0.0117} / \textbf{0.0119} & \textbf{0.174} / \textbf{0.172} & \textbf{0.155} / \textbf{0.134} & 1.4 \\
LSMCMC V2  & 0.0228 / 0.0354 & 0.331 / 0.306 & 0.346 / 0.344 & 1.4 \\
\bottomrule
\end{tabular}
\end{table}

\begin{figure}[h!]
\centering
\includegraphics[width=0.95\textwidth]{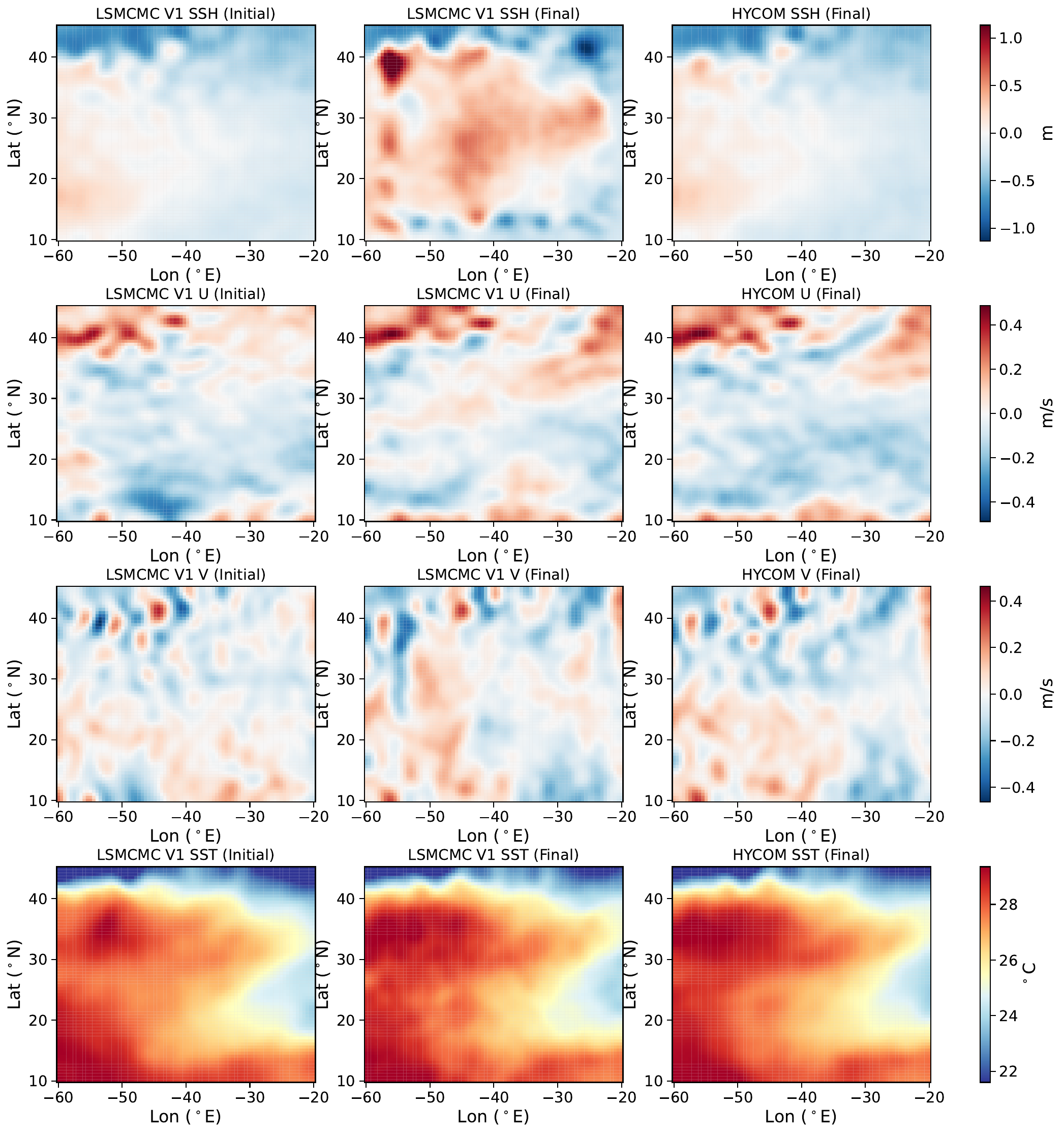}
\caption{MLSWE linear observation model: initial and final LSMCMC~V1 analysis fields (left, centre) and final HYCOM reanalysis (right). Rows show SSH anomaly, eastward velocity~$u$, northward velocity~$v$, and SST. Each row shares a single colour scale; the SSH scale is the average of the analysis and HYCOM ranges.}
\label{fig:ldata_fields} 
\end{figure}

All three methods run at approximately $1.2{-}1.4$\,s per cycle, making them comparable in wall-clock time. All three methods completed the full 240 cycles without blow-up events.  Figure~\ref{fig:ldata_fields} displays the initial and final V1 analysis fields alongside the HYCOM reanalysis.  The velocity and SST fields show good qualitative agreement with HYCOM, capturing the large-scale North Atlantic circulation and its mesoscale eddies.  The SSH field exhibits noticeable differences, which are expected: the isopycnal MLSWE is a reduced-physics model that does not resolve the full primitive equations, and the presence of islands (e.g., the Azores, Canary Islands) introduces bathymetric effects that the Cartesian solver cannot represent faithfully.

\begin{figure}[h!]
\centering
\includegraphics[width=0.55\textwidth]{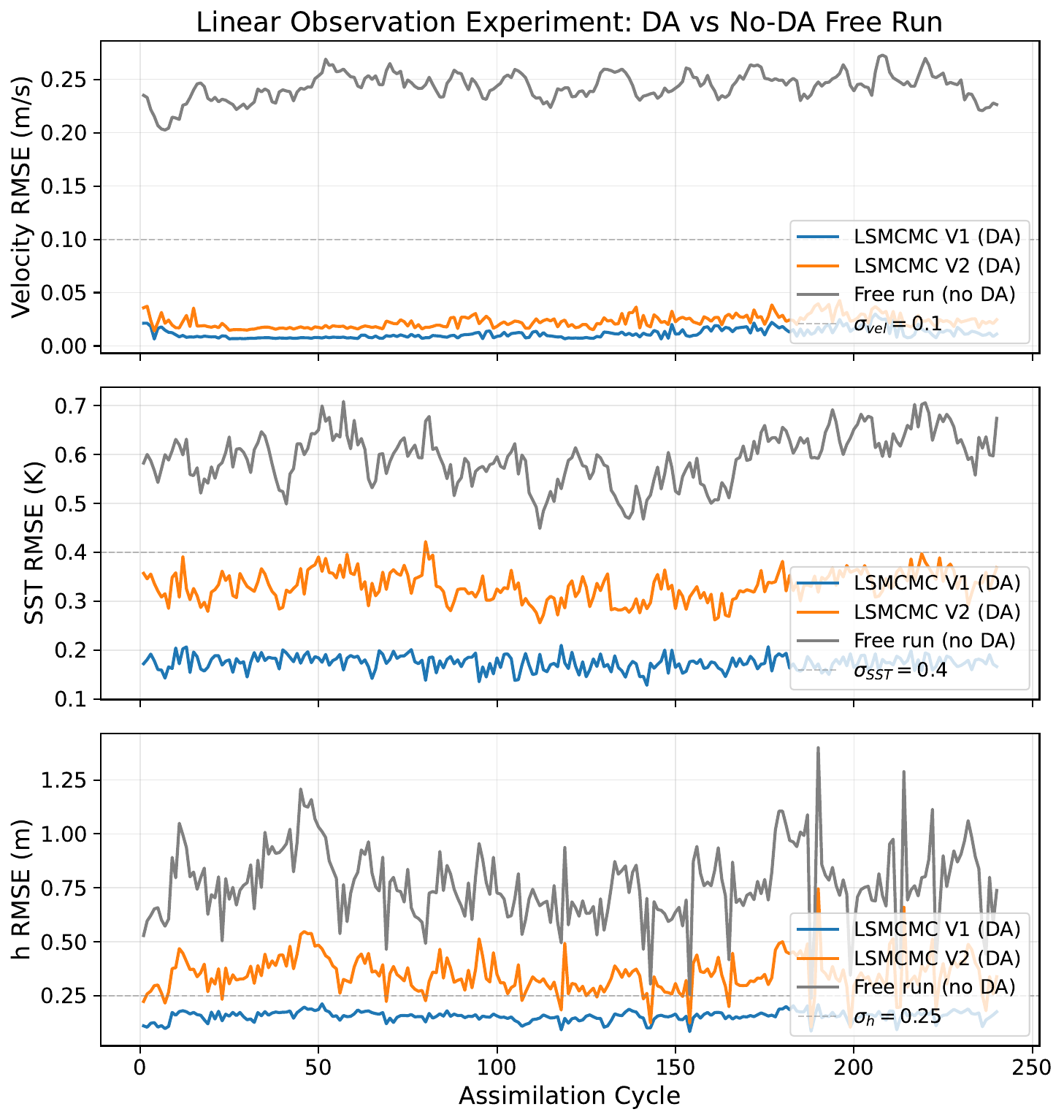}
\caption{MLSWE with linear-observation experiment: LSMCMC DA versus no-DA free forecast. Time series of RMSE for velocity, SST, and $h$ over 240 cycles. The free run remains substantially less accurate than both LSMCMC variants.}
\label{fig:ldata_freerun}
\end{figure}

To isolate the effect of assimilation in the linear-observation experiment, we also ran a no-DA free forecast initialized from the same initial condition. Figure~\ref{fig:ldata_freerun} shows that both LSMCMC variants substantially outperform the free run across all variables where velocity RMSE decreases from $0.243$ (free run) to $0.012$ (V1) and $0.023$ (V2), SST RMSE from $0.593$ to $0.174$ (V1) and $0.331$ (V2), and $h$ RMSE from $0.764$ to $0.155$ (V1) and $0.346$ (V2).

\begin{figure}[h!]
\centering
\includegraphics[width=0.99\textwidth]{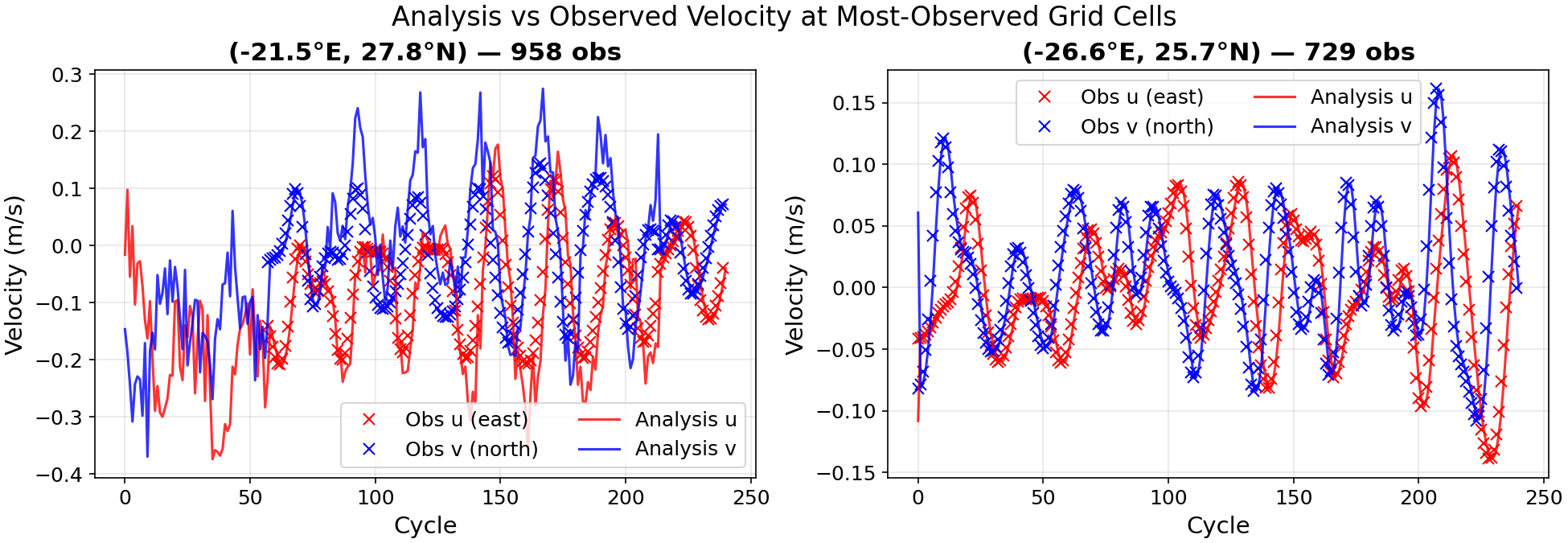}
\caption{LSMCMC~V2 analysis vs.\ drifter observations: eastward ($u$) and northward ($v$) velocity time series at the two most frequently observed grid cells over 240 cycles.}
\label{fig:ldata_vel_ts}
\end{figure}

\begin{figure}[h!]
\centering
\includegraphics[width=0.99\textwidth]{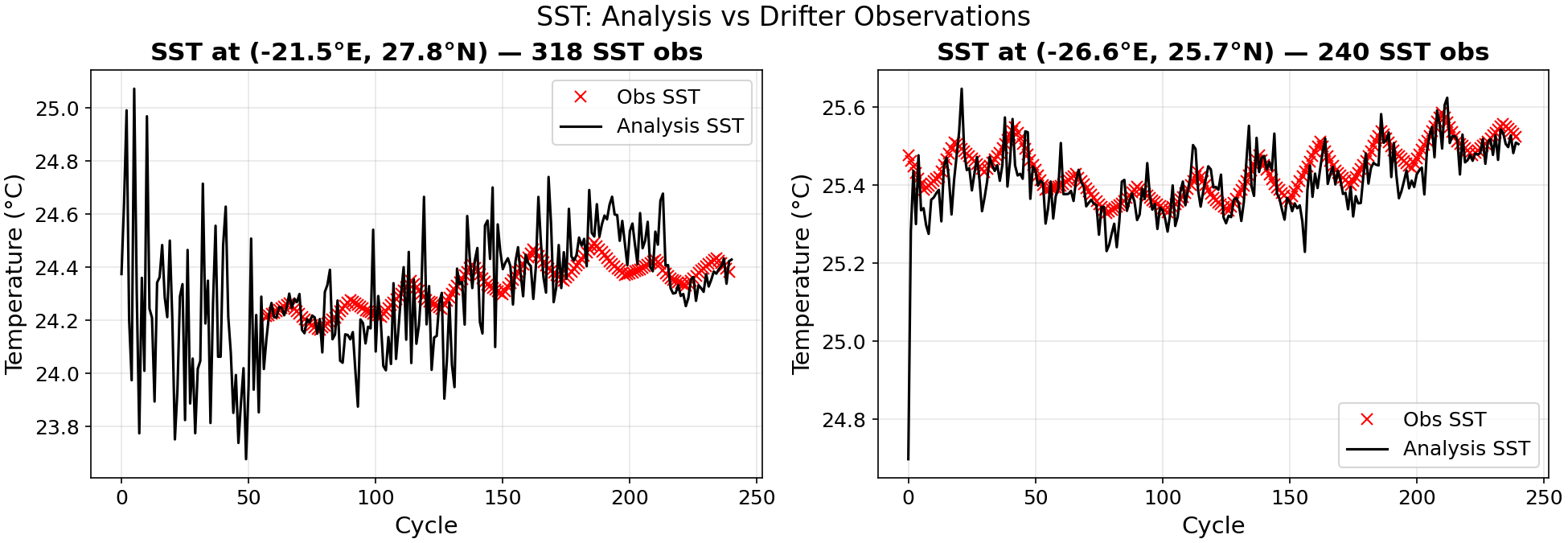}
\caption{LSMCMC~V2 analysis vs.\ drifter observations: SST time series at the two most frequently observed grid cells.}
\label{fig:ldata_sst_ts}
\end{figure}

\begin{figure}[h!]
\centering
\includegraphics[width=0.99\textwidth]{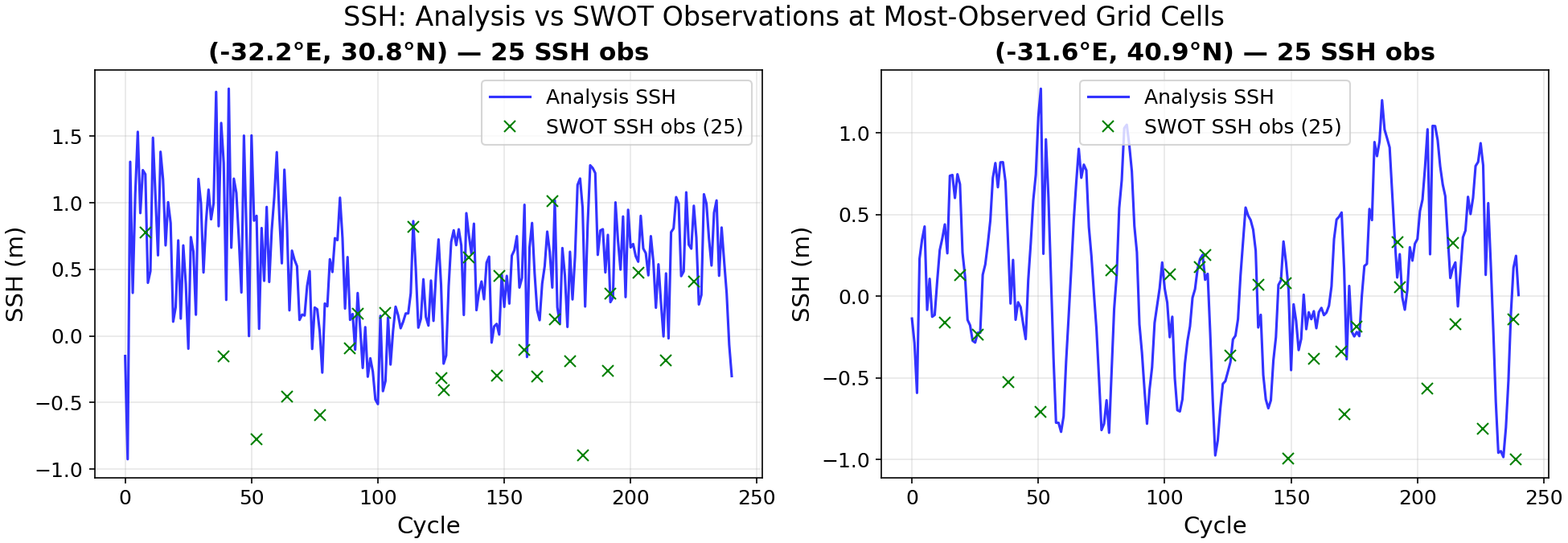}
\caption{LSMCMC~V2 analysis vs.\ SWOT observations: SSH time series at the two most frequently observed grid cells.}
\label{fig:ldata_ssh_ts}
\end{figure}

To complement the aggregate RMSE statistics, Figures~\ref{fig:ldata_vel_ts}--\ref{fig:ldata_ssh_ts} show the analysis time series at the most frequently observed grid cells for each variable. The velocity analysis (Figure~\ref{fig:ldata_vel_ts}) tracks the drifter observations closely at both cells, recovering both the mean flow and the higher-frequency fluctuations in eastward and northward velocity. The SST analysis (Figure~\ref{fig:ldata_sst_ts}) broadly follows the observed SST evolution, although the match is less tight than for velocity owing to the coarser SST observation density and the SST nudging time scale. For SSH (Figure~\ref{fig:ldata_ssh_ts}), the observations combine real SWOT altimetric data with synthetic observations drawn from HYCOM on cycles lacking real passes following realistic SWOT swath geometry. Even at the most frequently observed cells the total count remains modest (around 25 over 240 cycles), and neighboring cells often receive far fewer; the analysis tracks a subset of these observations while spreading the altimetric information spatially. We emphasize that the larger SSH error relative to velocity and SST does \emph{not} indicate filter failure or spurious gravity-wave generation, but reflects two effects. First, SSH is observed far more sparsely than the other fields. At the two most-observed cells the velocity record contains $958$ and $729$ observations and the SST record $318$ and $240$ observations over the 240 cycles, whereas SSH is constrained by only about $25$ observations in each cell. Despite this scarcity, Figure~\ref{fig:ldata_ssh_ts} shows that the analysis still follows the trend of the SSH observations wherever they are available. Second, as was mentioned before the isopycnal MLSWE solver has intrinsic difficulty representing the free-surface height in the presence of the numerous islands of the North Atlantic domain, whose coastlines induce sharp bathymetric gradients near their boundaries that a Cartesian solver cannot resolve faithfully. Although the forecast is initialized from HYCOM at $t_0$, the SSH field gradually departs from HYCOM, even though the velocity and SST fields remain in good agreement. The elevated SSH error is therefore a property of the forward model and the observation sparsity, shared by both LSMCMC and LETKF, rather than a breakdown of the assimilation update.

V2's embarrassingly parallel per-block processing trades slightly higher velocity RMSE for the best SST and SSH performance among all three methods. V2's architecture across $\Gamma = 1{,}400$ blocks makes it more scalable to larger grids.

\subsection{MLSWE with Nonlinear Gaussian Observation Model}\label{subsec:nldata_experiment}

\subsubsection{Problem setup}
We now consider the same MLSWE forward model but with a nonlinear data operator $\mathbf{Y}_{t_k} = \arctan(C_k \mathbf{Z}_{t_k}) + \mathbf{V}_{t_k}$, $\mathbf{V}_{t_k} \sim \mathcal{N}(0, R_k)$. The $\arctan$ function compresses the range of the observations and introduces nonlinearity, making the posterior non-Gaussian. This choice is not a physical instrument model; $C_k$ here is a dimensionless selection/scaling matrix, and $\arctan(C_k\mathbf{Z}_{t_k})$ is interpreted as a numerical transform rather than a physical measurement \cite{bao_2025, liang_2025}. In this setting, the exact Gaussian mixture sampling of Example~\ref{ex:linear_gaussian} is unavailable, and LSMCMC must use an MCMC kernel. We test two kernels for V1: the pCN kernel \cite{pCN} and HMC kernel \cite{HMC}, where the later exploits gradient information via leapfrog integration. V2 uses the pCN kernel throughout. We note that pCN is exact only for a Gaussian reference (prior) measure. The within-cycle prior here is the forecast distribution, which we approximate as Gaussian $\mathcal{N}(\mathbf{m}_f,\widehat{Q}_f)$ with mean and covariance estimated from the $N_f$ forecast samples (which each is assumed Gaussian; see \eqref{eq:signal_observed}); pCN leaves this Gaussian reference invariant while the fully general (probably non-Gaussian) likelihood enters through the acceptance ratio. As a cross-check we also run HMC, which does not require a Gaussian reference and yields matching results, indicating that the Gaussian-prior approximation is adequate for these experiments.

Because the $\arctan$ function saturates at $\pm\pi/2$, applying it to real observations is physically inconsistent: for example, real sea-surface temperature observations lie in the range $\approx 290$--$305$\,K, but $\arctan(300) \approx \pi/2$ regardless of the temperature value, erasing all observational information. The nonlinear observation model is therefore tested exclusively in a synthetic \emph{twin experiment} where observations are generated self-consistently from the arctan operator.

\subsubsection{Twin experiment}
In the twin experiment, a nature run of the MLSWE model generates the reference trajectory $\{\mathbf{Z}_{t_k}^{\text{nr}}\}$. Synthetic observations are created by $\mathbf{Y}_{t_k} = \arctan(C_k \mathbf{Z}_{t_k}^{\text{nr}}) + \mathbf{V}_{t_k}$. The filter starts from a perturbed initial condition. RMSE is computed over \emph{all grid cells} against the nature run, following standard practice for twin experiments (e.g., \cite{miyoshi,greybush}). We note that the posterior mean is the conditional expectation given all observations up to time~$t_k$ and need not coincide exactly with the nature run, which is a single realization of the stochastic dynamics; lower, non-diverging RMSE nevertheless indicates better filter performance. The forward model noise uses $\sigma_x^{\text{vel}} = 0.15$\,m/s, $\sigma_x^{\text{SSH}} = 0.50$\,m, and $\sigma_x^{\text{SST}} = 1.0$\,K.

Table~\ref{tab:nltwin_params} summarizes the key parameters. The pCN V1 variant uses $N_a = 1{,}000$ MCMC steps with $500$ burn-in and step size $\beta = 0.3$. It runs $P=10$ chains each chain has a 500 burn-in and $N_a = 100$. The HMC V1 variant uses $N_a = 500$ with $200$ burn-in, with $L = 10$ leapfrog steps and initial step size $\varepsilon_0 = 0.5$ (adapted to $\sim\!0.14$). V2 runs per-block pCN chains with $N_a = 2{,}000$, $500$ burn-in, and step size $\beta = 0.81$ (enabled by the smaller per-block dimension).

\begin{table}[ht]
\centering
\caption{Parameters for the MLSWE nonlinear twin experiment.}
\label{tab:nltwin_params}
\small
\begin{tabular}{lccc}
\toprule
\multicolumn{4}{c}{State dimension $d = 67{,}200$ \quad | \quad Assimilation cycles $T = 240$} \\
\midrule
Parameter & LSMCMC V1 (pCN) & LSMCMC V1 (HMC) & LSMCMC V2 (pCN) \\
\midrule
Blocks $\Gamma$ & 50 & 50 & 700 \\
Halo radius $r_h$ & --- & --- & 1.5 \\
$N_f$ / $N_a$ & 25 / 1{,}000 & 25 / 500 & 25 / 2{,}000 \\
Burn-in & 500 & 200 & 500 \\
MCMC kernel & pCN & HMC & pCN \\
Step size & $\beta = 0.3$ & $\varepsilon_0 = 0.5$, $L = 10$ & $\beta = 0.81$ \\
Adaptive tuning & target $0.40$ & target $0.65$ & target $0.35$ \\
RTPS $\alpha$ & --- & --- & 0.0 \\
$\sigma_{\text{vel}},\sigma_{\text{SSH}},\sigma_{\text{SST}}$ & \multicolumn{3}{c}{0.10, 0.50, 0.40} \\
$\sigma_x^{\text{vel}},\sigma_x^{\text{SSH}},\sigma_x^{\text{SST}}$ & \multicolumn{3}{c}{0.15, 0.50, 1.0} \\
\bottomrule
\end{tabular}
\end{table}

\begin{figure}[h!]
\centering
\includegraphics[width=0.7\textwidth]{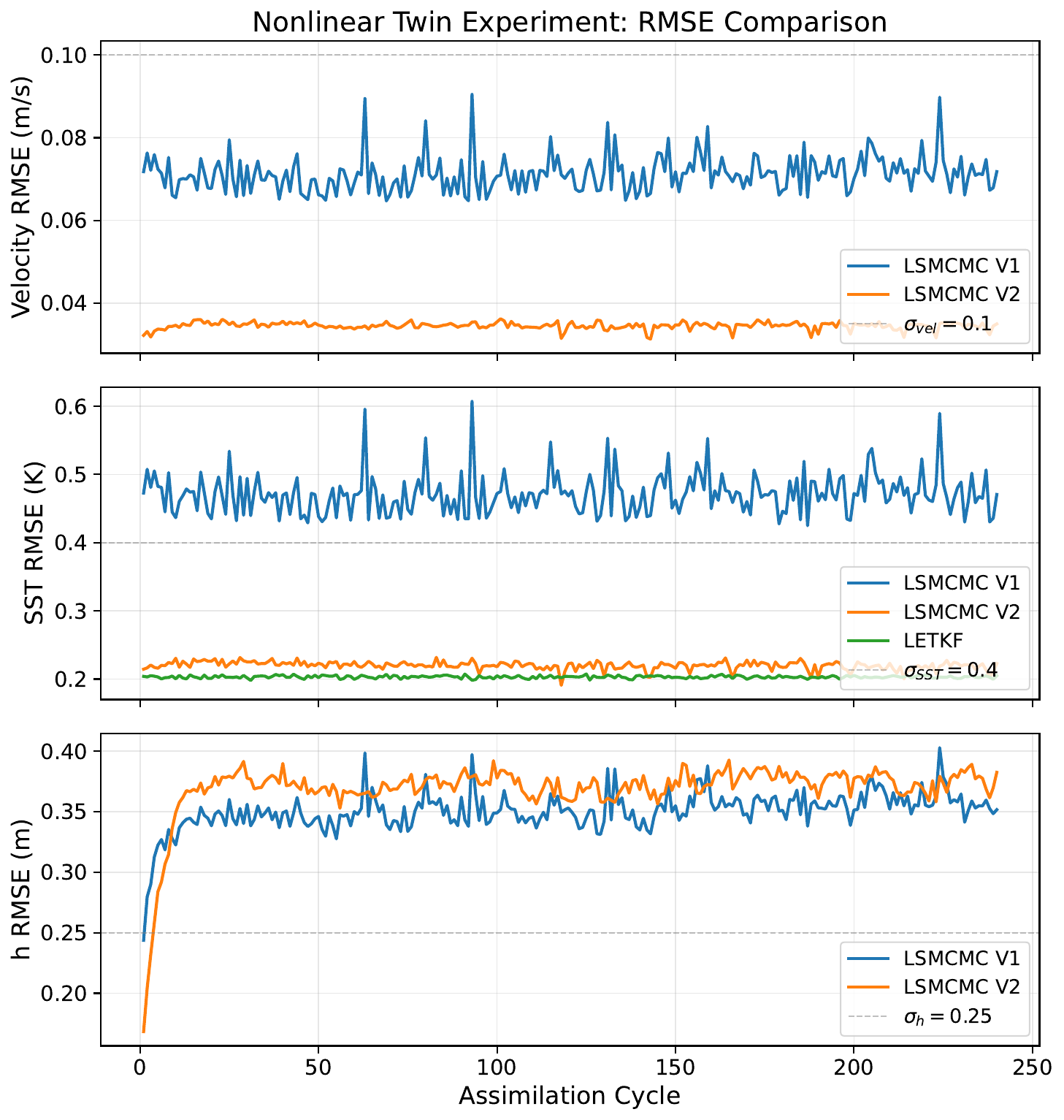}
\caption{MLSWE nonlinear twin experiment (truth/reference here is the nature run used to generate the synthetic observations): velocity, SST, and layer-thickness $h$ RMSE comparison of LSMCMC V1, V2, and LETKF against the nature run over 240 cycles. The LETKF $h$ RMSE $\approx 146$\,m is off-scale and not shown; it remains constant because the $\arctan$ operator maps surface-layer thickness values $h_0$ to $\approx \pi/2$, producing zero observational information for this field.}
\label{fig:nltwin_rmse}
\end{figure}

Table~\ref{tab:nltwin_results} and Figure~\ref{fig:nltwin_rmse} present the results. The HMC V1 variant outperforms pCN V1 across all three fields (velocity $0.0591$ vs $0.0702$\,m/s, SST $0.389$ vs $0.465$\,K, layer thickness $h$ $0.295$ vs $0.346$\,m), demonstrating that HMC's gradient-guided proposals explore the posterior more effectively in this $\sim\!20{,}000$-dimensional reduced domain. HMC achieves this with $2\times$ fewer MCMC iterations ($N_a{=}500$ vs $1{,}000$). V2 outperforms both V1 variants in velocity RMSE ($0.0546$\,m/s, a 22\% improvement over pCN V1) and SST RMSE ($0.362$\,K), while V1 HMC achieves the best $h$ RMSE ($0.295$\,m). V2 is the fastest at $2.0$\,s per cycle thanks to per-block parallelism.

We also ran LETKF ($K{=}25$, $h_{\text{loc}}{=}100$\,km, RTPS${=}0.50$, where $h_{\text{loc}}$ and RTPS are obtained from a sensitivity analysis test) on the same twin problem. The LETKF completed all 240 cycles without divergence, achieving a mean velocity RMSE of $0.072$\,m/s (but not stable; velocity RMSE started around 0.05 and reached 0.1 in the last 20 cycles) and SST RMSE of $0.203$\,K---comparable to LSMCMC for this field because of the nudging term. However, the LETKF \textbf{completely fails to update the layer thickness $h$}, producing a mean $h$ RMSE of $146.66$\,m that remains constant throughout the assimilation window. This failure is explained by the $\arctan$ saturation effect. Although the LETKF does not explicitly linearize the observation operator, it forms the observation-space ensemble perturbation matrix $\mathbf{Y}_b$ by applying $\mathcal{O}_{t_k}$ to each ensemble member and subtracting the ensemble mean. Because the observed surface-layer thickness is $h_0 \approx \mathcal{O}(100\,\text{m})$ (and total thickness satisfies $\sum_k h_k \in [500, 6800]$\,m), every ensemble member maps to $\arctan(h) \approx \pi/2$, so the ensemble spread in observation space collapses: $\mathbf{Y}_b \approx \mathbf{0}$. The LETKF's analysis increment is proportional to $\mathbf{Y}_b^{\mathrm{T}}(\mathbf{Y}_b\mathbf{Y}_b^{\mathrm{T}} + \mathbf{R})^{-1}\mathbf{d}$; with $\mathbf{Y}_b \approx \mathbf{0}$ the denominator is dominated by $\mathbf{R}$ and the Kalman gain vanishes. In contrast, LSMCMC evaluates the full nonlinear likelihood $p(\mathbf{Y} \,|\, \arctan(C\mathbf{Z}))$ via MCMC sampling, which does not require local linearization and can still extract information even from the saturated regime.

\begin{figure}[h!]
\centering
\includegraphics[width=0.9\textwidth]{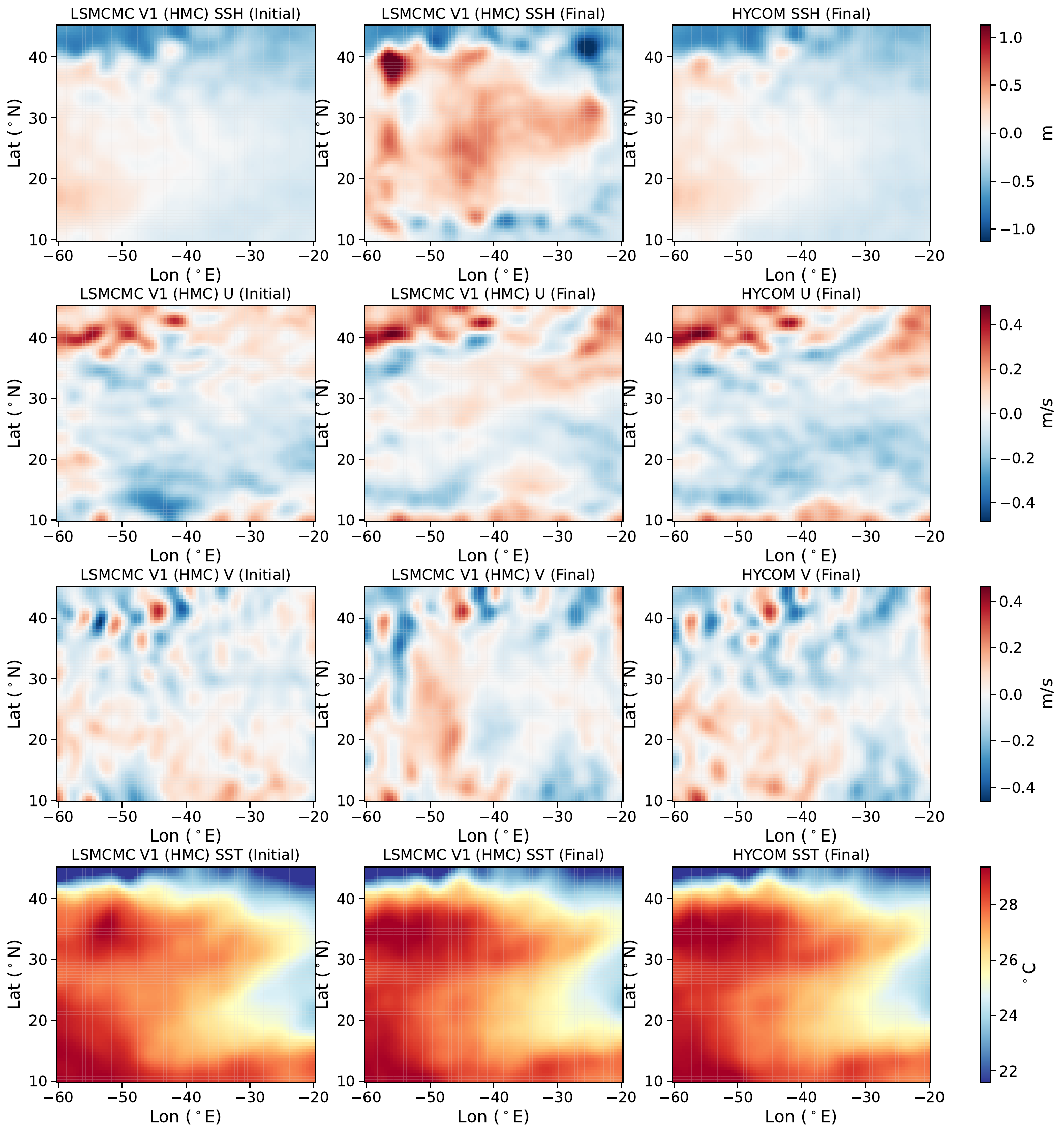}
\caption{MLSWE NL twin: initial and final LSMCMC~V1 (HMC) analysis fields (left, centre) and final HYCOM reanalysis (right). Rows show SSH anomaly, $u$, $v$, and SST.}
\label{fig:nltwin_v1_fields}
\end{figure}

For velocity ($|u|, |v| < 2$\,m/s), the ensemble members map to distinct $\arctan$ values, preserving meaningful spread in $\mathbf{Y}_b$, and the LETKF performs well. For SST ($T \approx 300$\,K), $\arctan(T) \approx \pi/2$ for all members, so $\mathbf{Y}_b$ again collapses and SST observations are uninformative; the LETKF's low SST RMSE of $0.203$\,K is driven entirely by the SST nudging term in the model, not by the data assimilation update.  Figure~\ref{fig:nltwin_v1_fields} shows the HMC V1 analysis fields at the initial and final assimilation times alongside the HYCOM reanalysis, confirming that LSMCMC recovers the large-scale ocean structure despite the nonlinear observation operator.

\begin{table}[ht]
\centering
\caption{MLSWE NL twin experiment: mean RMSE over 240 cycles (vs nature run). The field reported as $h$ is the surface-layer thickness (the $\arctan$ input, $\mathcal{O}(100\,\text{m})$); the physical SSH anomaly is $\mathcal{O}(1\,\text{m})$.}
\label{tab:nltwin_results}
\small
\begin{tabular}{lcccc}
\toprule
Method & Vel RMSE (m/s) & SST RMSE (K) & h RMSE (m) & Time/cycle (s) \\
\midrule
LSMCMC V1 pCN ($\Gamma{=}50$) & 0.0702 & 0.465 & 0.346 & 3.4 \\
LSMCMC V1 HMC ($\Gamma{=}50$) & 0.0591 & 0.389 & \textbf{0.295} & 4.0 \\
LSMCMC V2 ($\Gamma{=}700$) & \textbf{0.0546} & \textbf{0.362} & 0.421 & \textbf{2.0} \\
LETKF ($K{=}25$) & 0.072 & 0.203$^{\dagger}$ & 146.66 & 1.2 \\
\bottomrule
\end{tabular}
\\[2pt]
{\footnotesize $^{\dagger}$\,SST RMSE from model nudging, not data assimilation (arctan SST observations are uninformative).}
\end{table}

The pCN kernel achieves acceptance rates of approximately 35\% for V1 and 64\% for V2, consistent with V2's smaller per-block dimension. The HMC kernel achieves acceptance rates of $69.9\%$ with the adaptive leapfrog step size ($L{=}10$ steps, $\varepsilon$ adapted from $0.5$ to $\sim\!0.14$), reflecting the efficiency of gradient information in the $\sim\!20{,}000$-dimensional reduced domain.

\subsection{MLSWE with Nonlinear Observations and Non-Gaussian Noise}\label{subsec:nlgamma_experiment}

The preceding experiment used a nonlinear observation operator but retained Gaussian observation noise.  In many geophysical applications, however, observation errors exhibit heavy tails that are poorly described by a Gaussian distribution.  \cite{elipot2016} showed that Argos satellite positioning errors for NOAA Global Drifter Program buoys \cite{noaa_hourly, noaa} are non-Gaussian: Kolmogorov--Smirnov tests reject normality for every Argos location class (Section~2.3, Table~4 therein), while Student-$t$ location-scale distributions provide an adequate fit with shape parameters $\nu \approx 3.4$--$5.6$ depending on coordinate and class (Table~3 therein).  Since drifter velocities are derived from these positions, typically by interpolating sparse Argos fixes (sampled every $\sim$1--2\,h) onto a regular hourly grid, the heavy-tailed location errors propagate directly into the velocity estimates.  Indeed, fitting the hourly interpolation residuals to Student-$t$ distributions yields even heavier tails ($\nu \approx 1.9$--$2.4$; Table~8 therein).  Standard ensemble Kalman methods, which implicitly assume Gaussian errors, can be severely degraded by such outliers \cite{pires, anderson-nongauss}.  LSMCMC, by contrast, evaluates the full non-Gaussian likelihood via MCMC and is therefore naturally suited to heavy-tailed noise models.

\subsubsection{Problem setup}
We retain the $\arctan$ observation operator from Section~\ref{subsec:nldata_experiment} but replace the Gaussian observation noise with \emph{Cauchy} (Student-$t$, $\nu{=}1$) noise: $\mathbf{Y}_{t_k} = \arctan(C_k\,\mathbf{Z}_{t_k}) + \boldsymbol{\varepsilon}_{t_k}$, $\varepsilon_{t_k}^{(j)} \sim \sigma_j \cdot t_{1}$, where $t_1$ denotes the standard Cauchy distribution. The Cauchy distribution has undefined mean and variance and exhibits very heavy tails: roughly 30\% of draws exceed $2\sigma$ in magnitude (compared to 5\% for a Gaussian), so the observations are contaminated by frequent, large outliers. Combined with the $\arctan$ saturation, this creates a \emph{doubly challenging} problem: the nonlinear operator makes the posterior non-Gaussian, and the Cauchy noise means that even the observation residuals have infinite variance, violating the fundamental assumptions of all ensemble Kalman methods.

We use a uniform observation noise scale $\sigma_y = 0.05$ for all variables (velocity, SST, and SSH), which is smaller than in Section~\ref{subsec:nldata_experiment}.  This is appropriate because the $\arctan$ operator compresses all observations into $[-\pi/2, \pi/2]$, so the relevant observation-space scale is $\mathcal{O}(1)$, not the physical units of each variable.

This experiment is again a synthetic twin: the nature run generates a reference trajectory, synthetic observations are created by applying the $\arctan$ operator and adding Cauchy noise, and the filter starts from a perturbed initial condition. As in the Gaussian-noise twin, RMSE is computed against the nature run.  The forward-model noise is the same as in Section~\ref{subsec:nldata_experiment}: $\sigma_x^{\text{vel}} = 0.15$\,m/s, $\sigma_x^{\text{SSH}} = 0.50$\,m, $\sigma_x^{\text{SST}} = 1.0$\,K.

\subsubsection{Filter parameters}
Table~\ref{tab:nlgamma_params} summarizes the LSMCMC and LETKF configurations.  We run V1 with both the pCN and HMC kernels.  The LSMCMC pCN kernel evaluates the Cauchy log-likelihood $\log p(\mathbf{y} \mid \mathbf{z}) = \sum_{j} \log t_{1}\bigl(\frac{y_j - \arctan(C z_j)}{\sigma_j}\bigr)$, which naturally down-weights large residuals (outliers) through the polynomial tails of the Cauchy density.  V1 pCN uses $N_a = 2{,}000$ MCMC steps with $500$ burn-in and $P = 10$ parallel chains to improve posterior sampling and run faster; V1 HMC runs with $200$ burn-in and $N_a=300$; and V2 pCN uses $N_a = 2{,}000$ with $500$ burn-in.

Both V1 and V2 employ Robbins--Monro adaptive step-size tuning: at each MCMC step~$s$, the log-step-size is updated by $\log\beta_{s+1} = \log\beta_s + \gamma_s\,(\mathbf{1}_{\text{accept}} - \alpha^*)$ with $\gamma_s = 0.5/(1+s)^{0.6}$, targeting an acceptance rate of $\alpha^* = 0.35$.  

\begin{table}[h!]
\centering
\caption{Parameters for the arctan + Cauchy twin experiment.}
\label{tab:nlgamma_params}
\small
\begin{tabular}{lcccc}
\toprule
\multicolumn{4}{c}{State dimension $d = 67{,}200$ \quad | \quad Assimilation cycles $T = 240$} \\
\midrule
Parameter & LSMCMC V1 (pCN) & LSMCMC V1 (HMC) & LSMCMC V2 (pCN) & LETKF \\
\midrule
Blocks $\Gamma$ & 50 & 50 & 700 & --- \\
Halo $r_h$ / Loc.\ scale & --- & --- & 1.5 & 200\,km \\
$N_f$ / $N_a$ & 25 / 2{,}000 & 25 / 300 & 25 / 2{,}000 & 25 / 25 \\
Burn-in & 500 & 200 & 500 & --- \\
Parallel chains $P$ & 10 & --- & --- & --- \\
MCMC kernel & pCN & HMC & pCN & --- \\
Step size & $\beta_0 = 0.10$ & $\varepsilon = 0.16$, $L = 10$ & $\beta = 0.81$ & --- \\
Adaptive tuning & \multicolumn{2}{c}{Robbins--Monro, target $0.35$/$0.65$} & target $0.35$ & --- \\
RTPS / Inflation & --- & --- & 0.0 & $\alpha{=}0.95$ \\
$\sigma_y$ (all variables) & \multicolumn{4}{c}{0.05} \\
$\sigma_x^{\text{vel}},\sigma_x^{\text{SSH}},\sigma_x^{\text{SST}}$ & \multicolumn{4}{c}{0.15,\; 0.50,\; 1.0} \\
\bottomrule
\end{tabular}
\end{table}

\subsubsection{Results}

\paragraph{LETKF divergence.}
The LETKF diverges catastrophically.  The velocity RMSE at cycle~1 exceeds $12$\,m/s and the layer-thickness $h$ RMSE is effectively infinite from the start.  The combination of $\arctan$ saturation and Cauchy noise is devastating: the Cauchy outliers produce observation residuals that are orders of magnitude larger than expected, and the LETKF's Gaussian assumption treats these as informative signals rather than noise, causing catastrophic Kalman gain updates.

\begin{figure}[h!]
\centering
\includegraphics[width=0.75\textwidth]{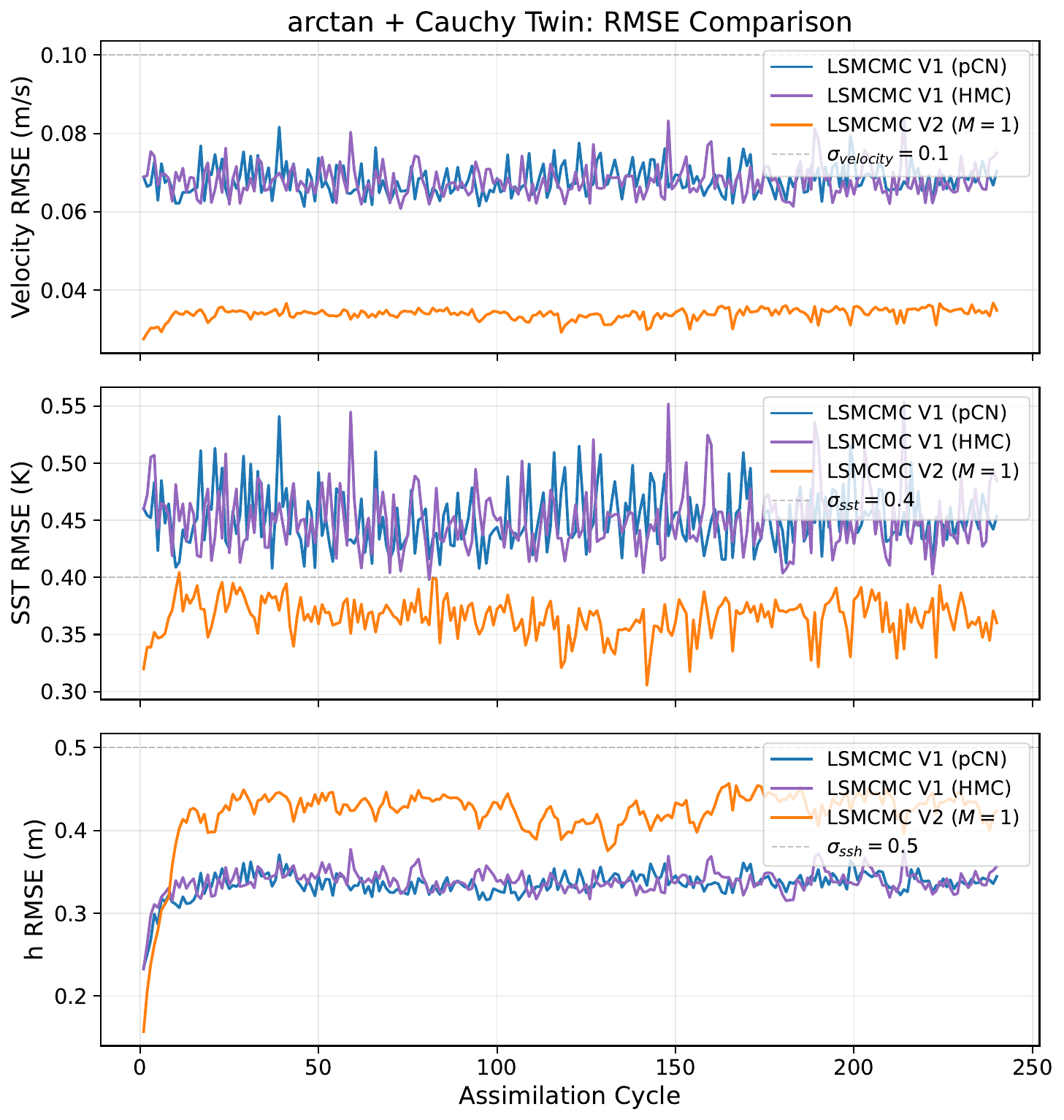}
\caption{Arctan + Cauchy twin experiment (truth/reference here is the nature run used to generate the synthetic observations): velocity, SST, and layer-thickness $h$ RMSE comparison of LSMCMC V1 and V2 ($M{=}1$) against the nature run over 240 cycles. V2 outperforms V1 in velocity and SST, while V1 achieves better $h$ RMSE. We notice here that LETKF diverges and hence its RMSE is not showed.}
\label{fig:nlgamma_rmse}
\end{figure}

\paragraph{LSMCMC V1 and V2.}
Table~\ref{tab:nlgamma_results} and Figure~\ref{fig:nlgamma_rmse} present the RMSE results.  V1~(pCN) achieves a mean velocity RMSE of $0.068$\,m/s, SST RMSE of $0.451$\,K, and layer-thickness $h$ RMSE of $\mathbf{0.335}$\,m over 240 cycles.  V1~(HMC) achieves nearly identical accuracy---velocity $0.068$\,m/s, SST $0.450$\,K, $h$ $0.338$\,m---despite requiring only $N_a = 300$ post-burn-in samples (vs $2{,}000$ for pCN) and no parallel chains.  The HMC kernel's gradient information reduces the per-cycle cost from $8.3$\,s to $3.8$\,s while maintaining the same RMSE, making it a compelling alternative to pCN for V1's high-dimensional reduced domain.  A single V2 run achieves velocity RMSE of $\mathbf{0.056}$\,m/s, SST RMSE of $\mathbf{0.365}$\,K, and $h$ RMSE of $0.421$\,m, outperforming both V1 variants in velocity and SST while being faster per cycle ($2.2$\,s).

The pCN acceptance rates are approximately $32\%$ for V1 (with Robbins--Monro adaptation stabilizing at $\beta \approx 0.06$) and $\sim\!35\%$ across V2 blocks, close to the target of $35\%$.  The HMC kernel achieves acceptance rates of $\sim\!69\%$ with final step size $\varepsilon \approx 0.16$ and $L = 10$ leapfrog steps, confirming that gradient information is highly beneficial in V1's $\sim\!20{,}000$-dimensional domain.  The longer burn-in ($1{,}000$ steps) and $P = 10$ parallel chains in V1~(pCN) ensure adequate exploration of the posterior despite the challenging Cauchy likelihood; V1~(HMC) achieves comparable exploration with only $200$ burn-in steps and a single chain.

\begin{table}[h!]
\centering
\caption{Arctan + Cauchy twin experiment: mean RMSE over 240 cycles (vs nature run). The field reported as $h$ is the surface-layer thickness (the $\arctan$ input, $\mathcal{O}(100\,\text{m})$); the physical SSH anomaly is $\mathcal{O}(1\,\text{m})$.}
\label{tab:nlgamma_results}
\small
\begin{tabular}{lcccc}
\toprule
Method & Vel RMSE (m/s) & SST RMSE (K) & h RMSE (m) & Time/cycle (s) \\
\midrule
LSMCMC V1 pCN ($\Gamma{=}50$, $P{=}10$) & 0.068 & 0.451 & \textbf{0.335} & 8.3 \\
LSMCMC V1 HMC ($\Gamma{=}50$) & 0.068 & 0.450 & 0.338 & 3.8 \\
LSMCMC V2 ($\Gamma{=}700$) & \textbf{0.056} & \textbf{0.365} & 0.421 & \textbf{2.2} \\
LETKF ($K{=}25$) & \multicolumn{4}{c}{Diverged (vel RMSE $> 12$\,m/s at cycle 1)} \\
\bottomrule
\end{tabular}
\end{table}

\begin{figure}[h!]
\centering
\includegraphics[width=0.95\textwidth]{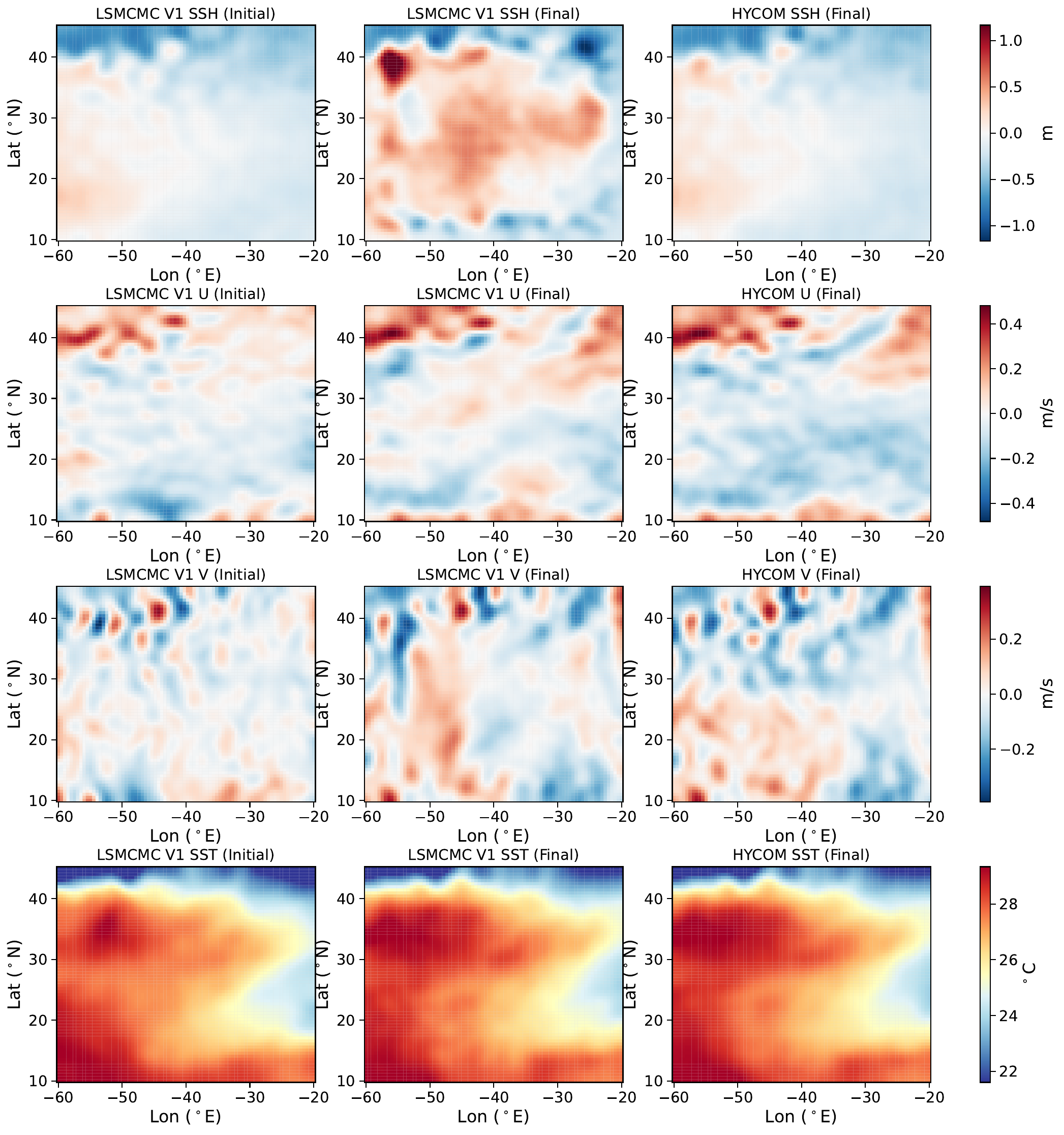}
\caption{Arctan + Cauchy twin: LSMCMC~V1 analysis fields at the initial and final (cycle~240) assimilation times, compared with the HYCOM reanalysis at the final time. Rows show SSH anomaly, eastward velocity~($U$), northward velocity~($V$), and SST.}
\label{fig:nlgamma_v1_fields}
\end{figure}

\begin{figure}[h!]
\centering
\includegraphics[width=0.95\textwidth]{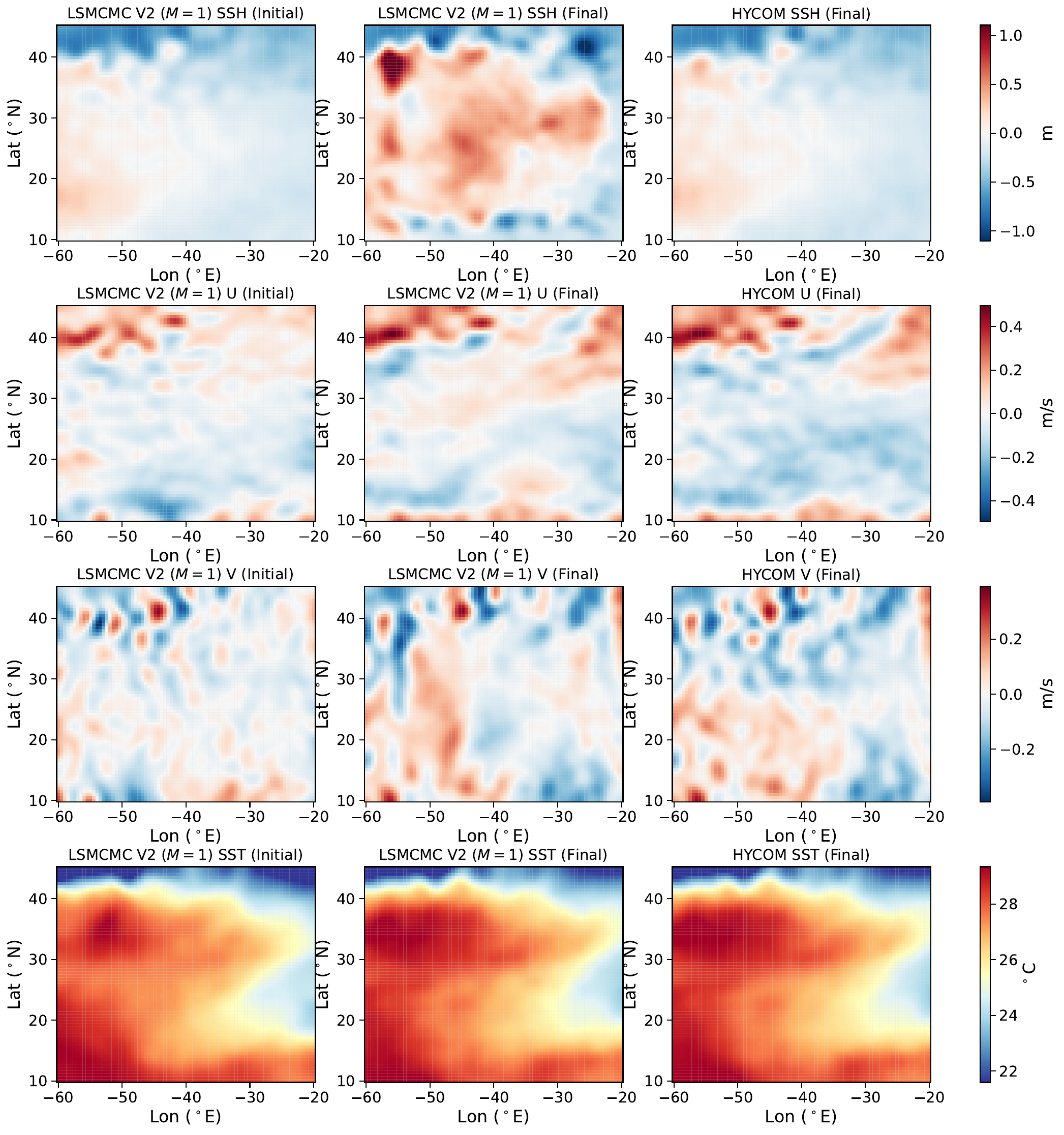}
\caption{As in Figure~\ref{fig:nlgamma_v1_fields}, but for LSMCMC~V2. The per-block localization produces slightly smoother fields.}
\label{fig:nlgamma_v2_fields}
\end{figure}

\begin{figure}[h!]
\centering
\includegraphics[width=0.99\textwidth]{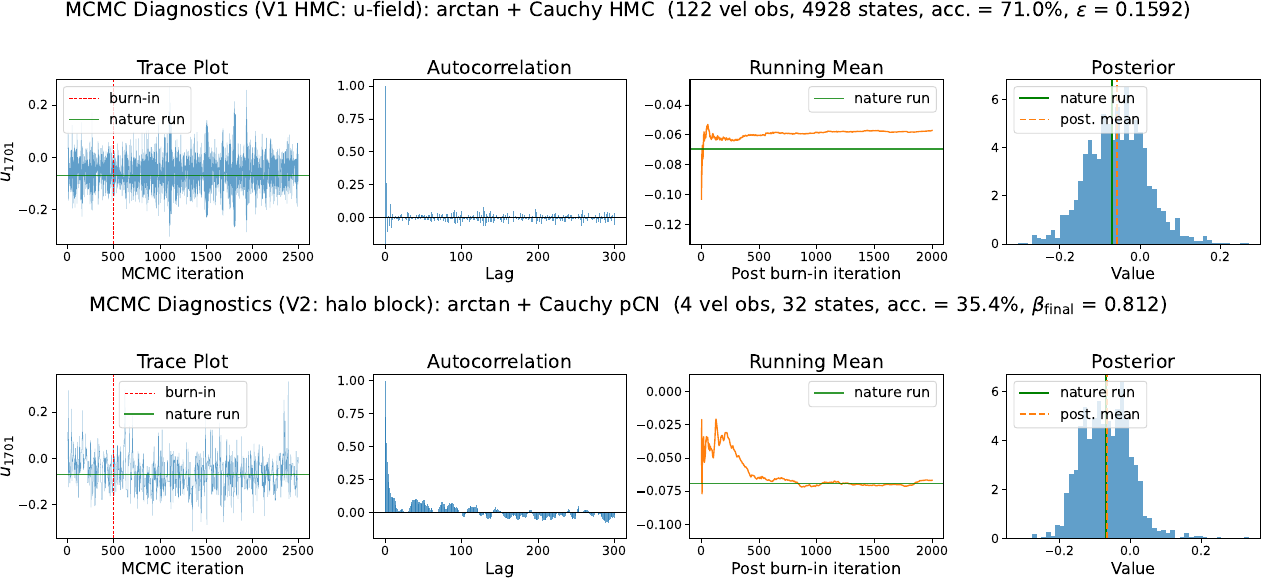}
\caption{MCMC diagnostics for the arctan + Cauchy twin experiment at cycle~120. Each row shows the $u$-velocity component at the most-observed grid cell: trace plot, autocorrelation function, running posterior mean, and marginal posterior histogram.  Top: V1 HMC chain over the block-partition reduced domain ($4{,}928$ state dimensions from $44$ observed partition blocks, $122$ velocity observations, $L{=}10$ leapfrog steps, $\varepsilon_{\text{init}}{=}0.005$, $\varepsilon_{\text{final}}\approx 0.16$, acceptance $\approx 71\%$).  Bottom: V2 pCN chain on a halo-localized block ($32$ layer-0 state components, $4$ velocity observations, acceptance $\approx 35\%$, $\beta_{\text{final}}\approx 0.81$).  The green line marks the nature-run value for reference; the chains converge to the filtering posterior, which need not coincide with the nature run.}
\label{fig:nlgamma_mcmc_diag}
\end{figure}

\begin{figure}[h!]
\centering
\includegraphics[width=0.99\textwidth]{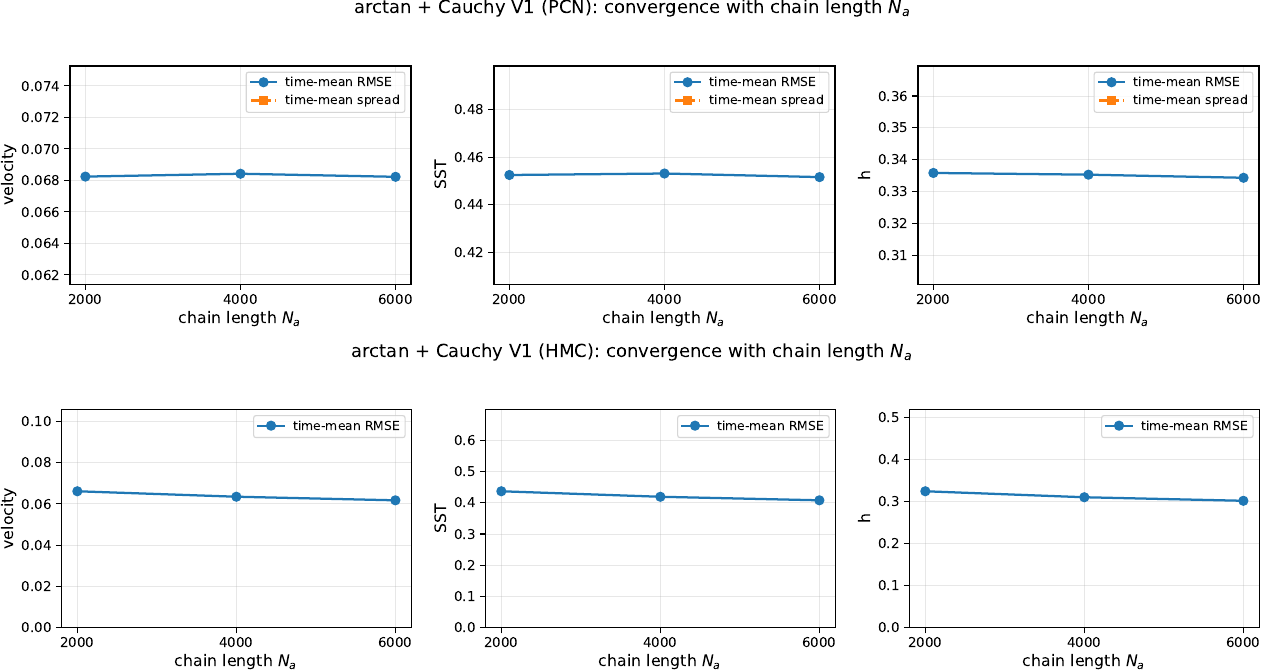}
\caption{Convergence of the time-averaged analysis RMSE (velocity, SST, and $h$) with the per-cycle MCMC chain length $N_a\in\{2{,}000,\,4{,}000,\,6{,}000\}$ for the arctan + Cauchy twin experiment using V1 algorithm. Top: pCN kernel; bottom: HMC kernel. Doubling or tripling the chain length changes the RMSE by only a few percent, indicating that the chains have already converged at $N_a\sim\mathcal{O}(10^3)$.}
\label{fig:nlgamma_convergence_rmse}
\end{figure}

\paragraph{Discussion.}
This experiment demonstrates that LSMCMC handles the combination of a nonlinear observation operator ($\arctan$) and Cauchy noise ($\nu{=}1$, the heaviest possible Student-$t$ distribution with infinite mean) without any modification to the algorithm: both the pCN and HMC kernels target the correct non-Gaussian posterior by evaluating the Cauchy log-likelihood directly.  The LETKF fails catastrophically because the $\arctan$ saturation collapses observation-space ensemble perturbations (making the Kalman gain vanish) and the Gaussian error assumption is strongly violated by the heavy-tailed Cauchy noise.  Figures~\ref{fig:nlgamma_v1_fields} and~\ref{fig:nlgamma_v2_fields} show the analysis fields for V1 and V2, respectively, alongside the HYCOM reanalysis, confirming that the LSMCMC analysis captures the large-scale ocean structure despite the extreme observation noise.

The close agreement between pCN and HMC V1 results (Table~\ref{tab:nlgamma_results}) demonstrates that both kernels converge to the same posterior.  The HMC kernel achieves this with $6.7\times$ fewer MCMC iterations ($300$ vs $2{,}000$) and $2.2\times$ lower per-cycle cost ($3.8$\,s vs $8.3$\,s), making it the more efficient choice for V1's high-dimensional reduced domain.  The V2 advantage over V1 in velocity and SST RMSE (18\% and 19\% improvement, respectively) mirrors the pattern observed in the Gaussian-noise arctan experiment (Section~\ref{subsec:nldata_experiment}), confirming that V2's per-block localization benefits from higher effective acceptance rates in the smaller-dimensional blocks.  V1 retains a 20\% advantage in $h$, consistent with its ability to capture cross-block correlations in the joint observed-block chain.

Figure~\ref{fig:nlgamma_mcmc_diag} presents MCMC diagnostics for V1 (HMC) and V2 (pCN) at cycle~120, showing the $u$-velocity component at the most-observed grid cell.  The V1 HMC chain runs over the block-partition reduced domain with $4{,}928$ state dimensions and $122$ velocity observations ($\varGamma{=}50$), using $L{=}10$ leapfrog steps with adaptively tuned step size $\varepsilon_{\text{final}}\approx 0.16$ and acceptance $\approx 68\%$.  The V2 pCN chain runs on a halo-localized block with $32$ state dimensions and $4$ velocity observations ($\varGamma{=}700$, $r_{\text{loc}}{=}1.5$), with $\approx 35\%$ acceptance and $\beta_{\text{final}}\approx 0.81$.  All running means settle near the filtering posterior mean (the nature-run value is shown for reference), and the HMC chains exhibit rapid decorrelation thanks to the gradient-guided leapfrog dynamics. Figure~\ref{fig:nlgamma_convergence_rmse} shows the time-averaged analysis RMSE (velocity, SST, and $h$) versus the per-cycle chain length $N_a$ for both kernels, which stays essentially flat as $N_a$ increases from $2{,}000$ to $6{,}000$.  This study is done for V1, whose state dimension is much larger than in V2, where the MCMC acts only on a small halo-localized state.

We note that the Cauchy ($\nu{=}1$) noise used here is more extreme than the heavy-tailed errors observed in real drifter data: \cite{elipot2016} report $\nu \approx 3.4$--$5.6$ for Argos location errors and $\nu \approx 1.9$--$2.4$ for hourly interpolation residuals, all of which have finite variance.  The Cauchy distribution, by contrast, has infinite variance, making this a stringent stress test.  The fact that LSMCMC handles $\nu{=}1$ noise without modification confirms its robustness to arbitrary heavy-tailed distributions.

\subsection{Summary and Discussion}\label{subsec:summary}

Table~\ref{tab:overall_summary} provides a comprehensive comparison across all experiments.

The last two experiments are synthetic twin experiments: (i) nonlinear observations with Gaussian noise (Section~\ref{subsec:nldata_experiment}) and (ii) nonlinear observations with non-Gaussian (Cauchy) noise (Section~\ref{subsec:nlgamma_experiment}).
\begin{table}[ht]
\centering
\caption{Summary of all experiments. For the arctan twin experiments, the reported field is the surface-layer thickness $h$ (the $\arctan$ input, $\mathcal{O}(100\,\text{m})$); for the linear-observation experiment, the same reported scalar is the assimilated thickness variable, diagnostically equivalent to SSH anomaly through $\eta=(\sum_k h_k)-H_b$.}
\label{tab:overall_summary}
\small
\begin{tabular}{llcccr}
\toprule
Experiment & Method & Vel RMSE & SST RMSE & h RMSE & Time/cyc \\
\midrule
\multirow{3}{*}{\shortstack[l]{Linear Obs. +\\
Gaussian ($d{=}67{,}200$)}} 
 & LETKF & 0.0227 & 0.426 & 0.369 & 1.2\,s \\
 & LSMCMC V1 (direct sampling) & \textbf{0.0117} & \textbf{0.174} & \textbf{0.155} & 1.4\,s \\
 & LSMCMC V2 (direct sampling) & 0.0228 & 0.331 & 0.346 & 1.4\,s \\
\midrule
\multirow{4}{*}{\shortstack[l] {NL Obs. arctan +\\Gaussian ($d{=}67{,}200$)}} 
 & LETKF & 0.072 & 0.203 & 146.66 & \textbf{1.2\,s} \\
 & LSMCMC V1 (pCN) & 0.0702 & 0.465 & 0.346 & 3.4\,s \\
 & LSMCMC V1 (HMC) & 0.0591 & 0.389 & \textbf{0.295} & 4.0\,s \\
 & LSMCMC V2 (pCN) & \textbf{0.0546} & \textbf{0.362} & 0.421 & 2.0\,s \\
\midrule
\multirow{4}{*}{\shortstack[l]{NL obs. arctan +\\Cauchy ($d{=}67{,}200$)}}
 & LETKF & \multicolumn{4}{c}{Diverged (vel RMSE $> 12$\,m/s at cycle 1)} \\
 & LSMCMC V1 (pCN) & 0.068 & 0.451 & \textbf{0.335} & 8.3\,s \\
 & LSMCMC V1 (HMC) & 0.068 & 0.450 & 0.338 & 3.8\,s \\
 & LSMCMC V2 (pCN) & \textbf{0.056} & \textbf{0.365} & 0.421 & \textbf{2.2\,s} \\
\bottomrule
\end{tabular}
\end{table}

Several patterns emerge from these results:

\begin{enumerate}
\item \textbf{V1 vs V2 trade-off.} V1 generally achieves the best layer-thickness $h$ RMSE in a single-chain setting. V2 compensates with better velocity and SST RMSE and lower per-cycle cost. In the NL arctan twin experiment (Gaussian noise), V2 outperforms V1 in velocity and SST thanks to higher acceptance rates in the smaller per-block pCN chains. The same pattern holds in the arctan + Cauchy experiment: V2 achieves 18\% lower velocity RMSE and 19\% lower SST RMSE than V1, while V1 retains a 20\% $h$ advantage.

\item \textbf{LSMCMC vs LETKF.} For linear observations, LSMCMC V1 outperforms LETKF in velocity RMSE, and V2 leads in SST, while LETKF achieves the best $h$ RMSE. All three methods completed the full 240 cycles without instability. For the arctan nonlinear operator with Gaussian noise, the LETKF \emph{completely fails} to update the layer thickness $h$---its $h$~RMSE of $146.66$\,m is identical to the prior, reflecting the fact that $\arctan$ saturation collapses the observation-space ensemble perturbations for the surface-layer thickness $h_0 \approx \mathcal{O}(100\,\text{m})$, driving the Kalman gain to zero. For the arctan operator with Cauchy noise, the LETKF \emph{diverges catastrophically} within the first cycle, whereas both LSMCMC variants complete all 240 cycles without difficulty.

\item \textbf{Non-Gaussian noise robustness.} The arctan + Cauchy experiment highlights a fundamental advantage of LSMCMC: because it evaluates the full (non-Gaussian) likelihood via MCMC, it naturally handles heavy-tailed observation noise without algorithmic modification. The Cauchy ($\nu{=}1$) noise has no finite mean or variance and produces frequent extreme outliers that cause the LETKF to produce catastrophic updates, while LSMCMC's log-likelihood evaluation automatically down-weights these outliers through the polynomial tails of the Cauchy density.

\item \textbf{Adaptive MCMC.}  The Robbins--Monro step-size adaptation used in the Cauchy experiment achieves stable acceptance rates of $\sim\!32$--$35\%$ (pCN) and $\sim\!68\%$ (HMC) across all cycles, demonstrating that the adaptive tuning converges reliably despite the frequent large outliers produced by the Cauchy observation noise

\item \textbf{pCN vs HMC.}  In both nonlinear experiments, the HMC kernel achieves lower RMSE than pCN with $2\times$ fewer MCMC iterations ($N_a{=}500$ vs $1{,}000$).  In the Cauchy experiment, HMC reduces V1 per-cycle cost by $2.2\times$ ($3.8$\,s vs $8.3$\,s).  HMC's gradient-guided proposals are particularly effective in V1's $\sim\!20{,}000$-dimensional reduced domain, where pCN's random-walk proposals require many more steps to decorrelate.  HMC is the more efficient choice for high-dimensional V1 problems.

\item \textbf{Computational cost.} In the linear case, all methods achieve $\sim 1.2{-}1.4$\,s per cycle since the dominant cost is the MLSWE forecast step. In the nonlinear cases, V1 costs $3.4$--$8.3$\,s per cycle (depending on the kernel and experiment), while V2 costs only $2.0$--$2.2$\,s (parallel per-block chains).
\end{enumerate}

\section{Conclusion}
\label{sec:conclusion}

We have presented two localization variants of the sequential MCMC (SMCMC) filter for high-dimensional data assimilation. Both variants exploit the spatial sparsity of observations by partitioning the domain into subdomains and restricting the MCMC updates to regions where observations exist, thereby reducing the effective state dimension from $d$ to $d'$ (with $d'<d$ in Variant~1 and $d' \ll d$ in Varient~2). Variant~1 (joint observed-block localization) collects all observed subdomains into a single reduced domain and runs parallel MCMC chains over this combined region. Variant~2 (halo-based per-block localization) decomposes the problem into independent blocks with Gaspari--Cohn tapering, enabling embarrassing parallelism across blocks.

We distinguished between the number of forecast samples $N_f$ and analysis samples $N_a$, allowing long MCMC chains ($N_a \gg N_f$) that thoroughly explore the posterior while economizing on the forecast step. When the observation model is linear and Gaussian, the filtering density reduces to a Gaussian mixture from which independent samples can be drawn without MCMC iterations.

Numerical experiments on a linear Gaussian model and the 3-layer MLSWE with linear-Gaussian, nonlinear-Gaussian, and nonlinear-non-Gaussian observation models confirmed that both LSMCMC variants match or outperform the LETKF under standard linear-Gaussian observations, while providing a sampling-based posterior that can represent non-Gaussian features. Under nonlinear and heavy-tailed observation models---where the LETKF's observation-space ensemble perturbations collapse or its Gaussian assumption is violated---LSMCMC remains stable and accurate throughout the full assimilation window because it evaluates the exact nonlinear likelihood. Analysis of real NOAA drifter data further motivates this capability, as ocean observation errors are well described by heavy-tailed Student-$t$ distributions \cite{elipot2016}. On the computational side, both the pCN and HMC MCMC kernels perform reliably: pCN provides dimension-robust proposals for V2's moderate per-block problems, while HMC exploits gradient information to dramatically reduce the iteration count for V1's high-dimensional joint domain. V2's independent per-block architecture enables embarrassing parallelism and straightforward multi-chain averaging for further variance reduction.

On balance, we recommend Variant~2 as the default choice for practitioners. V2 achieves superior velocity and SST accuracy across all nonlinear experiments (18--19\% lower RMSE than V1), runs 2--4$\times$ faster per cycle thanks to its embarrassingly parallel per-block architecture, and scales naturally to larger domains because each block is solved independently. Two further considerations reinforce this preference. First, because V2 decomposes the observed region into many independent blocks, each MCMC chain explores a much lower-dimensional space than V1's single combined domain; this smaller per-chain dimension yields higher acceptance rates and more efficient posterior exploration, which is the main reason V2 attains its accuracy with the dimension-robust pCN kernel and without gradient information. Second, V2 avoids the boundary discontinuity that V1 can introduce: in V1 the analysis is computed on a single combined observed region and grafted back onto the unobserved state, which may leave a jump between updated and retained variables, whereas V2's Gaspari--Cohn taper drives each block's increment smoothly to zero at its halo boundary, so the analysis transitions continuously into the surrounding forecast. This smooth, tapered update is also the standard remedy for the unbalanced gravity waves that abrupt update/retain discontinuities can excite in a shallow-water model; we nonetheless note that, as in localized particle filters, exact dynamical balance is not guaranteed, and for balance-critical (e.g.\ high-resolution atmospheric) applications the analysis increment should be combined with an incremental analysis update or digital-filter initialization.

In future work, we plan to incorporate adaptive localization schemes that dynamically adjust the block partition and halo radius based on local observation density \cite{anderson}. We also aim to extend LSMCMC to operational-resolution grids ($\sim 1{,}000 \times 1{,}000$) and couple it with established physical models such as WRF, ROMS, HYCOM, and coupled atmosphere--ocean frameworks (e.g., CESM, EC-Earth).

\subsection*{Acknowledgements}
The work of HR and OK was supported by KAUST baseline fund. HC acknowledges support from Florida State University's CRC Seed Grant 047080. We thank the Editor, the Associate Editor, and the two anonymous reviewers for their careful reading and constructive comments, which substantially improved the manuscript.

\subsection*{Data Availability Statement}

The HYCOM GLBv0.08 reanalysis data used in this study for the RSWEs initial state and boundary conditions are openly available from the HYCOM Consortium at \url{https://www.hycom.org/dataserver/gofs-3pt1/reanalysis} \cite{chassignet2007,metzger2014}. The 6-hourly quality-controlled drifter data used in the second RSWEs model are openly available from the NOAA Global Drifter Program at \url{https://doi.org/10.25921/7ntx-z961}. The SWOT Level~2 Low Rate Sea Surface Height data used in the second RSWEs model are openly available from NASA's Physical Oceanography Distributed Active Archive Center (PO.DAAC) at \url{https://doi.org/10.5067/SWOT-SSH-1.0}. 
The Jupyter Notebook to execute the analysis in this paper can be run interactively at \url{https://mybinder.org/v2/gh/ruzayqat/LSMCMC/HEAD?labpath=LSMCMC.ipynb}, is hosted on GitHub at \url{https://github.com/ruzayqat/LSMCMC}, and is preserved at Zenodo (\url{https://doi.org/10.5281/zenodo.18888945}, Version 1.0.0, MIT License).

\appendix
\section{The Pseudocode for the Original SMCMC Algorithm}
The original algorithm of the SMCMC filter \cite{smcmc} is presented below. Convergence of the SMCMC filter is established in \cite{martin}. Note that in the original formulation, the same number of samples $N$ is used for both the forecast and analysis steps.

\setcounter{algorithm}{0}
\renewcommand{\thealgorithm}{A.\arabic{algorithm}}
\begin{flushleft}
\captionsetup[algorithm]{style=algori}
\captionof{algorithm}{Pseudocode for the original SMCMC filtering method \protect\cite{smcmc} for $T$ observational time steps.}
\label{alg:smcmc}

 \textbf{Input:} The initial state $\breve{\mathbf{Z}}_0=\mathbf{z}_0$, the observations $\{\mathbf{Y}_{t_k} = \mathbf{y}_{t_k}\}_{k\geq 1}$, the number of samples $N$, and the number of time discretization steps (observational time-lag) $L$. Set $\tau_k=(t_k-t_{k-1})/L$. 

\begin{enumerate}
\item Initialize: For $l=0,\cdots,L-1$ compute $\breve{\mathbf{Z}}_{(l+1)\tau_1} = \Phi(\breve{\mathbf{Z}}_{l\tau_1},l\tau_1;(l+1)\tau_1)$. Compute $\tilde{\mathbf{Z}}_{t_1}:=\breve{\mathbf{Z}}_{t_1}+ \mathbf{W}_{t_1}$, where $\mathbf{W}_{t_1}\sim \mathcal{N}_d(0,Q_1)$. Run an MCMC kernel of choice initialised at $\tilde{\mathbf{Z}}_{t_1}$ to generate $N$ samples $\{\mathbf{Z}_{t_1}^{(i)}\}_{i=1}^{N}$ from $\pi_1$ in \eqref{eq:pi_1}. Set $\widehat{\pi}_{1}^{N}(\varphi) \leftarrow \frac{1}{N} \sum_{i=1}^{N} \varphi(\mathbf{Z}_{t_1}^{(i)})$.

\item For $k=2,\ldots,T$: 
\begin{enumerate}
\item For $i=1,\cdots, N$: compute 
\begin{equation*}
\breve{\mathbf{Z}}_{(l+1)\tau_k+t_{k-1}}^{(i)} = \Phi(\breve{\mathbf{Z}}_{l\tau_k+t_{k-1}}^{(i)},l\tau_k+t_{k-1};(l+1)\tau_k+t_{k-1}),
\end{equation*}
where $0\leq l \leq L-1$ and $\breve{\mathbf{Z}}^{(i)}_{t_{k-1}}= \mathbf{Z}^{(i)}_{t_{k-1}}$.  
 
\item Run an MCMC kernel of choice to return $N$ samples $\{\mathbf{Z}_{t_k}^{(i)}\}_{i=1}^{N}$. Set $\widehat{\pi}_{k}^{N}(\varphi) \leftarrow \frac{1}{N} \sum_{i=1}^{N} \varphi(\mathbf{Z}_{t_k}^{(i)})$.
\end{enumerate}
\end{enumerate}
\textbf{Output:} Return $\{\widehat{\pi}_k^{N}(\varphi)\}_{k\in\{1,\cdots,T\}}$.

\vspace{-0.1cm}
\hrulefill
\vspace{0.2cm}
\end{flushleft}


\begin{thebibliography}{99}
\bibitem{ades}
{\sc Ades, M. \& van Leeuwen, P. J. }~(2013). An exploration of the equivalent weights particle filter. \emph{Q. J. R. Meteorol. Soc.} {\bf 139}(672): 820--840.

\bibitem{etopo1}
{\sc Amante, C. \& Eakins, B. W.}~(2009). ETOPO1 1 Arc-Minute Global Relief Model: Procedures, Data Sources and Analysis. \emph{NOAA Technical Memorandum NESDIS NGDC-24}. National Geophysical Data Center, NOAA. \url{https://doi.org/10.7289/V5C8276M}

\bibitem{anderson03}
{\sc Anderson, J. L.}~(2003). A local least squares framework for ensemble filtering. \emph{Mon. Wea. Rev.}, {\bf 131}: 634--642.

\bibitem{anderson}
{\sc Anderson, J. L .}~(2007). Exploring the need for localization in ensemble data assimilation using a hierarchical ensemble filter. \emph{Physica D} {\bf 230}: 99--111.

\bibitem{anderson12}
{\sc Anderson, J. L.}~(2012). Localization and sampling error correction in ensemble Kalman filter data assimilation. \emph{Mon. Wea. Rev.}, {\bf 140}: 2359--2371.

\bibitem{anderson-nongauss}
{\sc Anderson, J. L. \& Anderson, S. L.}~(1999). A Monte Carlo implementation of the nonlinear filtering problem to produce ensemble assimilations and forecasts. \emph{Mon. Wea. Rev.}, {\bf 127}: 2741--2758.

\bibitem{bain}
{\sc Bain}, A., \& {\sc Crisan}, D.~(2009). \emph{Fundamentals of Stochastic Filtering}. Springer: New York.

\bibitem{bao_2025}
{\sc Bao, F., Chipilski, H. G., Liang, S., Zhang, G. \& Whitaker, J. S.}~(2025). Nonlinear Ensemble Filtering with Diffusion Models: Application to the Surface Quasigeostrophic Dynamics. \emph{Mon. Wea. Rev.}, {\bf 153}(7): 1155--1169.

\bibitem{bengtsson}
{\sc Bengtsson, T., Bickel, P. \& Li, B.}~(2008). Curse-of-dimensionality
revisited: collapse of the particle filter in very large scale systems.
\emph{IMS Collections: Prob. Stat. Essays Honor David A. Freedman}. {\bf 2}:316--334.

\bibitem{berzuini}
{\sc Berzuini, C., Best, N. G., Gilks, W. R. \& Larizza, C.}~(1997). Dynamic conditional independence models and Markov chain Monte Carlo methods. \emph{J. Amer. Statist. Soc.} {\bf 92}:1403--1412.

\bibitem{pCN}
{\sc Cotter, S. L., Roberts, G. O., Stuart, A. M. \& White, D.}~(2013). MCMC Methods for Functions: Modifying Old Algorithms to Make Them Faster. \emph{Statist. Sci.} {\bf 28}(3): 424--446.

\bibitem{cappe}
{\sc Capp\'e}, O., {\sc Ryden}, T, \& {\sc Moulines}, \'E.~(2005). \emph{Inference in Hidden Markov Models}. Springer: New York.

\bibitem{carmi} 
{\sc Carmi, A., Septier, F., \& Godsill, S. J.}~(2012). The Gaussian mixture MCMC particle algorithm for dynamic cluster tracking. \emph{Automatica}, {\bf 48}(10): 2454--2467.

\bibitem{carrassi}
{\sc Carrassi, A., Grudzien, C., Bocquet, M., Demaeyer, J., Raanes, P. \&  Vannitsem, S.} ~(2020). \emph{`Data assimilation for chaotic systems', in Data Assimilation for Atmospheric, Oceanic and Hydrological Applications}, edited by S.-K. Park and X. Liang. Springer Science \& Business.

\bibitem{chassignet2007}
{\sc Chassignet, E. P., Hurlburt, H. E., Smedstad, O. M., Halliwell, G. R., Hogan, P. J., Wallcraft, A. J., Baraille, R. \& Bleck, R.}~(2007). The HYCOM (HYbrid Coordinate Ocean Model) data assimilative system. \emph{J. Mar. Syst.} {\bf 65}(1--4): 60--83.

\bibitem{chipilski_2025}
{\sc Chipilski, H. G.}~(2025). Exact Nonlinear State Estimation. \emph{Journal of the Atmospheric Sciences}, {\bf 82}(4): 809--827.

\bibitem{cohn_1997}
{\sc Cohn, S.~E.}~(1997). An Introduction to Estimation Theory. Data Assimilation Office, Goddard Space Flight Center, Greenbelt, Maryland. Internal Report. Available at: \url{https://gmao.gsfc.nasa.gov/pubs/docs/Cohn192.pdf}.


\bibitem{cotter}
{\sc Cotter, C., Crisan, D., Holm, D.,  Pan, W., \& Shevchenko, I.}~(2020). A Particle Filter for Stochastic Advection by Lie Transport: A Case Study for the Damped and Forced Incompressible Two-Dimensional Euler Equation. \emph{SIAM/ASA J. Uncert. Quant.}, {\bf 8}:1446--1492.

\bibitem{delmoral2004}
{\sc Del Moral, P.}~(2004).
\emph{Feynman--Kac Formulae: Genealogical and Interacting Particle Systems with Applications}.
Springer, New York.

\bibitem{delm13}
{\sc Del Moral}, P.~(2013). \emph{Mean Field Simulation for Monte Carlo Integration}. Chapman \& Hall: London.


\bibitem{elipot2016}
{\sc Elipot, S., Lumpkin, R., Perez, R.~C., Lilly, J.~M., Early, J.~J. \& Sykulski, A.~M.}~(2016). A global surface drifter data set at hourly resolution. \emph{Journal of Geophysical Research: Oceans}, {\bf 121}(5):2937--2966. 

\bibitem{noaa_hourly}
{\sc Elipot, S., Sykulski, A., Lumpkin, R., Centurioni, L. \& Pazos, M.}~(2022). Hourly location, current velocity, and temperature collected from Global Drifter Program drifters world-wide. Version 2.00. \emph{NOAA National Centers for Environmental Information.} Dataset. \url{https://doi.org/10.25921/x46c-3620}



 
\bibitem{evensen1}
{\sc Evensen, G.}~(1994). Sequential data assimilation with a nonlinear quasi-geostrophic model using Monte Carlo methods to forecast error statistics. \emph{J. Geophys. Res.}, {\bf 99}(C5):10143--10162.

\bibitem{evensen2}
{\sc Evensen, G., \& Van Leeuwen, P. J.}~(2000). An ensemble Kalman smoother for nonlinear dynamics. \emph{Mon. Wea. Rev.}, {\bf 128}: 1852--1867.

\bibitem{farchi}
{\sc Farchi, A. \& Bocquet, M.}~(2019). Review article: Comparison of local particle filters and new implementations. \emph{Nonlin. Processes Geophys.}, {\bf 25}: 765--807.

\bibitem{frolov_2024}
{\sc Frolov, S.}, {\sc Shlyaeva, A.}, {\sc Huang, W.}, {\sc Sluka, T.}, {\sc Draper, C.}, {\sc Huang, B.}, {\sc Martin, C.}, {\sc Elless, T.}, {\sc Bhargava, K.}, and {\sc Whitaker, J.}~(2024). Local volume solvers for {Earth} system data assimilation: {I}mplementation in the framework for {Joint Effort for Data Assimilation Integration}. \emph{Journal of Advances in Modeling Earth Systems}, {\bf 16}: e2023MS003692.

\bibitem{gaspari}
{\sc Gaspari, G. \& Cohn, S.E.}~(1999). Construction of correlation functions in two and three dimensions. \emph{Q.J.R. Meteorol. Soc.}, {\bf 125}(554): 723--757.

\bibitem{ghil}
{\sc Ghil, M. \& Malanotte-Rizzoli, P.}~(1991). Data assimilation in meteorology and oceanography. \emph{Adv. Geophys.}, {\bf 33}: 141--266.

\bibitem{PF1}
{\sc Gordon, N.J., Salmond, D.J. \& Smith, A.F.M.}~(1993). Novel approach to nonlinear/non-Gaussian Bayesian state estimation. IEE-Proceedings-F, {\bf 140}(2):107--113.

\bibitem{greybush}
{\sc Greybush, S. J., Kalnay, E., Miyoshi, T., Ide, K. \& Hunt, B. R.}~(2011). Balance and ensemble Kalman filter localization techniques. \emph{Mon. Wea. Rev.}, {\bf 139}: 511--522.

\bibitem{hamill}
{\sc Hamill, T. M., Whitaker, J. S. \&  Snyder, C.}~(2001). Distance-Dependent Filtering of Background Error Covariance Estimates in an Ensemble Kalman Filter. \emph{Mon. Wea. Rev.}, {\bf 129}: 2776--2790.

\bibitem{houtekamer}
{\sc Houtekamer, P. L., \& Mitchell, H. L.}~(1998). Data assimilation using an ensemble Kalman filter technique. \emph{Mon. Wea. Rev.}, {\bf 126}:796--811.

\bibitem{houtekamer01}
{\sc Houtekamer, P. L. \& Mitchell, H. L.}~(2001). A sequential ensemble Kalman filter for atmospheric data assimilation. \emph{Mon. Wea. Rev.}, {\bf 129}: 123--137.

\bibitem{hunt_2007}
{\sc Hunt, B.~R.}, {\sc Kostelich, E.~J.}, and {\sc Szunyogh, I.}~(2007).
Efficient data assimilation for spatiotemporal chaos: A local ensemble transform Kalman filter.
\emph{Physica D: Nonlinear Phenomena}, {\bf 230}(1--2): 112--126.

\bibitem{jazwinski_1970}
{\sc Jazwinski, A.~H.}~(1970). \emph{Stochastic Processes and Filtering Theory}. Volume 64, Mathematics in Science and Engineering series. Academic Press, New York. ISBN 978-0-12-381550-7.

\bibitem{kalnay_mote_da_2024}
{\sc Kalnay, E.}, {\sc Mote, S.}, and {\sc Da, C.}~(2024). \emph{Earth System Modeling, Data Assimilation and Predictability: Atmosphere, Oceans, Land and Human Systems}. 2nd ed. Cambridge University Press. ISBN 978-1-107-00900-4 (hardback) / 978-1-107-40146-4 (paperback). Available from Cambridge University Press.

\bibitem{kantas}
{\sc Kantas, N., Beskos, A., \&  Jasra, A.}~(2014). Sequential Monte Carlo for inverse problems: a case study for the Navier Stokes equation. \emph{SIAM/ASA JUQ}, {\bf 2}, 464--489.

\bibitem{liang_2025}
{\sc Liang, S., Tran, H., Bao, F., Chipilski, H. G., van Leeuwen, P. J. \& Zhang, G.}~(2025). Ensemble Score Filter with Image Inpainting for Data Assimilation in Tracking Surface Quasi-Geostrophic Dynamics with Partial Observations. \emph{arXiv preprint} arXiv:2501.12419.

\bibitem{noaa}
{\sc Lumpkin, R. \& Centurioni, L.} ~(2019). Global Drifter Program quality-controlled 6-hour interpolated data from ocean surface drifting buoys. \emph{NOAA National Centers for Environmental Information.} Dataset. \url{https://doi.org/10.25921/7ntx-z961}. Accessed Aug. 2024.


\bibitem{martin}
{\sc Martin, J. S.,  Jasra, A., \&  McCoy}, E.~(2013). Inference for a class of partially observed point process models. \emph{Ann. Inst. Stat. Math.}, {\bf 65}: 413--437.


\bibitem{metzger2014}
{\sc Metzger, E. J., Smedstad, O. M., Thoppil, P. G., Hurlburt, H. E., Cummings, J. A., Wallcraft, A. J., Zamudio, L., Franklin, D. S., Posey, P. G., Phelps, M. W., Hogan, P. J., Bub, F. L. \& DeHaan, C. J.}~(2014). US Navy operational global ocean and Arctic ice prediction systems. \emph{Oceanography} {\bf 27}(3): 32--43.

\bibitem{miyoshi}
{\sc Miyoshi, T. \& Kunii, M.}~(2012). The Local Ensemble Transform Kalman Filter with the Weather Research and Forecasting Model: Experiments with Real Observations. \emph{Pure Appl. Geophys.} {\bf 169}: 321--333.


\bibitem{tsunami}
{\sc Mulia, I. E., Inazu, D., Waseda, T. \& Gusman, A. R.}~(2017). Preparing for the future Nankai Trough tsunami: A data assimilation and inversion analysis from various observational systems. \emph{Journal of Geophysical Research: Oceans}, {\bf 122}(10): 7924--7937.

\bibitem{HMC}
{\sc Neal, R. M.}~(2011). MCMC using Hamiltonian dynamics. In \emph{Handbook of Markov Chain Monte Carlo}, eds.\ S. Brooks, A. Gelman, G. L. Jones, X.-L. Meng, Chapman \& Hall/CRC, pp.\ 113--162.


\bibitem{ott}
{\sc Ott, E., Hunt, B.R., Szunyogh, I., Zimin, A.V., Kostelich, E.J., Corazza, M., Kalnay, E., Patil, D.J. \& Yorke, J.A.}~(2004). A local ensemble Kalman filter for atmospheric data assimilation. \emph{Tellus A}, {\bf 56}: 415--428.

\bibitem{pires}
{\sc Pires, C. A., Talagrand, O. \& Bocquet, M.}~(2010). Diagnosis and impacts of non-Gaussianity of innovations in data assimilation. \emph{Physica D}, {\bf 239}:1701--1717.


\bibitem{poterjoy}
{\sc Poterjoy, J.}~(2016). A Localized Particle Filter for High-Dimensional Nonlinear Systems.\emph{Mon. Wea. Rev.}, {\bf 144}: 59--76.

\bibitem{swot}
{\sc Rodriguez, E., Fernandez, D. E., Peral, E., Chen, C. W., De Bleser, J.-W., \& Williams, B.}~(2017). ``Wide-swath altimetry: a review'', in \emph{Satellite Altimetry Over Oceans and Land Surfaces}, eds D. Stammer and A. Cazenave (Boca Raton, FL: CRC Press), 71--112.

\bibitem{MALA}
{\sc Roberts, G. O. \& Tweedie, R. L.}~(1996). Exponential convergence of Langevin distributions and their discrete approximations. \emph{Bernoulli}, {\bf 2}(4): 341--363.

\bibitem{laggedPF}
{\sc Ruzayqat, H. , Er-Raiy,  A., Beskos, A., Crisan, D., Jasra, A. \& Kantas, N.}~(2022). A Lagged Particle Filter for Stable Filtering of certain High-Dimensional State-Space Models. \emph{SIAM/ASA JUQ} {\bf 10}:1130--1161.

\bibitem{smcmc}
{\sc Ruzayqat, H.,  Beskos, A., Crisan, D.,  Jasra, A., \& Kantas, N.}~(2024). 
Sequential Markov chain Monte Carlo for Lagrangian data assimilation with applications to unknown data locations. \emph{Q. J. R. Meteorol. Soc.}, {\bf 150}(761):2418--2439.


\bibitem{schraff_2016}
{\sc Schraff, C.}, {\sc Reich, H.}, {\sc Rhodin, A.}, {\sc Schomburg, A.}, 
{\sc Stephan, K.}, {\sc Peri\'a\~nez, A.}, and {\sc Potthast, R.}~(2016).
Kilometre-scale ensemble data assimilation for the {COSMO} model ({KENDA}).
\emph{Quarterly Journal of the Royal Meteorological Society}, {\bf 142}: 1453--1472.

\bibitem{septier}
{\sc Septier, F. \& Peters, G. W.}~ (2016). Langevin and Hamiltonian Based Sequential MCMC for Efficient Bayesian Filtering in High-Dimensional Spaces. \emph{IEEE J. Sel. Topics Signal Process.}, {\bf 10}(2): 312--327.

\bibitem{storto}
{\sc Storto, A., Alvera-Azc\'{a}rate, A., Balmaseda, M. A., Barth, A., Chevallier, M., Counillon, F., Domingues, C. M., Drevillon, M., Drillet, Y. \& Forget, G.}~(2019). Ocean reanalyses: recent advances and unsolved challenges. \emph{Frontiers in Marine Science}, {\bf 6}(418).

\bibitem{leeuwen}
{\sc Van Leeuwen, P.J.}~(2009). Particle Filtering in Geophysical Systems. \emph{Mon. Wea. Rev.}, {\bf 137}: 4089--4114.



\bibitem{vetra}
{\sc Vetra-Carvalho, S., Van Leeuwen, P. J., Nerger, L., Barth, A., Altaf, M. U., Brasseur, P., Kirchgessner, P. \& Beckers, J.-M.}~(2018). State-of-the-art stochastic data assimilation methods for high-dimensional non-Gaussian problems. \emph{Tellus A: Dynamic Meteorology and Oceanography}, {\bf 70}(1): 1--43.

\bibitem{whitaker02}
{\sc Whitaker, J. S. \& Hamill, T. M.}~(2002). Ensemble data assimilation without perturbed observations. \emph{Mon. Wea. Rev.}, {\bf 130}: 1913--1924.


\bibitem{nasa}
\url{https://swot.jpl.nasa.gov/}






\end{thebibliography}
\end{document}